\documentclass[journal]{IEEEtran}

\usepackage{cite}
\usepackage{amsmath,amssymb,amsfonts}
\usepackage{amsthm}
\usepackage{bm}
\usepackage{graphicx}
\usepackage{xcolor}
\usepackage{booktabs}
\usepackage{multirow}
\usepackage{microtype}
\usepackage{url}

\allowdisplaybreaks
\makeatletter
\g@addto@macro\normalsize{%
  \setlength{\abovedisplayskip}{4.5pt}%
  \setlength{\belowdisplayskip}{4.5pt}%
  \setlength{\abovedisplayshortskip}{4.5pt}%
  \setlength{\belowdisplayshortskip}{4.5pt}%
}
\renewenvironment{cases}{%
  \matrix@check\cases\env@cases
}{%
  \endarray\right.%
}
\def\env@cases{%
  \let\@ifnextchar\new@ifnextchar
  \left\lbrace
  \def\arraystretch{1.08}%
  \array{@{}l@{\;\;}l@{}}%
}
\makeatother
\newenvironment{ncases}{\left\lbrace\!\def\arraystretch{1.08}\begin{array}{@{}l@{\,}l@{}}}{\end{array}\right.}
\newcommand{\ccond}[1]{\text{\footnotesize $#1$}}
\newcommand{\tcond}[1]{\text{\footnotesize #1}}
\newcommand{\sublbl}[2]{\noindent\makebox[\linewidth][l]{\hspace*{#1\linewidth}\makebox[0pt][c]{\scriptsize #2}}}

\theoremstyle{plain}
\newtheorem{theorem}{Theorem}
\newtheorem{example}{Example}

\begin{document}

\title{Double-Bounce Multi-Type Landmarks Batch Radio SLAM via MCMC and Graph-Based Optimization}

\author{Zhibo~Duan, Jiancun~Fan, Yulin~Fu, Shiyu~Zhai, and~Jie~Luo%
\thanks{\strut\\\strut}%
\thanks{Zhibo Duan, Jiancun Fan, Yulin Fu, Shiyu Zhai, and Jie Luo are with the School of Information and Communications Engineering, Xi'an Jiaotong University, Xi'an 710049, China (Corresponding author: Jiancun Fan, e-mail: fanjc0114@gmail.com).}%
}

\maketitle

\begin{abstract}
Radio simultaneous localization and mapping (SLAM) exploits multipath geometry to jointly estimate a mobile receiver's trajectory and environment map. Most existing radio SLAM methods rely solely on line-of-sight (LoS) and single-bounce paths while discarding double-bounce reflections, losing geometric constraints and occluded-landmark sensing under deep occlusion. Although some snapshot and sequential filtering methods exploit double-bounce paths, they struggle to aggregate intermittent detections across scans and typically model only one reflector type. By contrast, batch SLAM jointly estimates the full trajectory and map to aggregate intermittent observations and exploit cross-scan correlations; however, incorporating double-bounce paths across planar walls and point scatterers introduces inter-landmark coupling, occluded-landmark identifiability issues, and cross-scan data association (DA) uncertainty. To address these challenges, we develop a double-bounce multi-type batch radio SLAM framework within random finite set (RFS) theory, formulating a unified multipath observation model and batch joint posterior, and deriving geometric identifiability conditions for occluded landmarks. To resolve cross-scan DA and type uncertainty under coupling, we design a pre-clustering-initialized two-stage Markov chain Monte Carlo (MCMC) sampler. Alternating with the sampler, a conditional graph-based SLAM (GraphSLAM) solver incorporating double-bounce constraints jointly optimizes the trajectory and coupled landmark states, and cross-sample marginalization outputs the global trajectory and multi-type Poisson multi-Bernoulli (PMB) map. Ray-tracing simulations and 60~GHz indoor channel measurements show that our method outperforms snapshot double-bounce, single-bounce batch, and sequential double-bounce filtering baselines, closely approaches the Cram\'er-Rao lower bound (CRLB) under deep occlusion, and recovers single-bounce-occluded landmarks.
\end{abstract}


\section{Introduction} \label{sec:intro}

\IEEEPARstart{R}{adio} simultaneous localization and mapping (radio SLAM) is an important research topic in integrated sensing and communications (ISAC) and digital twins, with valuable applications in indoor navigation, autonomous driving, and warehouse robotics~\cite{DiTaranto2014SPM,Witrisal2016SPM,Ge2023ISAC,Ge2025ComMag}. Conventional radio localization primarily relies on the line-of-sight (LoS) path while treating non-line-of-sight (NLoS) multipath components as interference. With the development of wide-bandwidth millimeter-wave (mmWave) systems and multiple-input multiple-output (MIMO) antenna arrays, transceivers can extract high-resolution geometric channel parameters such as time of arrival (ToA), angle of arrival (AoA), and angle of departure (AoD), providing a basis for multipath-assisted localization of the mobile receiver and mapping of its surrounding environment~\cite{Mendrzik2019JSTSP,Li2019TWC,Ge2024TVT,RastorguevaFoi2024JSAC,Liang2025TSP}.

Existing radio SLAM methods fall into two main categories: \emph{filtering-based} and \emph{batch-based} methods. Filtering-based methods recursively estimate the sensor state and environment map from current and past observations, including probability hypothesis density (PHD), Poisson multi-Bernoulli mixture (PMBM), and related filters grounded in random finite set (RFS) theory~\cite{GarciaFernandez2018TAES,Kim2020TWC,Ge2020Sensors,Ge2022JSAC,Kaltiokallio2024TRO,Kim2024TSP}, as well as belief propagation (BP-SLAM) filters running message passing on factor graphs~\cite{Mendrzik2019JSTSP,Leitinger2019TWC}. Because filtering-based methods use only information up to the current time step without leveraging future measurements to update past poses, early data association (DA) errors are difficult to correct retrospectively, and cross-time correlations between the sensor trajectory and map features are difficult to propagate. By contrast, batch-based methods, such as GraphSLAM~\cite{Thrun2006IJRR}, jointly estimate the entire trajectory and map over the full measurement sequence, combining geometric constraints across time to improve global consistency and better exploit trajectory-map correlations. Recently,~\cite{Ge2025TSP} combined RFS modeling, Markov chain Monte Carlo (MCMC) DA sampling~\cite{Xia2024TAES}, and GraphSLAM into a batch Graph PMBM-SLAM method, reducing the reliance of conventional graph-based SLAM backends on heuristic DA gating.

However, most existing radio SLAM methods, including~\cite{Ge2022JSAC,Leitinger2019TWC,Ge2025TSP}, exploit only single-bounce (1-bounce) paths and discard double-bounce (2-bounce) and higher-order reflections as clutter. Indoor channel measurements show that 2-bounce reflections carry appreciable energy~\cite{Steinboeck2016TAP,Xu2025AWPL}: exploiting them adds geometric constraints and enables sensing of 1-bounce-occluded landmarks. Existing 2-bounce studies fall into two categories. One line of work develops snapshot or iterative geometric localization and mapping methods~\cite{RastorguevaFoi2024JSAC,Nazari2023TVT,Li2023WCL,Zhang2026TWC}, which process single-epoch measurements independently and cannot fuse multi-scan motion priors under intermittent observations and clutter. The other incorporates master virtual anchors (MVAs) or surface feature vectors into BP-SLAM filters, fusing 1-bounce and 2-bounce reflections via ray tracing~\cite{Leitinger2023TSP} or multipath amplitude statistics~\cite{Li2022TWC,Li2026TWC}, though mostly modeling planar walls (virtual anchors, VAs). Meanwhile, multi-type radio SLAM that jointly models planar walls (VAs) and point scatterers (reflection points, RPs, e.g., pillars or edges) has also been studied~\cite{Kim2020TWC,Kim2022TVT,Li2024ICC,Zhai2025TWC}, yet remains largely confined to 1-bounce paths and sequential filtering. Batch radio SLAM jointly handling 2-bounce paths and multi-type landmarks remains largely unexplored.

However, integrating 2-bounce paths and multi-type landmarks into a batch framework is not a straightforward extension of existing single-bounce batch or double-bounce filtering methods, as it requires addressing two key challenges:
\begin{itemize}
    \item \emph{Inter-landmark coupling and identifiability:} In a single-bounce batch model, each landmark generates only its own 1-bounce measurement cluster, so landmark clusters are conditionally independent given the trajectory and their marginal likelihoods integrate separately per cluster. In a double-bounce multi-type setting, however, a landmark can participate in both its 1-bounce path and 2-bounce paths formed with other landmarks (VA-VA, VA-RP, and RP-VA), coupling landmark positions and reflection types across clusters, breaking cluster-wise conditional independence given the trajectory, and raising landmark identifiability questions under 2-bounce observations.
    \item \emph{Cross-scan association of sparse, intermittent 2-bounce paths:} Owing to finite reflector dimensions and dual-reflection loss, 2-bounce paths often appear sparsely and intermittently over time. Whereas sequential filters tend to prematurely prune low-existence candidates from isolated detections after consecutive missed-detection updates, retaining the full measurement sequence in batch processing still requires capturing cross-scan correlations to reliably aggregate distant, fragmented 2-bounce detections and distinguish them from clutter.
\end{itemize}

To address these challenges, this paper develops a double-bounce multi-type batch radio SLAM algorithm within the RFS framework that alternates double-bounce DA and landmark type sampling with conditional graph-based optimization to jointly estimate the full trajectory and environment map. The main contributions are summarized as follows:
\begin{enumerate}
    \item \emph{Double-bounce multi-type batch SLAM modeling and identifiability analysis:} We formulate a multi-type radio measurement model covering LoS, 1-bounce, and 2-bounce paths, analyze geometric identifiability conditions for 2-bounce landmark pairs when 1-bounce observations are absent, and derive the batch joint posterior distribution of the sensor trajectory and multi-type map within the RFS framework.
    \item \emph{Double-bounce data association and landmark type sampling:} To resolve DA uncertainty across coupled 1-bounce and 2-bounce paths, we design a Gibbs and split-merge DA and type sampler under marginal likelihood approximations, using trajectory-based pre-clustering to speed convergence.
    \item \emph{Conditional double-bounce GraphSLAM and cross-sample fusion:} Conditioned on sampled DAs and landmark types, we formulate a GraphSLAM solver incorporating 2-bounce geometric constraints to jointly estimate the sensor trajectory and multi-type landmark states, together with a post-processing fusion method producing the trajectory posterior and multi-type PMB map across samples.
\end{enumerate}

Evaluations on ray-tracing simulations and 60~GHz indoor measurements show that the proposed method exploits 2-bounce paths to improve localization and mapping accuracy and recover 1-bounce-occluded landmarks. In the remainder of this paper, Section~\ref{sec:system} presents the system and double-bounce measurement models; Section~\ref{sec:framework} formulates the batch joint posterior and alternating framework; Section~\ref{sec:da_sampling} details DA weight computation, MCMC and type sampling, and pre-clustering; Section~\ref{sec:graphslam} presents conditional GraphSLAM and cross-sample PMB fusion; Section~\ref{sec:results} provides simulation and experimental results; and Section~\ref{sec:conclusion} concludes the paper.

\emph{Notations:} Scalars are denoted by italic letters ($x$), column vectors by lowercase boldface ($\boldsymbol{x}$), matrices by uppercase boldface ($\boldsymbol{X}$), and sets by calligraphic letters ($\mathcal{X}$). $|\mathcal{X}|$ denotes set cardinality, and $\biguplus$ denotes disjoint set union. $(\cdot)^{\mathsf{T}}$, $\|\cdot\|$, $\operatorname{tr}(\cdot)$, and $|\boldsymbol{X}|$ denote transpose, Euclidean norm, trace, and determinant, respectively. $[\boldsymbol{x}]_i$ denotes the $i$-th entry of $\boldsymbol{x}$, $\mathbb{I}(\cdot)$ is the indicator function, $\mathrm{wrap}(\alpha) \triangleq \operatorname{atan2}(\sin\alpha, \cos\alpha) \in (-\pi, \pi]$ denotes angle wrapping, and $\mathcal{N}(\boldsymbol{x}; \boldsymbol{\mu}, \boldsymbol{\Sigma})$ denotes a multivariate Gaussian density with mean $\boldsymbol{\mu}$ and covariance $\boldsymbol{\Sigma}$.

\section{System Model and Double-Bounce Radio Measurement Model} \label{sec:system}

This section presents the sensor motion model, multi-type RFS map prior, six-path observation models, and detection/clutter models.

\begin{figure}[!t]
    \centering
    \includegraphics[width=\columnwidth]{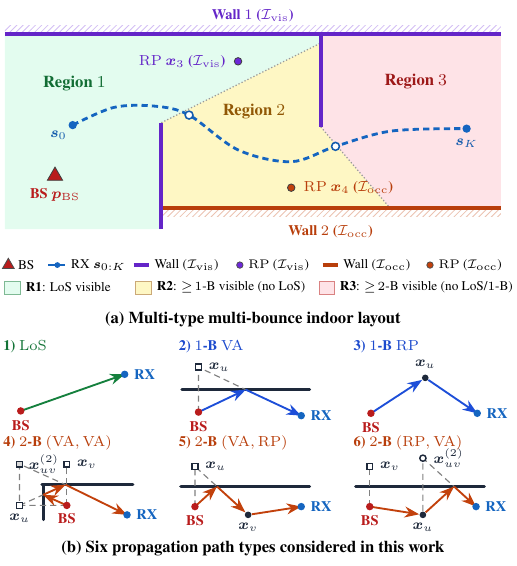}
    \caption{Multi-type double-bounce indoor radio SLAM scenario and propagation path models. (a) Indoor layout comprising a fixed BS, mobile RX, 1-bounce visible landmarks ($\mathcal{I}_{\mathrm{vis}}$), 1-bounce-occluded landmarks ($\mathcal{I}_{\mathrm{occ}}$), and three regions partitioned by minimum visible path order (R1: LoS visible; R2: No LoS, 1-bounce \& above visible; R3: No LoS, 2-bounce \& above visible). (b) Six propagation path geometries considered in this work: 0-bounce LoS, two 1-bounce paths (VA, RP), and three 2-bounce paths (VA-VA, VA-RP, RP-VA), where $\boldsymbol{x}_{uv}^{(2)} = \mathcal{M}(\boldsymbol{x}_u; \boldsymbol{x}_v)$ is the second-order virtual anchor across wall $v$.}
    \label{fig:scenario}
\end{figure}

\subsection{System Configuration and Sensor Motion Model}

Consider the indoor bistatic mmWave MIMO scenario in Fig.~\ref{fig:scenario}(a), where a fixed base station (BS) at known position $\boldsymbol{p}_{\mathrm{BS}} \in \mathbb{R}^2$ and orientation $\alpha_{\mathrm{BS}} \in (-\pi, \pi]$ transmits wireless signals, and a mobile receiver (RX) receives multipath signals reflected and scattered by the indoor environment~\cite{Leitinger2019TWC,Zhang2026TWC,Li2026TWC}. We assume that the BS and RX are located at the same height, with wireless signals propagating in the 2D horizontal plane. At discrete time steps $k \in \{0, 1, \dots, K\}$, the RX state is $\boldsymbol{s}_k = [\boldsymbol{p}_k^{\mathsf{T}}, \alpha_k]^{\mathsf{T}} = [x_k, y_k, \alpha_k]^{\mathsf{T}} \in \mathbb{R}^2 \times (-\pi, \pi]$, comprising 2D position and heading. The initial state follows a Gaussian prior $f(\boldsymbol{s}_0) = \mathcal{N}(\boldsymbol{s}_0; \boldsymbol{\epsilon}_{0,\mathrm{prior}}, \boldsymbol{P}_{0,\mathrm{prior}})$, and consecutive states evolve via a first-order Markov transition density:
\begin{align}
f(\boldsymbol{s}_k|\boldsymbol{s}_{k-1}) = \mathcal{N}\!\left(\boldsymbol{s}_k;\, \boldsymbol{v}(\boldsymbol{s}_{k-1}, \boldsymbol{u}_{k-1}),\, \boldsymbol{Q}_{k-1}\right), \label{eq:motion_density}
\end{align}
where $\boldsymbol{v}(\boldsymbol{s}_{k-1}, \boldsymbol{u}_{k-1})$ denotes the known state transition function driven by odometry input $\boldsymbol{u}_{k-1} = [\Delta x_{k-1}, \Delta y_{k-1}, \Delta\alpha_{k-1}]^{\mathsf{T}}$ (explicit form and Jacobian in Appendix~A of the Supplementary Material, hereafter Appendices~A--E), and $\boldsymbol{Q}_{k-1}$ is the process noise covariance.

Both BS and RX use antenna arrays for AoA/AoD estimation with synchronized clocks (clock offsets can be jointly tracked as in~\cite{Ge2022JSAC,Venus2024TWC,Fascista2025JSTSP}). At scan $k \ge 1$, a channel estimator~\cite{RastorguevaFoi2024JSAC,Li2022TWC} extracts $M_k$ multipath measurements $\mathcal{Z}_k = \{\boldsymbol{z}_k^1, \dots, \boldsymbol{z}_k^{M_k}\}$, where each $\boldsymbol{z}_k^m = [d_k^m, \theta_k^m, \phi_k^m]^{\mathsf{T}}$ comprises propagation distance $d_k^m$, RX local AoA $\theta_k^m$, and BS local AoD $\phi_k^m$.

\subsection{Multi-Type Landmark and RFS Map Modeling}

Static indoor environments typically contain both planar reflectors (e.g., walls and cabinets) and small point scatterers (e.g., pillars and edges)~\cite{Kim2020TWC,Kim2022TVT,Li2024ICC,Wen2021TWC}. Each landmark $i$ is represented as an augmented state $\tilde{\boldsymbol{x}}_i = (\boldsymbol{x}_i, \psi_i)$ with position parameter $\boldsymbol{x}_i \in \mathbb{R}^2$ and type label $\psi_i \in \{\mathrm{VA}, \mathrm{RP}\}$. For a point scatterer ($\psi_i = \mathrm{RP}$), $\boldsymbol{x}_i$ directly represents its 2D physical coordinate. For a planar wall ($\psi_i = \mathrm{VA}$), following the image-source model~\cite{Witrisal2016SPM}, the wall is parameterized by the first-order mirror image (virtual anchor) $\boldsymbol{x}_i$ of $\boldsymbol{p}_{\mathrm{BS}}$ across the wall line. The mirror reflection of a point $\boldsymbol{a} \in \mathbb{R}^2$ across wall $\boldsymbol{x}_i$ is then given by the Householder reflection operator
\begin{align}
\mathcal{M}(\boldsymbol{a}; \boldsymbol{x}_i) = \boldsymbol{x}_i + \boldsymbol{H}(\boldsymbol{d}_i)(\boldsymbol{a} - \boldsymbol{p}_{\mathrm{BS}}), \label{eq:mirror_op}
\end{align}
where $\boldsymbol{H}(\boldsymbol{d}_i) \triangleq \boldsymbol{I}_2 - 2\boldsymbol{n}_i\boldsymbol{n}_i^{\mathsf{T}}$, $\boldsymbol{d}_i \triangleq \boldsymbol{x}_i - \boldsymbol{p}_{\mathrm{BS}}$, and $\boldsymbol{n}_i \triangleq \boldsymbol{d}_i / \|\boldsymbol{d}_i\|$ is the unit wall normal.

Since both the number and states of landmarks are unknown, the static map is modeled as an RFS $\mathcal{X} = \{\tilde{\boldsymbol{x}}_1, \dots, \tilde{\boldsymbol{x}}_n\} = \mathcal{X}_{\mathrm{U}} \biguplus \mathcal{X}_{\mathrm{D}}$, partitioned over the batch horizon into undetected landmarks $\mathcal{X}_{\mathrm{U}}$ and landmarks detected at least once $\mathcal{X}_{\mathrm{D}}$. The map prior follows a marked Poisson point process (PPP) with intensities $\lambda_{\mathrm{VA}}(\boldsymbol{x}) \ge 0$ and $\lambda_{\mathrm{RP}}(\boldsymbol{x}) \ge 0$~\cite{GarciaFernandez2018TAES,Ge2025TSP,Kim2022TVT}:
\begin{align}
f_{\mathrm{P}}(\mathcal{X}) = e^{-\sum_{\psi \in \{\mathrm{VA}, \mathrm{RP}\}} \int_{\mathbb{R}^2} \lambda_\psi(\boldsymbol{x})\,\mathrm{d}\boldsymbol{x}} \prod_{(\boldsymbol{x}, \psi) \in \mathcal{X}} \lambda_\psi(\boldsymbol{x}). \label{eq:ppp_prior}
\end{align}

In the RFS framework, the detected map is a disjoint union of potential landmark subsets $\mathcal{X}_{\mathrm{D}} = \biguplus_{i=1}^N \mathcal{Y}_i$ ($|\mathcal{Y}_i| \le 1$), where each independent $\mathcal{Y}_i$ is a Bernoulli RFS component with existence probability $r_i \in [0, 1]$ ($\mathcal{Y}_i = \emptyset$ with probability $1 - r_i$, or $\{\tilde{\boldsymbol{x}}_i\}$ with density $r_i f_i(\tilde{\boldsymbol{x}}_i)$), forming a PMBM/PMB map with the undetected PPP $\mathcal{X}_{\mathrm{U}}$~\cite{GarciaFernandez2018TAES,Ge2025TSP}.

\subsection{Geometric Measurement Models for Six Path Types}

We model propagation paths up to second order. Owing to curved-wavefront divergence at small scatterers, double point scattering (RP-RP) incurs large signal attenuation~\cite{Witrisal2016SPM,Steinboeck2016TAP}, with echo power typically falling below the detection threshold, and is therefore omitted in this work. Accordingly, we define the 1-bounce and 2-bounce path type sets as $\mathcal{T}_1 \triangleq \{\mathrm{VA}, \mathrm{RP}\}$ and $\mathcal{T}_2 \triangleq \{(\mathrm{VA}, \mathrm{VA}), (\mathrm{VA}, \mathrm{RP}), (\mathrm{RP}, \mathrm{VA})\}$, forming six propagation path types $\mathcal{H} \triangleq \{\mathrm{LoS}\} \cup \mathcal{T}_1 \cup \mathcal{T}_2$ across bounce orders $0, 1, 2$ (Fig.~\ref{fig:scenario}(b)). For each path type $\eta \in \mathcal{H}$, let $\boldsymbol{h}_\eta(\cdot) \in \mathbb{R}^3$ denote the measurement function~\cite{Ge2025TSP,Zhang2026TWC,Li2026TWC}. Under zero-mean additive Gaussian measurement noise with covariance $\boldsymbol{R}_\eta$, the measurement likelihood functions (LHFs) are expressed by bounce order as follows.
\par 1)\hspace{\IEEEiednormlabelsep}\emph{LHF for 0-Bounce LoS Paths ($\eta = \mathrm{LoS}$):}
\begin{align}
f_0(\boldsymbol{z}_k^m|\boldsymbol{s}_k) = \mathcal{N}\!\left(\boldsymbol{z}_k^m;\, \boldsymbol{h}_{\mathrm{LoS}}(\boldsymbol{s}_k),\, \boldsymbol{R}_{\mathrm{LoS}}\right). \label{eq:lhf_los}
\end{align}
\par 2)\hspace{\IEEEiednormlabelsep}\emph{LHF for 1-Bounce Paths ($\eta = \psi_i \in \mathcal{T}_1$):}
\begin{align}
f_1(\boldsymbol{z}_k^m|\tilde{\boldsymbol{x}}_i, \boldsymbol{s}_k) = \mathcal{N}\!\left(\boldsymbol{z}_k^m;\, \boldsymbol{h}_{\psi_i}(\boldsymbol{x}_i, \boldsymbol{s}_k),\, \boldsymbol{R}_{\psi_i}\right). \label{eq:lhf_sb}
\end{align}
\par 3)\hspace{\IEEEiednormlabelsep}\emph{LHF for 2-Bounce Paths ($\eta = (\psi_u, \psi_v) \in \mathcal{T}_2$):}
\begin{align}
f_2(\boldsymbol{z}_k^m|\tilde{\boldsymbol{x}}_u, \tilde{\boldsymbol{x}}_v, \boldsymbol{s}_k) = \mathcal{N}\!\left(\boldsymbol{z}_k^m;\, \boldsymbol{h}_\eta(\boldsymbol{x}_u, \boldsymbol{x}_v, \boldsymbol{s}_k),\, \boldsymbol{R}_\eta\right). \label{eq:lhf_db}
\end{align}
Using~\eqref{eq:mirror_op}, all three 2-bounce measurement functions $\boldsymbol{h}_{\psi_u,\psi_v}$ can be expressed directly in terms of $(\boldsymbol{x}_u, \boldsymbol{x}_v)$ so that 1-bounce and 2-bounce paths share the same landmark state parameters; explicit observation formulas and Jacobians for all six path types are given in Appendix~A.

\subsection{Double-Bounce Detection and Clutter Statistical Model}

Due to path loss, blockage, thermal noise, and estimation errors, each scan $\mathcal{Z}_k$ may contain missed detections and clutter. Under the point-target model, given $(\boldsymbol{s}_k, \mathcal{X})$, each valid path generates at most one measurement per scan: the 0-bounce LoS detection probability is $p_{\mathrm{D},0}(\boldsymbol{s}_k)$; the 1-bounce detection probability of $\tilde{\boldsymbol{x}}_i \in \mathcal{X}$ is $p_{\mathrm{D},1}(\tilde{\boldsymbol{x}}_i, \boldsymbol{s}_k)$; the 2-bounce detection probability of an ordered pair $(\tilde{\boldsymbol{x}}_u, \tilde{\boldsymbol{x}}_v)$ with $(\psi_u, \psi_v) \in \mathcal{T}_2$ is $p_{\mathrm{D},2}(\tilde{\boldsymbol{x}}_u, \tilde{\boldsymbol{x}}_v, \boldsymbol{s}_k)$; clutter follows an independent PPP with intensity $c(\boldsymbol{z}) \ge 0$. Thus, a landmark may generate 1-bounce and multiple 2-bounce measurements, and landmarks with blocked 1-bounce paths remain observable via 2-bounce paths.

\section{Problem Formulation and Double-Bounce Batch SLAM Framework} \label{sec:framework}

This section derives the batch joint posterior distribution of the sensor trajectory and multi-type map and outlines the alternating MCMC-GraphSLAM framework.

\subsection{Joint Posterior Formulation}

\subsubsection{Trajectory and Multi-Type Map Posterior}

Let $\mathcal{Z}_{1:K} = (\mathcal{Z}_1, \dots, \mathcal{Z}_K)$ denote the measurement batch and $\boldsymbol{s}_{0:K} = [\boldsymbol{s}_0^{\mathsf{T}}, \dots, \boldsymbol{s}_K^{\mathsf{T}}]^{\mathsf{T}}$ the trajectory. By Bayes' rule,~\eqref{eq:motion_density}, and conditional scan independence, the joint posterior density (with map prior $f(\mathcal{X}) = f_{\mathrm{P}}(\mathcal{X})$ in~\eqref{eq:ppp_prior}) is
\begin{align}
f(\boldsymbol{s}_{0:K}, \mathcal{X}|\mathcal{Z}_{1:K}) &= \frac{1}{f(\mathcal{Z}_{1:K})}\,f(\boldsymbol{s}_0)f(\mathcal{X}) \notag\\*
&\quad\times \prod_{k=1}^K f(\boldsymbol{s}_k|\boldsymbol{s}_{k-1})g(\mathcal{Z}_{1:K}|\boldsymbol{s}_{1:K}, \mathcal{X}), \label{eq:bayes_joint}
\end{align}
where $g(\mathcal{Z}_{1:K}|\boldsymbol{s}_{1:K}, \mathcal{X}) = \prod_{k=1}^K g(\mathcal{Z}_k|\boldsymbol{s}_k, \mathcal{X})$ is the batch RFS likelihood and $f(\mathcal{Z}_{1:K})$ the normalizing constant.

Under double-bounce propagation, 1-bounce and 2-bounce clusters share landmarks, and 2-bounce paths depend on the reflection sequence between landmarks. Under global DA hypothesis $j \in \mathbb{J}$, we assign candidate landmarks an index set $\mathcal{I}^j = \mathcal{I}_{\mathrm{vis}}^j \biguplus \mathcal{I}_{\mathrm{occ}}^j$, where $\mathcal{I}_{\mathrm{vis}}^j \triangleq \{1, \dots, N_{\mathrm{vis}}^j\}$ and $\mathcal{I}_{\mathrm{occ}}^j \triangleq \{N_{\mathrm{vis}}^j + 1, \dots, N^j\}$ ($N^j = N_{\mathrm{vis}}^j + N_{\mathrm{occ}}^j$) are each indexed internally by the global index of their earliest measurement (ordered by scan $k$ and then within-scan index $m$) to remove label symmetry: $\mathcal{I}_{\mathrm{vis}}^j$ indexes \emph{visible candidate landmarks} with non-empty 1-bounce clusters, whereas $\mathcal{I}_{\mathrm{occ}}^j$ indexes \emph{occluded candidate landmarks} with empty 1-bounce clusters ($\mathcal{Z}_v^{j,(1)} = \emptyset$), observed solely via 2-bounce paths.

\begin{figure}[!t]
    \centering
    \includegraphics[width=\columnwidth]{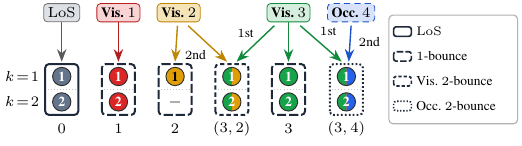}
    \caption{Cluster box representation ($K = 2$ scans) for LoS, $\mathrm{Vis.\ }1\text{--}3 \in \mathcal{I}_{\mathrm{vis}}$, and $\mathrm{Occ.\ }4 \in \mathcal{I}_{\mathrm{occ}}$. Rows denote scans $k = 1, 2$ (``$-$'' = miss); border styles indicate cluster classes; ball colors identify landmarks, with bicolor balls denoting 1st- (left) and 2nd-bounce (right) landmarks.}
    \label{fig:clusters}
\end{figure}

Given $\mathcal{I}^j$, a global DA hypothesis $\mathcal{A}^j \in \mathbb{J}$ partitions each scan $\mathcal{Z}_k$ ($k = 1, \dots, K$) by bounce order $0, 1, 2$ into disjoint subsets $\mathcal{Z}_k = \mathcal{Z}_{0,k}^j \allowbreak\biguplus (\biguplus_{i \in \mathcal{I}_{\mathrm{vis}}^j} \mathcal{Z}_{i,k}^{j,(1)}) \allowbreak\biguplus (\biguplus_{u \neq v \in \mathcal{I}^j} \mathcal{Z}_{uv,k}^{j,(2)})$, forming across time the 0-bounce LoS cluster $\mathcal{Z}_0^j = (\mathcal{Z}_{0,1}^j, \dots, \mathcal{Z}_{0,K}^j)$, 1-bounce clusters $\mathcal{Z}_i^{j,(1)} = (\mathcal{Z}_{i,1}^{j,(1)}, \dots, \mathcal{Z}_{i,K}^{j,(1)})$, and ordered 2-bounce clusters $\mathcal{Z}_{uv}^{j,(2)} = (\mathcal{Z}_{uv,1}^{j,(2)}, \dots, \mathcal{Z}_{uv,K}^{j,(2)})$. Non-empty 2-bounce clusters are divided by endpoint visibility into \emph{visible 2-bounce clusters} (both endpoints in $\mathcal{I}_{\mathrm{vis}}^j$) and \emph{occluded 2-bounce clusters} (one endpoint in $\mathcal{I}_{\mathrm{vis}}^j$ and the other in $\mathcal{I}_{\mathrm{occ}}^j$; see Fig.~\ref{fig:clusters}).

A valid global DA $\mathcal{A}^j \in \mathbb{J}$ satisfies three constraints: (i) at most one measurement per cluster per scan, $|\mathcal{Z}_{0,k}^j|, |\mathcal{Z}_{i,k}^{j,(1)}|, |\mathcal{Z}_{uv,k}^{j,(2)}| \le 1$; (ii) for tractable marginal likelihood factorization, each occluded candidate $v \in \mathcal{I}_{\mathrm{occ}}^j$ associates with one non-empty occluded 2-bounce cluster (multi-anchor paths from the same object are merged in Section~\ref{sec:marg}); and (iii) every $\mathcal{Z}_{uv}^{j,(2)} \neq \emptyset$ has at least one endpoint in $\mathcal{I}_{\mathrm{vis}}^j$. Constraint~(iii) stems from geometric identifiability: without 1-bounce observations at either endpoint, 2-bounce measurements suffer Jacobian rank deficiency for wall-wall paths and bounce-order/mirror ambiguity for wall-point paths; conversely, when at least one endpoint has 1-bounce observations, the other endpoint's position and type are uniquely identifiable across two or more non-collinear scans (proof in Appendix~D).

For a path of bounce order $b \in \{0, 1, 2\}$, combining $p_{\mathrm{D},b}(\cdot)$ with $f_b(\boldsymbol{z}|\cdot)$ in~\eqref{eq:lhf_los}--\eqref{eq:lhf_db}, define the single-scan factor
\begin{align}
\ell_b(\mathcal{Z}|\cdot) \triangleq
\begin{cases}
1 - p_{\mathrm{D},b}(\cdot), & \ccond{\mathcal{Z} = \emptyset,} \\
p_{\mathrm{D},b}(\cdot)\, f_b(\boldsymbol{z}|\cdot), & \ccond{\mathcal{Z} = \{\boldsymbol{z}\},} \\
0, & \tcond{otherwise.}
\end{cases} \label{eq:ell_eta}
\end{align}
\begingroup\predisplaypenalty=0
Partitioning $\mathcal{X}_{\mathrm{D}} = \mathcal{Y}_1^j \biguplus \dots \biguplus \mathcal{Y}_{N^j}^j$ ($|\mathcal{Y}_i^j| \le 1$) with singletons $\mathcal{Y}_i^j = \{\tilde{\boldsymbol{x}}_i\}$ ($\tilde{\boldsymbol{x}}_i = (\boldsymbol{x}_i, \psi_i)$, $\eta_{uv} \triangleq (\psi_u, \psi_v)$), the 0-bounce factor is $L_0(\mathcal{Z}_0^j|\boldsymbol{s}_{1:K}) \triangleq \prod_{k=1}^K \ell_0(\mathcal{Z}_{0,k}^j|\boldsymbol{s}_k)$, and the 1- and 2-bounce factors (under Constraints~(ii)--(iii)) are
\begin{align}
&L_1(\mathcal{Z}_i^{j,(1)}|\boldsymbol{s}_{1:K}, \mathcal{Y}_i^j) \notag\\*
&= \begin{ncases}
c(\boldsymbol{z}), & \ccond{|\mathcal{Z}_i^{j,(1)}|=1,\,\mathcal{Y}_i^j=\emptyset,} \\
\prod_{k=1}^K \ell_1(\mathcal{Z}_{i,k}^{j,(1)}|\tilde{\boldsymbol{x}}_i, \boldsymbol{s}_k), & \ccond{\mathcal{Y}_i^j=\{\tilde{\boldsymbol{x}}_i\},} \\
0, & \tcond{otherwise,}
\end{ncases} \label{eq:L1_a} \\[2pt]
&L_2(\mathcal{Z}_{uv}^{j,(2)}|\boldsymbol{s}_{1:K}, \mathcal{Y}_u^j, \mathcal{Y}_v^j) \notag\\*
&= \begin{ncases}
\prod_{k=1}^K\! \ell_2(\mathcal{Z}_{uv,k}^{j,(2)}\!|\tilde{\boldsymbol{x}}_u{,}\tilde{\boldsymbol{x}}_v{,}\boldsymbol{s}_k), & \ccond{|\mathcal{Y}_u^j|\!=\!|\mathcal{Y}_v^j|\!=\!1{,}\eta_{uv}\!\in\!\mathcal{T}_2,} \\
1, & \tcond{other $(\mathcal{Y}_u^j{,}\mathcal{Y}_v^j)$, $\mathcal{Z}_{uv}^{j,(2)}\!=\!\emptyset$,} \\
0, & \tcond{otherwise.}
\end{ncases} \label{eq:L2_a}
\end{align}
\endgroup
Clutter is modeled via singleton 1-bounce clusters in~\eqref{eq:L1_a} ($|\mathcal{Z}_i^{j,(1)}|=1, \mathcal{Y}_i^j=\emptyset$) avoiding double counting (clutter misassigned to a 2-bounce singleton yields low $\hat{r}_m$ and is rejected by $\vartheta_{\mathrm{det}}$ in Section~\ref{sec:marg}), while $|\mathcal{Z}_{uv,k}^{j,(2)}| \le 1$ follows from~\eqref{eq:ell_eta}.

Undetected landmarks $\mathcal{X}_{\mathrm{U}}$ generate no measurements over scans $1:K$ and thus enter the batch likelihood expansion solely through their 1-bounce missed-detection factors $1 - p_{\mathrm{D},1}$ and pairwise 2-bounce missed-detection factors $1 - p_{\mathrm{D},2}$ with other landmarks. Since the 2-bounce detection probability $p_{\mathrm{D},2}$ involving weak undetected landmarks in $\mathcal{X}_{\mathrm{U}}$ is very small ($1 - p_{\mathrm{D},2} \approx 1$), approximating the 2-bounce missed-detection factors involving $\mathcal{X}_{\mathrm{U}}$ as $1$ decouples $\mathcal{X}_{\mathrm{U}}$ from $\mathcal{X}_{\mathrm{D}}$ and preserves a marked PPP posterior form for $\mathcal{X}_{\mathrm{U}}$, yielding Theorem~\ref{thm:posterior} (Appendix~B).

\begin{theorem}[Batch Double-Bounce Multi-Type Joint Posterior] \label{thm:posterior}
Under the above DA constraints and approximation, the joint posterior density~\eqref{eq:bayes_joint} is given by
\begin{align}
&f(\boldsymbol{s}_{0:K}, \mathcal{X}|\mathcal{Z}_{1:K}) = \frac{e^{-\sum_{\psi \in \mathcal{T}_1} \int \lambda_\psi(\boldsymbol{x})\,\mathrm{d}\boldsymbol{x} - K\int c(\boldsymbol{z})\,\mathrm{d}\boldsymbol{z}}}{f(\mathcal{Z}_{1:K})} \notag\\*[-4pt]
&\;\times f(\boldsymbol{s}_0)\prod_{k=1}^K f(\boldsymbol{s}_k|\boldsymbol{s}_{k-1}) \sum_{j \in \mathbb{J}} L_0(\mathcal{Z}_0^j|\boldsymbol{s}_{1:K}) \notag\\*[-1.5pt]
&\;\times \sum_{\mathcal{X}_{\mathrm{U}} \biguplus \mathcal{Y}_1^j \biguplus \dots \biguplus \mathcal{Y}_{N^j}^j = \mathcal{X}} \prod_{\tilde{\boldsymbol{x}} \in \mathcal{X}_{\mathrm{U}}} p_{\mathrm{U}}(\tilde{\boldsymbol{x}}, \boldsymbol{s}_{1:K})\lambda_\psi(\boldsymbol{x}) \notag\\*[-1pt]
&\;\times \prod_{i=1}^{N^j} L_1(\mathcal{Z}_i^{j,(1)}|\boldsymbol{s}_{1:K}, \mathcal{Y}_i^j)\lambda(\mathcal{Y}_i^j) \notag\\*[-1pt]
&\;\times \prod_{u \neq v}^{N^j} L_2(\mathcal{Z}_{uv}^{j,(2)}|\boldsymbol{s}_{1:K}, \mathcal{Y}_u^j, \mathcal{Y}_v^j), \label{eq:joint_post}
\end{align}
where $\tilde{\boldsymbol{x}} = (\boldsymbol{x}, \psi)$, $p_{\mathrm{U}}(\tilde{\boldsymbol{x}}, \boldsymbol{s}_{1:K}) \triangleq \prod_{k=1}^K (1 - p_{\mathrm{D},1}(\tilde{\boldsymbol{x}}, \boldsymbol{s}_k))$, and $\lambda(\emptyset) = 1$, $\lambda(\{\tilde{\boldsymbol{x}}_i\}) = \lambda_{\psi_i}(\boldsymbol{x}_i)$ (zero otherwise).
\end{theorem}

In~\eqref{eq:joint_post}, the 2-bounce factor $L_2$ depends jointly on pairwise sets $(\mathcal{Y}_u^j, \mathcal{Y}_v^j)$, so potential landmark components $\mathcal{Y}_i^j$ are no longer conditionally independent given global hypothesis $j$; only when 2-bounce paths are absent ($L_2 \equiv 1$) does~\eqref{eq:joint_post} reduce to a product of independent multi-type Bernoulli components (a standard multi-type PMBM posterior).

\subsubsection{Joint Posterior With Augmented Data Association}

\begingroup\predisplaypenalty=0
To distinguish clutter from landmarks of different types, introduce a ternary variable $\psi_i^j \in \{0, \mathrm{VA}, \mathrm{RP}\}$ for each $i \in \mathcal{I}^j$, where $\psi_i^j = 0$ denotes clutter ($\mathcal{Y}_i^j = \emptyset$) and $\psi_i^j \in \mathcal{T}_1$ denotes a landmark $\tilde{\boldsymbol{x}}_i^j \triangleq (\boldsymbol{x}_i, \psi_i^j)$ ($\mathcal{Y}_i^j = \{\tilde{\boldsymbol{x}}_i^j\}$). With $\eta_{uv}^j \triangleq (\psi_u^j, \psi_v^j)$, the 1- and 2-bounce factors in~\eqref{eq:joint_post} rewrite as
\begin{align}
&\tilde{L}_1(\mathcal{Z}_i^{j,(1)}, \mathcal{Y}_i^j|\boldsymbol{s}_{1:K}, \psi_i^j) \notag\\*
&\!\triangleq\!\begin{ncases}
c(\boldsymbol{z}), & \begin{array}{@{}l@{}}\ccond{\psi_i^j\!=\!0,\,\mathcal{Y}_i^j\!=\!\emptyset,}\\[-3pt]\ccond{|\mathcal{Z}_i^{j,(1)}|\!=\!1,}\end{array} \\[4pt]
\lambda_{\psi_i^j}(\boldsymbol{x}_i)\!\prod_{k=1}^K\! \ell_1(\mathcal{Z}_{i,k}^{j,(1)}\!|\tilde{\boldsymbol{x}}_i^j{,}\boldsymbol{s}_k), & \ccond{\psi_i^j\!\in\!\mathcal{T}_1{,}\,\mathcal{Y}_i^j\!=\!\{\tilde{\boldsymbol{x}}_i^j\},} \\
0, & \tcond{otherwise,}
\end{ncases} \label{eq:L1_aug} \\[1pt]
&\tilde{L}_2(\mathcal{Z}_{uv}^{j,(2)}|\boldsymbol{s}_{1:K}, \mathcal{Y}_u^j, \mathcal{Y}_v^j; \eta_{uv}^j) \notag\\*
&\!\triangleq\!\begin{ncases}
\prod_{k=1}^K\! \ell_2(\mathcal{Z}_{uv,k}^{j,(2)}\!|\tilde{\boldsymbol{x}}_u^j{,}\tilde{\boldsymbol{x}}_v^j{,}\boldsymbol{s}_k), & \ccond{|\mathcal{Y}_u^j|\!=\!|\mathcal{Y}_v^j|\!=\!1{,}\eta_{uv}^j\!\in\!\mathcal{T}_2,} \\
1, & \tcond{other states, $\mathcal{Z}_{uv}^{j,(2)}\!=\!\emptyset$,} \\
0, & \tcond{otherwise.}
\end{ncases} \label{eq:L2_aug}
\end{align}
\endgroup
With $\boldsymbol{\psi}^j \triangleq [\psi_1^j, \dots, \psi_{N^j}^j]^{\mathsf{T}}$ and augmented DA $\tilde{\mathcal{A}} \triangleq (\mathcal{A}^j, \boldsymbol{\psi}^j)$,~\eqref{eq:L1_aug}--\eqref{eq:L2_aug} in~\eqref{eq:joint_post} yield the augmented posterior (summing over valid $\tilde{\mathcal{A}}$ recovers~\eqref{eq:joint_post}):
\begin{align}
&f(\boldsymbol{s}_{0:K}, \mathcal{X}, \tilde{\mathcal{A}}|\mathcal{Z}_{1:K}) = \frac{e^{-\sum_{\psi \in \mathcal{T}_1} \int \lambda_\psi(\boldsymbol{x})\,\mathrm{d}\boldsymbol{x} - K\int c(\boldsymbol{z})\,\mathrm{d}\boldsymbol{z}}}{f(\mathcal{Z}_{1:K})} \notag\\*[-5pt]
&\;\times \underbrace{f(\boldsymbol{s}_0)\prod_{k=1}^K f(\boldsymbol{s}_k|\boldsymbol{s}_{k-1})}_{\text{trajectory motion prior}}\,
\underbrace{L_0(\mathcal{Z}_0^j|\boldsymbol{s}_{1:K})}_{\text{0-bounce LoS}} \notag\\*[-1pt]
&\;\times \sum_{\mathcal{X}_{\mathrm{U}} \biguplus \mathcal{Y}_1^j \biguplus \dots \biguplus \mathcal{Y}_{N^j}^j = \mathcal{X}}
\underbrace{\prod_{\tilde{\boldsymbol{x}} \in \mathcal{X}_{\mathrm{U}}} p_{\mathrm{U}}(\tilde{\boldsymbol{x}}, \boldsymbol{s}_{1:K})\lambda_\psi(\boldsymbol{x})}_{\text{undetected landmarks}} \notag\\*[-1pt]
&\;\times \underbrace{\prod_{i=1}^{N^j} \tilde{L}_1(\mathcal{Z}_i^{j,(1)}, \mathcal{Y}_i^j|\boldsymbol{s}_{1:K}, \psi_i^j)}_{\text{1-bounce clusters \& clutter}} \notag\\*[-1pt]
&\;\times \underbrace{\prod_{u \neq v}^{N^j} \tilde{L}_2(\mathcal{Z}_{uv}^{j,(2)}|\boldsymbol{s}_{1:K}, \mathcal{Y}_u^j, \mathcal{Y}_v^j; \eta_{uv}^j)}_{\text{2-bounce clusters}}. \label{eq:aug_post}
\end{align}

\subsection{Overall Alternating Batch Framework}

\begin{figure*}[!t]
    \centering
    \includegraphics[width=\textwidth]{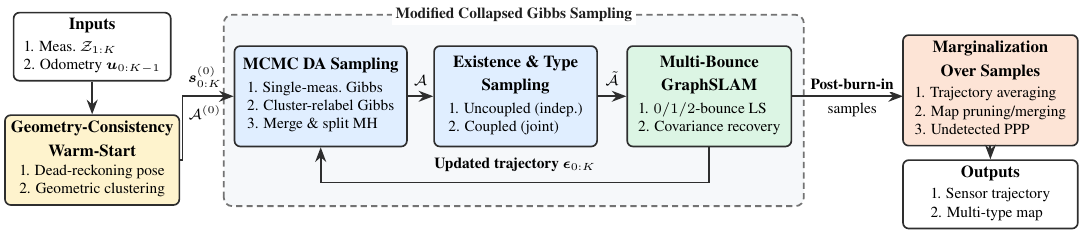}
    \caption{Overall flowchart of the proposed double-bounce multi-type batch radio SLAM algorithm. Starting from measurement batch $\mathcal{Z}_{1:K}$ and odometry $\boldsymbol{u}_{0:K-1}$, warm-start pre-clustering generates initial trajectory $\boldsymbol{s}_{0:K}^{(0)}$ and partition $\mathcal{A}^{(0)}$. Within the modified collapsed Gibbs loop, it alternates MCMC DA sampling of $\mathcal{A}$, landmark existence and type sampling to obtain $\tilde{\mathcal{A}}$, and double-bounce GraphSLAM optimization with covariance recovery, feeding updated $\boldsymbol{\epsilon}_{0:K}$ back to the sampler. Finally, post-burn-in samples are marginalized to yield the trajectory and multi-type map.}
    \label{fig:flowchart}
\end{figure*}

Targeting~\eqref{eq:aug_post}, we adopt a modified collapsed Gibbs sampling scheme~\cite{Ge2025TSP} alternating between discrete association sampling and continuous GraphSLAM optimization (Fig.~\ref{fig:flowchart}), where ``collapsed'' indicates integrating out the unknown map $\mathcal{X}$ when sampling $\tilde{\mathcal{A}}$, and ``modified'' indicates replacing random trajectory sampling with GraphSLAM maximum a posteriori (MAP) estimation conditioned on $\tilde{\mathcal{A}}$. Following the section order below, the overall method comprises four modules:
\begin{enumerate}
    \item \emph{Two-Stage Double-Bounce DA and Type Sampling (Sections~\ref{sec:da_weight}--\ref{sec:type_sampling}):} Conditioned on the current trajectory estimate $\boldsymbol{s}_{0:K}$, sequentially sample the cluster partition $\mathcal{A}$ and landmark state vector $\boldsymbol{\psi}$ from the collapsed marginal posterior.
    \item \emph{Warm-Start Pre-Clustering (Section~\ref{sec:warmstart}):} Prior to iteration~1, perform preliminary geometric clustering on the measurement batch using the dead-reckoning trajectory to construct an initial partition $\mathcal{A}^{(0)}$ and shorten burn-in.
    \item \emph{Conditional Double-Bounce GraphSLAM (Section~\ref{sec:cond_slam}):} Fixing $\tilde{\mathcal{A}} = (\mathcal{A}, \boldsymbol{\psi})$ and discarding clutter, solve nonlinear least squares (LS) to estimate the sensor trajectory and detected landmarks and recover the posterior covariance.
    \item \emph{Marginalization Over Samples (Section~\ref{sec:marg}):} Alternate DA sampling and GraphSLAM optimization, fuse post-burn-in trajectory and multi-type PMB map samples, and update the undetected landmark intensity.
\end{enumerate}

\section{Double-Bounce Data Association Sampling and Warm-Start Pre-Clustering} \label{sec:da_sampling}

Fixing $\boldsymbol{s}_{0:K}$ and taking the set integral of~\eqref{eq:aug_post} over $\mathcal{X}$ gives the collapsed marginal posterior $p(\tilde{\mathcal{A}}|\boldsymbol{s}_{0:K}, \mathcal{Z}_{1:K}) \propto \int f(\boldsymbol{s}_{0:K}, \mathcal{X}, \tilde{\mathcal{A}}|\mathcal{Z}_{1:K})\,\delta\mathcal{X}$, which factorizes into two sequential sampling steps~\cite{Ge2025TSP,Xia2024TAES} (omitting superscript $j$ hereafter):
\begin{equation}
p\mkern-1.5mu(\mkern-1mu\mathcal{A}{,}\boldsymbol{\psi}|\boldsymbol{s}_{0\mkern-1mu:\mkern-1mu K}{,}\mathcal{Z}_{1\mkern-1mu:\mkern-1mu K}\mkern-1.5mu)\mkern-2mu=\mkern-2mup\mkern-1.5mu(\mkern-1mu\mathcal{A}|\boldsymbol{s}_{0\mkern-1mu:\mkern-1mu K}{,}\mathcal{Z}_{1\mkern-1mu:\mkern-1mu K}\mkern-1.5mu)p\mkern-1.5mu(\mkern-1mu\boldsymbol{\psi}|\mathcal{A}{,}\boldsymbol{s}_{0\mkern-1mu:\mkern-1mu K}{,}\mathcal{Z}_{1\mkern-1mu:\mkern-1mu K}\mkern-1.5mu). \label{eq:da_factor}
\end{equation}

\subsection{Double-Bounce Data Association Weight Calculation} \label{sec:da_weight}

Summing $p(\tilde{\mathcal{A}}|\boldsymbol{s}_{0:K}, \mathcal{Z}_{1:K})$ over $\boldsymbol{\psi}$ recovers the $\mathcal{A}$-term of~\eqref{eq:joint_post}. Because $\mathcal{X}_{\mathrm{U}}$ decouples from $\mathcal{Y}_1, \dots, \mathcal{Y}_N$, its set integral depends only on $\boldsymbol{s}_{1:K}$ (Appendix~C) and is absorbed into the proportionality constant, yielding the partition marginal posterior $p(\mathcal{A}|\boldsymbol{s}_{0:K}, \mathcal{Z}_{1:K}) \propto w(\mathcal{A})$, where the unnormalized association weight $w(\mathcal{A})$ is defined as
\begin{align}
&w(\mathcal{A}) \triangleq \mathcal{L}_0(\mathcal{Z}_0) \int\!\cdots\!\int \prod_{i=1}^N \bigl(L_1(\mathcal{Z}_i^{(1)}|\boldsymbol{s}_{1:K}, \mathcal{Y}_i)\lambda(\mathcal{Y}_i)\bigr) \notag\\
&\quad \times \prod_{u \neq v}^N L_2(\mathcal{Z}_{uv}^{(2)}|\boldsymbol{s}_{1:K}, \mathcal{Y}_u, \mathcal{Y}_v)\,\delta\mathcal{Y}_1 \cdots \delta\mathcal{Y}_N, \label{eq:weight_exact}
\end{align}
with $\mathcal{L}_0(\mathcal{Z}_0) \triangleq L_0(\mathcal{Z}_0|\boldsymbol{s}_{1:K})$. Because $L_2$ couples $(\mathcal{Y}_u, \mathcal{Y}_v)$,~\eqref{eq:weight_exact} cannot split directly into $N$ independent set integrals. Define the 1-bounce marginal likelihood and 1-bounce posterior of candidate $i \in \mathcal{I}$:
\begin{align}
&\mathcal{L}_1(\mathcal{Z}_i^{(1)}) \triangleq \int L_1(\mathcal{Z}_i^{(1)}|\boldsymbol{s}_{1:K}, \mathcal{Y}_i)\lambda(\mathcal{Y}_i)\,\delta\mathcal{Y}_i, \notag\\
&p(\mathcal{Y}_i|\mathcal{Z}_i^{(1)}, \boldsymbol{s}_{1:K}) \triangleq \frac{L_1(\mathcal{Z}_i^{(1)}|\boldsymbol{s}_{1:K}, \mathcal{Y}_i)\lambda(\mathcal{Y}_i)}{\mathcal{L}_1(\mathcal{Z}_i^{(1)})}. \label{eq:L1_post}
\end{align}
Rewriting~\eqref{eq:L1_post} as $L_1(\mathcal{Z}_i^{(1)}|\boldsymbol{s}_{1:K}, \mathcal{Y}_i)\lambda(\mathcal{Y}_i) = \mathcal{L}_1(\mathcal{Z}_i^{(1)})\allowbreak\,p(\mathcal{Y}_i|\mathcal{Z}_i^{(1)}, \boldsymbol{s}_{1:K})$ and substituting it into~\eqref{eq:weight_exact} incorporates each candidate's 1-bounce posterior $p(\mathcal{Y}_i|\mathcal{Z}_i^{(1)}, \boldsymbol{s}_{1:K})$ directly into the 2-bounce integral. Under 3D joint measurements, the 1-bounce posterior of a visible candidate is sharply concentrated around the true landmark position and decays rapidly; combined with the fact that each occluded candidate participates in only a single non-empty 2-bounce cluster, we approximate~\eqref{eq:weight_exact} as a product of pairwise conditional marginal likelihoods (details in Appendix~C):
\begin{align}
w(\mathcal{A}) \approx \hat{w}(\mathcal{A}) &\triangleq \mathcal{L}_0(\mathcal{Z}_0) \prod_{i=1}^N \mathcal{L}_1(\mathcal{Z}_i^{(1)}) \notag\\*
&\quad \times \prod_{u \neq v}^N \mathcal{L}_2(\mathcal{Z}_{uv}^{(2)}|\mathcal{Z}_u^{(1)}, \mathcal{Z}_v^{(1)}), \label{eq:weight_approx}
\end{align}
with pairwise conditional 2-bounce marginal likelihood
\begin{align}
&\mathcal{L}_2(\mathcal{Z}_{uv}^{(2)}|\mathcal{Z}_u^{(1)}, \mathcal{Z}_v^{(1)}) \triangleq \iint L_2(\mathcal{Z}_{uv}^{(2)}|\boldsymbol{s}_{1:K}, \mathcal{Y}_u, \mathcal{Y}_v) \notag\\
&\quad \times p(\mathcal{Y}_u|\mathcal{Z}_u^{(1)}, \boldsymbol{s}_{1:K})\, p(\mathcal{Y}_v|\mathcal{Z}_v^{(1)}, \boldsymbol{s}_{1:K})\,\delta\mathcal{Y}_u\delta\mathcal{Y}_v. \label{eq:L2_cond}
\end{align}

The three factor classes in~\eqref{eq:weight_approx} are evaluated via Laplace approximation (closed-form expressions in Appendix~C):
\begin{enumerate}
    \item \emph{0-bounce LoS factor $\mathcal{L}_0(\mathcal{Z}_0)$:} Computed as $\prod_{k=1}^K \ell_0(\mathcal{Z}_{0,k}|\boldsymbol{s}_k)$.
    \item \emph{1-bounce factor $\mathcal{L}_1(\mathcal{Z}_i^{(1)})$:} By cluster cardinality,
    \begin{align}
    &\mathcal{L}_1(\mathcal{Z}_i^{(1)}) \notag\\*
    &= \begin{ncases}
    c(\boldsymbol{z}) + \sum_{\psi\in\mathcal{T}_1}\mathcal{L}_1(\mathcal{Z}_i^{(1)};\psi), & \ccond{|\mathcal{Z}_i^{(1)}|=1,} \\
    \sum_{\psi\in\mathcal{T}_1}\mathcal{L}_1(\mathcal{Z}_i^{(1)};\psi), & \ccond{|\mathcal{Z}_i^{(1)}|>1,} \\
    \sum_{\psi\in\mathcal{T}_1}\!\int p_{\mathrm{U}}(\tilde{\boldsymbol{x}}_i, \boldsymbol{s}_{1:K})\lambda_\psi(\boldsymbol{x}_i)\,\mathrm{d}\boldsymbol{x}_i, & \ccond{\mathcal{Z}_i^{(1)}=\emptyset,}
    \end{ncases} \label{eq:L1_cases}
    \end{align}
    where $\mathcal{L}_1(\mathcal{Z}_i^{(1)}; \psi) \triangleq \int \lambda_\psi(\boldsymbol{x}_i)\allowbreak \prod_{k=1}^K \ell_1(\mathcal{Z}_{i,k}^{(1)}|\tilde{\boldsymbol{x}}_i, \boldsymbol{s}_k)\allowbreak\,\mathrm{d}\boldsymbol{x}_i$.
    \item \emph{Conditional 2-bounce factor $\mathcal{L}_2(\mathcal{Z}_{uv}^{(2)}|\mathcal{Z}_u^{(1)}, \mathcal{Z}_v^{(1)})$:} Approximating~\eqref{eq:L1_post} as Gaussian with $\pi_{i,0} \triangleq \mathbb{I}(|\mathcal{Z}_i^{(1)}|=1)\allowbreak c(\boldsymbol{z})/\allowbreak\mathcal{L}_1(\mathcal{Z}_i^{(1)})$ and $\pi_{i,\psi} \triangleq \mathcal{L}_1(\mathcal{Z}_i^{(1)}; \psi)/\allowbreak\mathcal{L}_1(\mathcal{Z}_i^{(1)})$ ($i \in \mathcal{I}_{\mathrm{vis}}$) gives
    \begin{align}
    &p(\mathcal{Y}_i|\mathcal{Z}_i^{(1)}, \boldsymbol{s}_{1:K}) \notag\\*
    &\approx \begin{cases}
    \pi_{i,0}, & \ccond{\mathcal{Z}_i^{(1)}\!\neq\!\emptyset,\,\mathcal{Y}_i\!=\!\emptyset,} \\
    \pi_{i,\psi}\mathcal{N}(\boldsymbol{x}_i; \hat{\boldsymbol{x}}_{i,\psi}, \hat{\boldsymbol{\Sigma}}_{i,\psi}), & \ccond{\mathcal{Z}_i^{(1)}\!\neq\!\emptyset,\,\mathcal{Y}_i\!=\!\{\tilde{\boldsymbol{x}}_i\},} \\
    p_{\mathrm{U}}(\tilde{\boldsymbol{x}}_i, \boldsymbol{s}_{1:K})\lambda_\psi(\boldsymbol{x}_i)/\mathcal{L}_1(\emptyset), & \ccond{\mathcal{Z}_i^{(1)}\!=\!\emptyset,\,\mathcal{Y}_i\!=\!\{\tilde{\boldsymbol{x}}_i\},} \\
    0, & \tcond{otherwise.}
    \end{cases} \label{eq:L1_gauss}
    \end{align}
    Substituting~\eqref{eq:L1_gauss} into~\eqref{eq:L2_cond} yields three cases according to cluster emptiness and endpoint visibility:
    \par (a)\hspace{\IEEEiednormlabelsep}\emph{Empty 2-bounce pair} ($\mathcal{Z}_{uv}^{(2)} = \emptyset$, $u,v \in \mathcal{I}_{\mathrm{vis}}$): With $\hat{\tilde{\boldsymbol{x}}}_{u,\psi_u} \triangleq (\hat{\boldsymbol{x}}_{u,\psi_u}, \psi_u)$,
    \begin{align}
    &\mathcal{L}_2(\emptyset|\mathcal{Z}_u^{(1)}, \mathcal{Z}_v^{(1)}) \approx 1 - \sum_{\eta_{uv} \in \mathcal{T}_2} \pi_{u,\psi_u}\pi_{v,\psi_v} \notag\\
    &\quad \times \left[1 - \prod_{k=1}^K \big(1 - p_{\mathrm{D},2}(\hat{\tilde{\boldsymbol{x}}}_{u,\psi_u}, \hat{\tilde{\boldsymbol{x}}}_{v,\psi_v}, \boldsymbol{s}_k)\big)\right]. \label{eq:L2_empty}
    \end{align}
    \par (b)\hspace{\IEEEiednormlabelsep}\emph{Non-empty visible 2-bounce pair} ($\mathcal{Z}_{uv}^{(2)} \neq \emptyset$, $u,v \in \mathcal{I}_{\mathrm{vis}}$):
    \begin{align}
    &\mathcal{L}_2(\mathcal{Z}_{uv}^{(2)}|\mathcal{Z}_u^{(1)}, \mathcal{Z}_v^{(1)}) \notag\\*
    &\quad \approx \sum_{\eta_{uv}\in\mathcal{T}_2} \pi_{u,\psi_u}\pi_{v,\psi_v}\mathcal{L}_2(\mathcal{Z}_{uv}^{(2)}|\mathcal{Z}_u^{(1)}, \mathcal{Z}_v^{(1)}; \eta_{uv}). \label{eq:L2_coupled}
    \end{align}
    \par (c)\hspace{\IEEEiednormlabelsep}\emph{Non-empty occluded 2-bounce pair} ($\mathcal{Z}_{uv}^{(2)} \neq \emptyset$, $u \in \mathcal{I}_{\mathrm{vis}}, v \in \mathcal{I}_{\mathrm{occ}}$; symmetric for $(v,u)$): Using the third branch of~\eqref{eq:L1_gauss} in~\eqref{eq:L2_cond} and factoring out $1/\mathcal{L}_1(\emptyset)$ gives $\mathcal{L}_{2,\mathrm{occ}}(\mathcal{Z}_{uv}^{(2)}|\mathcal{Z}_u^{(1)}) \triangleq \mathcal{L}_1(\emptyset)\,\mathcal{L}_2(\mathcal{Z}_{uv}^{(2)}|\mathcal{Z}_u^{(1)}, \emptyset)$:
    \begin{align}
    \mathcal{L}_{2,\mathrm{occ}}(\mathcal{Z}_{uv}^{(2)}\!|\mathcal{Z}_u^{(1)})\!\approx\!\!\sum_{\eta_{uv}\in\mathcal{T}_2}\!\!\pi_{u,\psi_u}\mathcal{L}_{2,\mathrm{occ}}(\mathcal{Z}_{uv}^{(2)}\!|\mathcal{Z}_u^{(1)}{;}\eta_{uv}), \label{eq:L2_occ}
    \end{align}
    where $\mathcal{L}_{2,\mathrm{occ}}(\mathcal{Z}_{uv}^{(2)}|\mathcal{Z}_u^{(1)}; \eta_{uv})$ is the type-conditioned occluded 2-bounce integral approximated in Appendix~C.
\end{enumerate}

Substituting~\eqref{eq:L1_cases} and~\eqref{eq:L2_empty}--\eqref{eq:L2_occ} into~\eqref{eq:weight_approx}, the empty 1-bounce factor $\mathcal{L}_1(\emptyset)$ for each $v \in \mathcal{I}_{\mathrm{occ}}$ cancels against the denominator $\mathcal{L}_1(\emptyset)$ in $\mathcal{L}_2(\mathcal{Z}_{uv}^{(2)}|\mathcal{Z}_u^{(1)}, \emptyset) = \mathcal{L}_{2,\mathrm{occ}}(\mathcal{Z}_{uv}^{(2)}|\mathcal{Z}_u^{(1)})/\mathcal{L}_1(\emptyset)$. Using $\mathcal{L}_{2,\mathrm{occ}}(v; \mathcal{A})$ as shorthand for the factor~\eqref{eq:L2_occ} uniquely associated with candidate $v$ under partition $\mathcal{A}$, the approximated DA weight simplifies to
\begin{align}
\hat{w}(\mathcal{A}) &= \mathcal{L}_0(\mathcal{Z}_0)
\underbrace{\left[\prod_{i \in \mathcal{I}_{\mathrm{vis}}} \mathcal{L}_1(\mathcal{Z}_i^{(1)})\right]}_{\text{visible 1-bounce terms}}
\underbrace{\left[\prod_{v \in \mathcal{I}_{\mathrm{occ}}} \mathcal{L}_{2,\mathrm{occ}}(v; \mathcal{A})\right]}_{\text{occluded 2-bounce terms}} \notag\\*[2pt]
&\quad \times \underbrace{\left[\prod_{u, v \in \mathcal{I}_{\mathrm{vis}},\, u \neq v} \mathcal{L}_2(\mathcal{Z}_{uv}^{(2)}|\mathcal{Z}_u^{(1)}, \mathcal{Z}_v^{(1)})\right]}_{\text{visible 2-bounce terms}}. \label{eq:weight_clean}
\end{align}

\subsection{Double-Bounce MCMC Data Association Sampling} \label{sec:mcmc}

Direct enumeration of~\eqref{eq:weight_clean} over $\mathbb{J}$ is computationally infeasible. Combining Gibbs and split-merge sampling~\cite{Ge2025TSP,Xia2024TAES,Fatemi2017TSP}, we construct a three-step MCMC sampler comprising single-measurement Gibbs, cluster-relabeling Gibbs, and merge-and-split Metropolis-Hastings (MH) updates (Fig.~\ref{fig:mcmc}; pseudocode in Appendix~E): all three steps run during burn-in to accelerate clustering, whereas only Steps~1 and~2 satisfying detailed balance are executed after burn-in. To cancel invariant factors in local weight ratios, group terms involving $u \in \mathcal{I}_{\mathrm{vis}}$ as ($C_u \equiv \mathcal{Z}_u^{(1)}$, occluded set $\mathcal{B}_u \subseteq \mathcal{I}_{\mathrm{occ}}$):
\begin{align}
&\Phi_u(C_u) \triangleq \mathcal{L}_1(C_u)\prod_{w \in \mathcal{B}_u} \mathcal{L}_{2,\mathrm{occ}}(w; C_u) \notag\\
&\quad\times \prod_{v \in \mathcal{I}_{\mathrm{vis}} \setminus \{u\}} \mathcal{L}_2(C_{uv}|C_u, C_v)\,\mathcal{L}_2(C_{vu}|C_v, C_u). \label{eq:Phi_u}
\end{align}
Local weight ratio formulas for each transition are detailed in Appendix~C.

\begin{figure*}[!t]
    \centering
    \includegraphics[width=\textwidth]{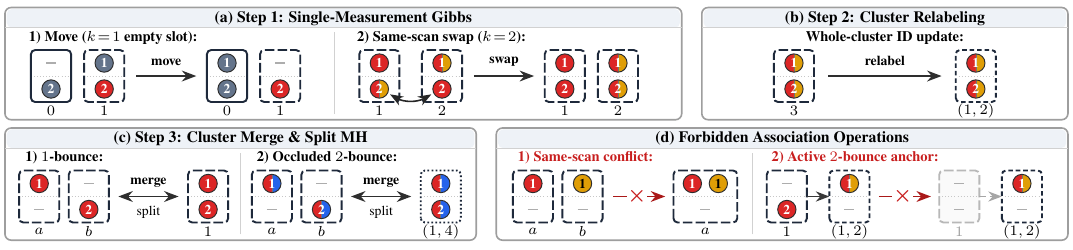}
    \caption{Transition operators and forbidden operations of the double-bounce MCMC DA sampler (box styles as in Fig.~\ref{fig:clusters}). (a) Step~1 (single-measurement Gibbs): \texttt{move} shifts a measurement into an existing or new cluster with an empty scan-$k$ slot, while \texttt{swap} exchanges two same-scan measurements between clusters. (b) Step~2 (cluster-relabeling Gibbs): updates the category and index of an entire cluster (\texttt{relabel}, e.g., from 1-bounce cluster $3$ to visible 2-bounce cluster $(1,2)$) without altering its measurements. (c) Step~3 (cluster merge \& split MH): merges scan-disjoint clusters $a,b$ into a 1-bounce or occluded 2-bounce cluster and reverses via splitting. (d) Forbidden operations: 1)~same-scan conflict, preventing two scan-$1$ measurements inside one slot; 2)~active 2-bounce anchor constraint, preventing anchor cluster $1$ from being emptied or relabeled while linked to non-empty 2-bounce cluster $(1,2)$.}
    \label{fig:mcmc}
\end{figure*}

\subsubsection{Step 1: Single-Measurement Gibbs Sampling}
When visiting a scan-$k$ measurement $\boldsymbol{m} \in C_\beta$, because single-point moves and same-scan swaps have distinct forward and reverse candidate sets (Appendix~C), we execute them sequentially as two conditional Gibbs sub-steps:
\begin{itemize}
    \item \emph{Single-point move:} Fixing all other measurements, form the candidate move set $\mathcal{M}_{\mathrm{move}}(\boldsymbol{m}; \mathcal{A}^{(\iota)})$ comprising: (i) remaining in $C_\beta$; (ii) moving into an existing cluster whose scan-$k$ slot is empty; (iii) creating a new 1-bounce singleton $\{\boldsymbol{m}\}$; and (iv) creating an occluded 2-bounce singleton $(u, v_{\mathrm{new}})$ or $(v_{\mathrm{new}}, u)$ anchored to $u \in \mathcal{I}_{\mathrm{vis}}$ (identical resulting partitions are counted once; if $C_\beta$ is a singleton 1-bounce cluster with non-empty 2-bounce children, only operation~(i) is permitted). Sample proportional to $\rho_{\beta\to\gamma}^{(\iota)}(\boldsymbol{m}) \triangleq \hat{w}(\mathcal{A}_{\beta\to\gamma}^{(\iota)}(\boldsymbol{m})) / \hat{w}(\mathcal{A}^{(\iota)})$:
    \begin{align}
    &P_{\mathrm{move}}(\mathcal{A}_{\beta\to\gamma}^{(\iota)}(\boldsymbol{m})|\mathcal{A}^{(\iota)}, \boldsymbol{s}_{0:K}, \mathcal{Z}_{1:K}) \notag\\*
    &\qquad = \frac{\rho_{\beta\to\gamma}^{(\iota)}(\boldsymbol{m})}{\sum_{\gamma' \in \mathcal{M}_{\mathrm{move}}} \rho_{\beta\to\gamma'}^{(\iota)}(\boldsymbol{m})}. \label{eq:p_move}
    \end{align}
    \item \emph{Same-scan swap:} For each scan-$k$ measurement $\boldsymbol{m}' \in C_\gamma$ in another cluster, accept the swap with probability
    \begin{align}
    P_{\mathrm{swap}}(\mathcal{A}_{\beta\leftrightarrow\gamma}(\boldsymbol{m})|\mathcal{A}, \boldsymbol{s}_{0:K}, \mathcal{Z}_{1:K}) = \frac{\rho_{\beta\leftrightarrow\gamma}(\boldsymbol{m})}{1 + \rho_{\beta\leftrightarrow\gamma}(\boldsymbol{m})}, \label{eq:p_swap}
    \end{align}
    where $\rho_{\beta\leftrightarrow\gamma}(\boldsymbol{m}) \triangleq \hat{w}(\mathcal{A}_{\beta\leftrightarrow\gamma}(\boldsymbol{m})) / \hat{w}(\mathcal{A})$.
\end{itemize}

\begin{example}
Let $K=2$ with $C_1 = \{\boldsymbol{z}_1^1, \boldsymbol{z}_2^1\}$, $C_2 = \{\boldsymbol{z}_1^2, \boldsymbol{z}_2^2\}$, and $C_{12} = \emptyset$. For $\boldsymbol{z}_2^1 \in C_1$, \texttt{move} evaluates staying in $C_1$, moving into empty scan-$2$ slots $C_{12}$ or $C_{21}$, forming a 1-bounce singleton $\{\boldsymbol{z}_2^1\}$, or forming $(1, v_{\mathrm{new}}) = \{\boldsymbol{z}_2^1\}$; \texttt{swap} evaluates exchanging $\boldsymbol{z}_2^1 \in C_1$ with $\boldsymbol{z}_2^2 \in C_2$ to obtain $C_1' = \{\boldsymbol{z}_1^1, \boldsymbol{z}_2^2\}$ and $C_2' = \{\boldsymbol{z}_1^2, \boldsymbol{z}_2^1\}$.
\end{example}

\subsubsection{Step 2: Cluster-Relabeling Gibbs Sampling}
When a multi-scan cluster is assigned an incorrect bounce order or reflection direction, single-measurement moves must pass through low-likelihood intermediate states to reassign the cluster. Step~2 keeps the measurements inside each non-empty cluster $C_{(l)}$ unchanged (skipping 1-bounce clusters associated with non-empty 2-bounce children) and updates its label over candidate set $\mathcal{K}_l$ with relative weight $\tilde{w}(\gamma) \triangleq \hat{w}(\mathcal{A}_\gamma)/\hat{w}(\mathcal{A}_{-l})$:
\begin{align}
P_{\mathrm{relabel}}(\mathcal{A}_\gamma|\mathcal{A}_{-l}, \boldsymbol{s}_{0:K}, \mathcal{Z}_{1:K}) = \frac{\tilde{w}(\gamma)}{\sum_{\gamma' \in \mathcal{K}_l} \tilde{w}(\gamma')}, \label{eq:p_relabel}
\end{align}
where $\mathcal{K}_l$ comprises: (i) LoS cluster $C_0$ (if currently empty); (ii) standalone 1-bounce cluster $u_{\mathrm{new}}$; (iii) currently empty visible 2-bounce clusters $(u, v)$ ($u \neq v \in \mathcal{I}_{\mathrm{vis}}$); and (iv) occluded 2-bounce clusters $(u, v_{\mathrm{new}})$ or $(v_{\mathrm{new}}, u)$ anchored to $u \in \mathcal{I}_{\mathrm{vis}}$.

\begin{example}
Given 1-bounce clusters $C_1, C_2$ ($C_{12} = C_{21} = \emptyset$) and a two-scan cluster $C_{(l)} = \{\boldsymbol{z}_1^3, \boldsymbol{z}_2^3\}$ labeled as $C_3$, Step~2 compares the weights of $C_3$, $C_{12}$, $C_{21}$, $(1, v_{\mathrm{new}})$, $(v_{\mathrm{new}}, 1)$, $(2, v_{\mathrm{new}})$, and $(v_{\mathrm{new}}, 2)$ without splitting $\{\boldsymbol{z}_1^3, \boldsymbol{z}_2^3\}$.
\end{example}

\subsubsection{Step 3: Cluster Merge \& Split MH Sampling}
During burn-in, two merge-and-split proposals accelerate clustering convergence: (i) merging two scan-disjoint clusters into $C_\beta \cup C_\gamma$, paired with a \texttt{k-means++} split operator; and (ii) merging two scan-disjoint 1-bounce clusters $C_b, C_g$ with nearby inverted positions under anchor $u \in \mathcal{I}_{\mathrm{vis}}$ into an occluded 2-bounce cluster $(u, v_{\mathrm{new}}) = C_b \cup C_g$ (or $(v_{\mathrm{new}}, u)$), paired with reverse splitting. Following~\cite{Ge2025TSP}, we refer to this step as merge-and-split MH sampling, where proposals are accepted with simplified probability $\min[1, \hat{w}(\mathcal{A}')/\hat{w}(\mathcal{A})]$ and turned off after burn-in.

\begin{example}
When an occluded scatterer observed via wall~$1$ ($C_1$) across two scans initially forms two 1-bounce singletons $C_3 = \{\boldsymbol{z}_1^3\}$ and $C_4 = \{\boldsymbol{z}_2^3\}$ whose single-scan likelihoods each lie below $c(\boldsymbol{z})$, Step~3 can directly merge them into $(1, v_{\mathrm{new}}) = \{\boldsymbol{z}_1^3, \boldsymbol{z}_2^3\}$.
\end{example}

The post-burn-in transition kernel $\mathcal{K}$ formed by Steps~1 and~2 satisfies Theorem~\ref{thm:ergodic}~\cite{Xia2024TAES} (proved in Appendix~D).

\begin{theorem}[Ergodic Convergence of the Double-Bounce DA Sampler] \label{thm:ergodic}
Assume finite total measurements, clutter intensity $c(\boldsymbol{z}) > 0$, prior intensity $\lambda_\psi(\boldsymbol{x}) > 0$ over the active region, and valid path detection probabilities in $(0, 1)$. On the finite valid partition space $\mathbb{J}$, $\mathcal{K}$ is stationary, irreducible, and aperiodic, converging asymptotically from any initial partition $\mathcal{A}^{(0)} \in \mathbb{J}$ to the approximated marginal posterior distribution $\hat{p}(\mathcal{A}|\boldsymbol{s}_{0:K}, \mathcal{Z}_{1:K}) \propto \hat{w}(\mathcal{A})$.
\end{theorem}

\subsection{Conditional Sampling of Landmark Existence and Type} \label{sec:type_sampling}

Given partition sample $\mathcal{A}$, taking the set integral of~\eqref{eq:aug_post} over $\mathcal{X}$ yields the second conditional distribution in~\eqref{eq:da_factor}:
\begin{align}
&p(\boldsymbol{\psi}|\mathcal{A}, \boldsymbol{s}_{0:K}, \mathcal{Z}_{1:K}) \propto \int\!\cdots\!\int \prod_{i=1}^N \tilde{L}_1(\mathcal{Z}_i^{(1)}, \mathcal{Y}_i|\boldsymbol{s}_{1:K}, \psi_i) \notag\\
&\quad \times \prod_{u \neq v}^N \tilde{L}_2(\mathcal{Z}_{uv}^{(2)}|\boldsymbol{s}_{1:K}, \mathcal{Y}_u, \mathcal{Y}_v; \eta_{uv})\,\delta\mathcal{Y}_{1:N}, \label{eq:type_exact}
\end{align}
where $\delta\mathcal{Y}_{1:N} \triangleq \delta\mathcal{Y}_1 \cdots \delta\mathcal{Y}_N$. Owing to finite wall lengths and interior blockage unmodeled by infinite-line VAs, many landmark pairs generate no 2-bounce paths even within the nominal half-plane FoV; moreover, retaining $\tilde{L}_2(\emptyset|\cdot)$ on all empty pairs would couple all $N_{\mathrm{vis}}$ visible candidates into a complete graph ($3^{N_{\mathrm{vis}}}$ states) and over-penalize VA existence ($\psi_i \neq 0$). We thus set $\tilde{L}_2(\emptyset|\cdot) \approx 1$ during type sampling and retain only active links $\mathcal{E}_{\mathrm{vis}} \triangleq \{(u, v) : u \neq v \in \mathcal{I}_{\mathrm{vis}}, \mathcal{Z}_{uv}^{(2)} \neq \emptyset\}$:
\begin{align}
&p(\boldsymbol{\psi}|\mathcal{A}, \boldsymbol{s}_{0:K}, \mathcal{Z}_{1:K}) \propto \prod_{i \in \mathcal{I}_{\mathrm{vis}}} \mathcal{L}_1(\mathcal{Z}_i^{(1)}; \psi_i) \prod_{v \in \mathcal{I}_{\mathrm{occ}}} \mathcal{L}_{2,\mathrm{occ}}(v; \boldsymbol{\psi}) \notag\\
&\quad \times \prod_{(u,v) \in \mathcal{E}_{\mathrm{vis}}} \mathcal{L}_2(\mathcal{Z}_{uv}^{(2)}|\mathcal{Z}_u^{(1)}, \mathcal{Z}_v^{(1)}; \eta_{uv}), \label{eq:type_fact}
\end{align}
where $\mathcal{L}_1(\mathcal{Z}_i^{(1)}; 0) = \mathbb{I}(|\mathcal{Z}_i^{(1)}|=1)c(\boldsymbol{z})$, terms with $\eta_{uv} \notin \mathcal{T}_2$ take value $0$, and all evaluations reuse the results computed in Section~\ref{sec:da_weight}.

In~\eqref{eq:type_fact}, candidates not participating in any non-empty 2-bounce cluster are conditionally independent and sampled directly from their 1-bounce posterior probabilities $(\pi_{i,0},\allowbreak \pi_{i,\mathrm{VA}},\allowbreak \pi_{i,\mathrm{RP}})$, whereas candidates linked through non-empty 2-bounce clusters naturally partition into mutually uncoupled subsets $\mathcal{G}_1,\allowbreak \dots,\allowbreak \mathcal{G}_M$. For each coupled subset $\mathcal{G}_m$, $\boldsymbol{\psi}_{\mathcal{G}_m}$ is sampled over its compatible state combinations $\Psi(\mathcal{G}_m)$ proportional to the joint subset weight $W_{\mathcal{G}_m}(\boldsymbol{\psi}_{\mathcal{G}_m})$:
\begin{align}
P(\boldsymbol{\psi}_{\mathcal{G}_m}|\mathcal{A}, \boldsymbol{s}_{0:K}, \mathcal{Z}_{1:K}) = \frac{W_{\mathcal{G}_m}(\boldsymbol{\psi}_{\mathcal{G}_m})}{\sum_{\boldsymbol{\psi}_{\mathcal{G}_m}' \in \Psi(\mathcal{G}_m)} W_{\mathcal{G}_m}(\boldsymbol{\psi}_{\mathcal{G}_m}')}. \label{eq:type_sub}
\end{align}
Moreover, because each occluded candidate $v \in \mathcal{I}_{\mathrm{occ}}$ associates with only a single visible candidate $u \in \mathcal{I}_{\mathrm{vis}}$, $\psi_v$ can be summed out beforehand when evaluating~\eqref{eq:type_sub} to reduce subset dimension and then sampled conditionally given $\psi_u$ (fast elimination details in Appendix~D).

\subsection{Warm-Start Pre-Clustering} \label{sec:warmstart}

Prior to iteration~1, we perform preliminary geometric clustering on $\mathcal{Z}_{1:K}$ using the dead-reckoning trajectory $\boldsymbol{s}_{0:K}^{(0)}$ propagated from odometry to construct $\mathcal{A}^{(0)}$ (pseudocode in Appendix~E). Specifically, LoS measurements are first gated per scan, and each remaining measurement $[d, \theta, \phi]^{\mathsf{T}}$ is tested against 1-bounce reflection geometry: under initial pose $\boldsymbol{s}_k^{(0)} = [(\boldsymbol{p}_k^{(0)})^{\mathsf{T}}, \alpha_k^{(0)}]^{\mathsf{T}}$, with baseline $\boldsymbol{b}_k = \boldsymbol{p}_k^{(0)} - \boldsymbol{p}_{\mathrm{BS}}$ and RX direction unit vector $\boldsymbol{e}_{\mathrm{RX}} = [\cos(\alpha_k^{(0)}+\theta), \sin(\alpha_k^{(0)}+\theta)]^{\mathsf{T}}$, 1-bounce geometry yields the candidate reflection point $\boldsymbol{p}_{\mathrm{ref}}$ and AoD residual ($\angle\boldsymbol{v} \triangleq \operatorname{atan2}([\boldsymbol{v}]_2, [\boldsymbol{v}]_1)$):
\begin{align}
&r_2 = \frac{d^2 - \|\boldsymbol{b}_k\|^2}{2(d + \boldsymbol{b}_k^{\mathsf{T}}\boldsymbol{e}_{\mathrm{RX}})}, \qquad \boldsymbol{p}_{\mathrm{ref}} = \boldsymbol{p}_k^{(0)} + r_2\boldsymbol{e}_{\mathrm{RX}}, \notag\\
&e_{\phi,\mathrm{SB}} \triangleq \Big|\mathrm{wrap}\!\big(\phi - (\angle(\boldsymbol{p}_{\mathrm{ref}} - \boldsymbol{p}_{\mathrm{BS}}) - \alpha_{\mathrm{BS}})\big)\Big|. \label{eq:warmstart}
\end{align}
Measurements satisfying $e_{\phi,\mathrm{SB}} < \tau_{\phi,\mathrm{SB}}$ are spatially clustered into initial 1-bounce visible clusters; measurements inconsistent with 1-bounce geometry are subsequently matched against the resulting visible clusters to form initial 2-bounce clusters (Appendix~C), while any remaining unmatched measurements are initialized as 1-bounce singletons.

\section{Conditional Double-Bounce GraphSLAM and Sample Marginalization} \label{sec:graphslam}

\subsection{Conditional Double-Bounce GraphSLAM} \label{sec:cond_slam}

Fixing an augmented DA sample $\tilde{\mathcal{A}} = (\mathcal{A}, \boldsymbol{\psi})$ in~\eqref{eq:aug_post}, the conditional posterior $f(\boldsymbol{s}_{0:K}, \mathcal{X}|\mathcal{Z}_{1:K}, \tilde{\mathcal{A}})$ factorizes into the conditional marked PPP density of undetected landmarks $\mathcal{X}_{\mathrm{U}}$,
\begin{align}
f(\mathcal{X}_{\mathrm{U}}|\boldsymbol{s}_{0:K}, \mathcal{Z}_{1:K}) &= \exp\!\biggl(-\!\sum_{\psi \in \mathcal{T}_1}\!\int p_{\mathrm{U}}(\tilde{\boldsymbol{x}}, \boldsymbol{s}_{1:K})\lambda_\psi(\boldsymbol{x})\,\mathrm{d}\boldsymbol{x}\biggr) \notag\\
&\quad \times \prod_{\tilde{\boldsymbol{x}} \in \mathcal{X}_{\mathrm{U}}} p_{\mathrm{U}}(\tilde{\boldsymbol{x}}, \boldsymbol{s}_{1:K})\lambda_\psi(\boldsymbol{x}), \label{eq:undetected_ppp}
\end{align}
and the conditional posterior of the trajectory and confirmed landmarks. Removing clutter ($\psi_i = 0$) and reindexing the $\kappa \triangleq \sum_{i=1}^N \mathbb{I}(\psi_i \neq 0)$ confirmed landmarks as $\{1, \dots, \kappa\}$ (with confirmed visible subset $\mathcal{I}_{\mathrm{vis}}^+$) yields~\cite{Ge2025TSP}
\begin{align}
&f(\boldsymbol{s}_{0:K}, \boldsymbol{x}_1, \dots, \boldsymbol{x}_\kappa|\mathcal{Z}_{1:K}, \tilde{\mathcal{A}}) \propto f(\boldsymbol{s}_0)L_0(\mathcal{Z}_0|\boldsymbol{s}_{1:K}) \notag\\
&\;\times \prod_{k=1}^K f(\boldsymbol{s}_k|\boldsymbol{s}_{k-1}) \prod_{i=1}^\kappa \lambda_{\psi_i}(\boldsymbol{x}_i)\prod_{k=1}^K \ell_1(\mathcal{Z}_{i,k}^{(1)}|\tilde{\boldsymbol{x}}_i, \boldsymbol{s}_k) \notag\\
&\;\times \prod_{u \neq v}^\kappa \prod_{k=1}^K \ell_2(\mathcal{Z}_{uv,k}^{(2)}|\tilde{\boldsymbol{x}}_u, \tilde{\boldsymbol{x}}_v, \boldsymbol{s}_k). \label{eq:cond_map_post}
\end{align}

Following~\cite{Thrun2006IJRR,Ge2025TSP}, we perform MAP estimation on~\eqref{eq:cond_map_post} (updating~\eqref{eq:undetected_ppp} in Section~\ref{sec:marg}), approximate active in-FoV detection probabilities as locally constant, and replace $\lambda_{\psi_i}(\boldsymbol{x}_i)$ with a Gaussian prior $f_{\mathrm{prior}}(\boldsymbol{x}_i) = \mathcal{N}(\boldsymbol{x}_i; \boldsymbol{\mu}_{i,\mathrm{prior}}, \boldsymbol{C}_{i,\mathrm{prior}})$ centered at the earliest-detection inversion with large covariance $\boldsymbol{C}_{i,\mathrm{prior}}$. Stacking all continuous states into $\boldsymbol{q} \triangleq [\boldsymbol{s}_0^{\mathsf{T}}, \dots, \boldsymbol{s}_K^{\mathsf{T}}, \boldsymbol{x}_1^{\mathsf{T}}, \dots, \boldsymbol{x}_\kappa^{\mathsf{T}}]^{\mathsf{T}} \in \mathbb{R}^{3(K+1) + 2\kappa}$ and denoting the set of non-empty 2-bounce pairs by $\mathcal{E}_2$, maximizing~\eqref{eq:cond_map_post} reduces to nonlinear LS $\hat{\boldsymbol{q}} = \arg\min_{\boldsymbol{q}} \mathcal{E}(\boldsymbol{q})$, where
\begin{align}
\mathcal{E}(\boldsymbol{q}) &= (\boldsymbol{s}_0 - \boldsymbol{\epsilon}_{0,\mathrm{prior}})^{\mathsf{T}}\boldsymbol{P}_{0,\mathrm{prior}}^{-1}(\boldsymbol{s}_0 - \boldsymbol{\epsilon}_{0,\mathrm{prior}}) \notag\\[-3pt]
&\;+ \sum_{k=1}^K \boldsymbol{r}_{\mathrm{mot},k}^{\mathsf{T}}\boldsymbol{Q}_{k-1}^{-1}\boldsymbol{r}_{\mathrm{mot},k}
+ \sum_{k:\,\boldsymbol{z}_{0,k} \in \mathcal{Z}_{0,k}} \boldsymbol{r}_{0,k}^{\mathsf{T}}\boldsymbol{R}_{\mathrm{LoS}}^{-1}\boldsymbol{r}_{0,k} \notag\\[-1.5pt]
&\;+ \sum_{i=1}^\kappa (\boldsymbol{x}_i - \boldsymbol{\mu}_{i,\mathrm{prior}})^{\mathsf{T}}\boldsymbol{C}_{i,\mathrm{prior}}^{-1}(\boldsymbol{x}_i - \boldsymbol{\mu}_{i,\mathrm{prior}}) \notag\\[-2.5pt]
&\;+ \sum_{i \in \mathcal{I}_{\mathrm{vis}}^+} \sum_{k:\,\boldsymbol{z}_k \in \mathcal{Z}_{i,k}^{(1)}} \boldsymbol{r}_{1,i,k}^{\mathsf{T}}\boldsymbol{R}_{\psi_i}^{-1}\boldsymbol{r}_{1,i,k} \notag\\[-2pt]
&\;+ \sum_{(u,v) \in \mathcal{E}_2} \sum_{k:\,\boldsymbol{z}_k \in \mathcal{Z}_{uv,k}^{(2)}} \boldsymbol{r}_{2,uv,k}^{\mathsf{T}}\boldsymbol{R}_{\psi_u,\psi_v}^{-1}\boldsymbol{r}_{2,uv,k}, \label{eq:graphslam_cost}
\end{align}
with residuals $\boldsymbol{r}_{\mathrm{mot},k} \triangleq \boldsymbol{s}_k - \boldsymbol{v}(\boldsymbol{s}_{k-1}, \boldsymbol{u}_{k-1})$, $\boldsymbol{r}_{0,k} \triangleq \boldsymbol{z}_{0,k} - \boldsymbol{h}_{\mathrm{LoS}}(\boldsymbol{s}_k)$, $\boldsymbol{r}_{1,i,k} \triangleq \boldsymbol{z}_k - \boldsymbol{h}_{\psi_i}(\boldsymbol{x}_i, \boldsymbol{s}_k)$, and $\boldsymbol{r}_{2,uv,k} \triangleq \boldsymbol{z}_k - \boldsymbol{h}_{\psi_u,\psi_v}(\boldsymbol{x}_u, \boldsymbol{x}_v, \boldsymbol{s}_k)$ (wrapping angular residuals into $(-\pi, \pi]$).

Linearizing~\eqref{eq:graphslam_cost} with the Jacobians in Appendix~A and solving via Levenberg-Marquardt (LM) yields the state estimate $\hat{\boldsymbol{q}}$ and the joint information matrix $\boldsymbol{\Omega}$. Partitioning $\boldsymbol{\Omega}$ into trajectory and landmark blocks gives the trajectory block $\boldsymbol{\Omega}_{\mathrm{ss}} \in \mathbb{R}^{3(K+1)\times 3(K+1)}$, the landmark block $\boldsymbol{\Omega}_{\mathrm{xx}} \in \mathbb{R}^{2\kappa\times 2\kappa}$, and the trajectory-landmark cross-information block $\boldsymbol{\Omega}_{\mathrm{sx}} \in \mathbb{R}^{3(K+1)\times 2\kappa}$. Because 2-bounce factors couple only landmark pairs $(u,v) \in \mathcal{E}_2$, ordering confirmed landmarks by their 2-bounce coupled subsets $\mathcal{G}_1^+, \dots, \mathcal{G}_{M_+}^+$ renders the landmark block block-diagonal, i.e., $\boldsymbol{\Omega}_{\mathrm{xx}} = \operatorname{blkdiag}(\boldsymbol{\Omega}_{\mathrm{xx}}^{(1)}, \dots, \boldsymbol{\Omega}_{\mathrm{xx}}^{(M_+)})$ (Appendix~D), reducing the inversion cost of $\boldsymbol{\Omega}_{\mathrm{xx}}$ from $O(\kappa^3)$ to $O(\sum_{m=1}^{M_+} |\mathcal{G}_m^+|^3)$. Using the Schur complement $\tilde{\boldsymbol{\Omega}}_{\mathrm{ss}} \triangleq \boldsymbol{\Omega}_{\mathrm{ss}} - \boldsymbol{\Omega}_{\mathrm{sx}}\boldsymbol{\Omega}_{\mathrm{xx}}^{-1}\boldsymbol{\Omega}_{\mathrm{sx}}^{\mathsf{T}}$, we recover the trajectory mean $\boldsymbol{\epsilon}_{0:K} = [\hat{\boldsymbol{q}}]_{1:3(K+1)}$ and covariance $\boldsymbol{P}_{0:K} = \tilde{\boldsymbol{\Omega}}_{\mathrm{ss}}^{-1}$, and from $\boldsymbol{\mu}_{\mathrm{map}} = [\hat{\boldsymbol{q}}]_{3(K+1)+1:\mathrm{end}}$ and $\boldsymbol{C}_{\mathrm{map}} = \boldsymbol{\Omega}_{\mathrm{xx}}^{-1} + \boldsymbol{\Omega}_{\mathrm{xx}}^{-1}\allowbreak\boldsymbol{\Omega}_{\mathrm{sx}}^{\mathsf{T}}\allowbreak\boldsymbol{P}_{0:K}\allowbreak\boldsymbol{\Omega}_{\mathrm{sx}}\allowbreak\boldsymbol{\Omega}_{\mathrm{xx}}^{-1}$ we extract each landmark's marginal mean $\boldsymbol{\mu}_i = [\boldsymbol{\mu}_{\mathrm{map}}]_{2i-1:2i}$ and covariance $\boldsymbol{C}_i = [\boldsymbol{C}_{\mathrm{map}}]_{2i-1:2i,\,2i-1:2i}$ ($i = 1, \dots, \kappa$). The updated trajectory mean $\boldsymbol{\epsilon}_{0:K}$ is fed back to Section~\ref{sec:da_sampling} for the next iteration.

\subsection{Sample Marginalization and Multi-Type PMB Fusion} \label{sec:marg}

After burn-in, the conditional posterior estimates from the final $\Gamma$ iterations $\{\tilde{\mathcal{A}}^t\}_{t=1}^\Gamma$ are averaged to marginalize out the augmented DA~\cite{Ge2025TSP}:
\begin{align}
f(\boldsymbol{s}_{0:K}|\mathcal{Z}_{1:K}) &\approx \frac{1}{\Gamma}\sum_{t=1}^{\Gamma} f(\boldsymbol{s}_{0:K}|\mathcal{Z}_{1:K}, \tilde{\mathcal{A}}^t), \notag\\
f(\mathcal{X}_{\mathrm{D}}|\mathcal{Z}_{1:K}) &\approx \frac{1}{\Gamma}\sum_{t=1}^{\Gamma} f(\mathcal{X}_{\mathrm{D}}|\mathcal{Z}_{1:K}, \tilde{\mathcal{A}}^t). \label{eq:sample_avg}
\end{align}
From~\eqref{eq:sample_avg}, with per-sample pose mean $\boldsymbol{\epsilon}_k^t = [(\boldsymbol{p}_k^t)^{\mathsf{T}}, \alpha_k^t]^{\mathsf{T}}$, the fused trajectory mean and covariance are given by
\begin{align}
&\boldsymbol{p}_k = \frac{1}{\Gamma}\sum_{t=1}^{\Gamma} \boldsymbol{p}_k^t, \;\; \alpha_k = \operatorname{atan2}\!\left(\sum_{t=1}^{\Gamma}\sin\alpha_k^t,\, \sum_{t=1}^{\Gamma}\cos\alpha_k^t\right), \notag\\
&\boldsymbol{P}_{0:K} = \frac{1}{\Gamma}\sum_{t=1}^{\Gamma}\Big(\boldsymbol{P}_{0:K}^t + \Delta\boldsymbol{\epsilon}_{0:K}^t(\Delta\boldsymbol{\epsilon}_{0:K}^t)^{\mathsf{T}}\Big), \label{eq:traj_merge}
\end{align}
where $\Delta\boldsymbol{\epsilon}_k^t = [(\boldsymbol{p}_k^t - \boldsymbol{p}_k)^{\mathsf{T}}, \mathrm{wrap}(\alpha_k^t - \alpha_k)]^{\mathsf{T}}$.

For the detected map, since VA/RP coordinates and 2-bounce reflection-order roles differ geometrically, they cannot be averaged across types. We thus assign each confirmed landmark $i \in \{1, \dots, \kappa^t\}$ in sample $t$ a key combining bounce order, reflection role, type, and earliest index $m_{\mathrm{fir}}$:
\begin{equation}
\sigma_i^t\!=\!\left\{\!\begin{array}{@{}l@{\;}l@{}}
(1{,}\psi_i^t{,}m_{i,\mathrm{fir}}^t), & \ccond{i\!\in\!\mathcal{I}_{\mathrm{vis}}^{t,+},} \\[1.5pt]
(2{,}1{,}\psi_i^t{,}\psi_u^t{,}m_{i,\mathrm{fir}}^t{,}m_{u,\mathrm{fir}}^t), & \tcond{$i\!\in\!\mathcal{I}_{\mathrm{occ}}^{t}$, 1st in $(i{,}u)$,} \\[1.5pt]
(2{,}2{,}\psi_u^t{,}\psi_i^t{,}m_{u,\mathrm{fir}}^t{,}m_{i,\mathrm{fir}}^t), & \tcond{$i\!\in\!\mathcal{I}_{\mathrm{occ}}^{t}$, 2nd in $(u{,}i)$.}
\end{array}\right. \label{eq:lmk_key}
\end{equation}
Let $\Sigma_{\mathrm{all}} = \{\bar{\sigma}_1, \dots, \bar{\sigma}_{|\mathbb{I}|}\}$ collect all unique keys across the $\Gamma$ samples. To form the MB approximation in Section~II-B~\cite{Ge2025TSP}, each $\bar{\sigma}_i$ defines a Bernoulli component with indicator $\tilde{\sigma}^t(i) \in \{0, 1\}$, existence $r_i = \frac{1}{\Gamma}\sum_{t=1}^{\Gamma} \tilde{\sigma}^t(i)$, mean $\boldsymbol{\mu}_i = \frac{1}{\Gamma r_i}\sum_{t:\,\tilde{\sigma}^t(i)=1} \boldsymbol{\mu}_i^t$, and covariance $\boldsymbol{C}_i = \frac{1}{\Gamma r_i}\sum_{t:\,\tilde{\sigma}^t(i)=1} \bigl(\boldsymbol{C}_i^t + \allowbreak(\boldsymbol{\mu}_i^t - \boldsymbol{\mu}_i)(\boldsymbol{\mu}_i^t - \boldsymbol{\mu}_i)^{\mathsf{T}}\bigr)$. After pruning $r_i < \vartheta_{\mathrm{prune}}$, following~\cite{GarciaFernandez2018TAES,Ge2025TSP} to merge nearby same-type components across samples or distinct anchors, we iteratively pick the highest-existence unmerged component $i^\star$ and merge same-type components within gate $\gamma_{\mathrm{merge}}$ as $\mathcal{I}_{\mathrm{merge}}$ via $\bar{w}_i = r_i / \sum_{i' \in \mathcal{I}_{\mathrm{merge}}} r_{i'}$, yielding $\hat{\psi}_m = \psi_{i^\star}$, $\hat{r}_m = \frac{1}{\Gamma}\sum_{t=1}^{\Gamma} \max_{i \in \mathcal{I}_{\mathrm{merge}}} \tilde{\sigma}^t(i)$, $\hat{\boldsymbol{\mu}}_m = \sum_{i \in \mathcal{I}_{\mathrm{merge}}} \bar{w}_i\boldsymbol{\mu}_i$, and $\hat{\boldsymbol{C}}_m = \sum_{i \in \mathcal{I}_{\mathrm{merge}}} \bar{w}_i\bigl(\boldsymbol{C}_i + \allowbreak(\boldsymbol{\mu}_i - \hat{\boldsymbol{\mu}}_m)(\boldsymbol{\mu}_i - \hat{\boldsymbol{\mu}}_m)^{\mathsf{T}}\bigr)$. Averaging~\eqref{eq:undetected_ppp} gives the undetected intensity $\check{\lambda}_\psi(\boldsymbol{x}) \approx \bigl(\frac{1}{\Gamma}\sum_{t=1}^\Gamma p_{\mathrm{U}}(\tilde{\boldsymbol{x}}, \boldsymbol{\epsilon}_{1:K}^t)\bigr)\lambda_\psi(\boldsymbol{x})$, forming the $\hat{M}$-component multi-type PMB posterior:
\begin{align}
\hat{f}(\mathcal{X}|\mathcal{Z}_{1:K}) &= \sum_{\mathcal{X}_{\mathrm{U}} \biguplus \mathcal{X}_1 \biguplus \dots \biguplus \mathcal{X}_{\hat{M}} = \mathcal{X}} \hat{f}_{\mathrm{P}}(\mathcal{X}_{\mathrm{U}})\prod_{m=1}^{\hat{M}} \hat{f}_{\mathrm{B},m}(\mathcal{X}_m), \notag\\
\hat{f}_{\mathrm{B},m}(\mathcal{X}_m) &=
\begin{cases}
1 - \hat{r}_m, & \ccond{\mathcal{X}_m = \emptyset,} \\
\hat{r}_m\mathcal{N}(\boldsymbol{x}; \hat{\boldsymbol{\mu}}_m, \hat{\boldsymbol{C}}_m), & \ccond{\mathcal{X}_m = \{(\boldsymbol{x}, \hat{\psi}_m)\},} \\
0, & \tcond{otherwise,}
\end{cases} \label{eq:pmb_out}
\end{align}
where $\hat{f}_{\mathrm{P}}(\mathcal{X}_{\mathrm{U}})$ is the marked PPP density with intensity $\check{\lambda}_\psi(\boldsymbol{x})$, and a point map estimate can be extracted using an existence threshold $\hat{r}_m \ge \vartheta_{\mathrm{det}}$. The complete batch radio SLAM pseudocode is summarized in Appendix~E.

\section{Simulation and Experimental Results} \label{sec:results}

This section evaluates the proposed algorithm in an indoor ray-tracing (RT) scenario (Sections~\ref{sec:exp_setup}--\ref{sec:exp_rt_robust}) and on real 60~GHz indoor measurements (Section~\ref{sec:exp_real}).

\begin{figure}[!b]
\centering
\includegraphics[width=\columnwidth]{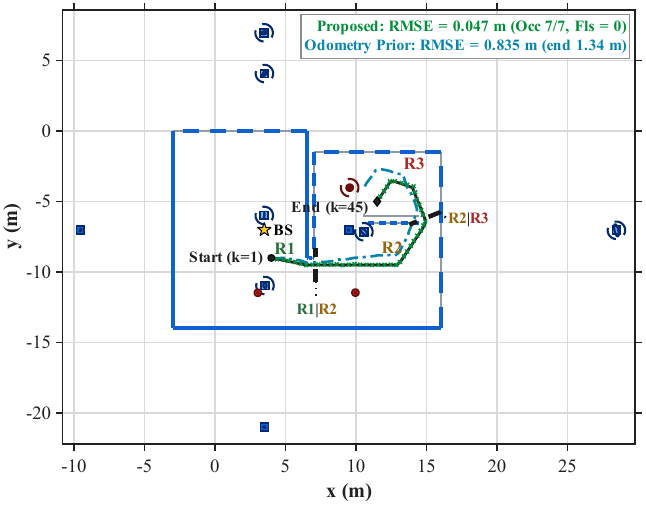}\\[2.5pt]
\includegraphics[width=\columnwidth]{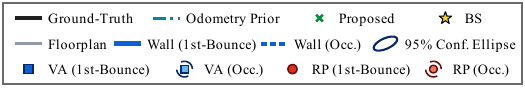}
\caption{Indoor RT simulation scenario and representative localization and mapping results of the proposed method under regime G1, showing the dead-reckoning odometry prior, three visibility stages (R1--R3), $95\%$ landmark confidence ellipses, and seven occluded landmarks observable via 2-bounce paths.}
\label{fig:rt_proposed}
\end{figure}

\begin{figure*}[!t]
\centering
\begin{minipage}[t]{0.320\textwidth}
\noindent\hspace*{0.136\linewidth}\makebox[0.844\linewidth][c]{\footnotesize\bfseries (a) Snapshot DB-SLAM ($k = 1\text{--}16$)}\\[2pt]
\noindent\includegraphics[width=\textwidth]{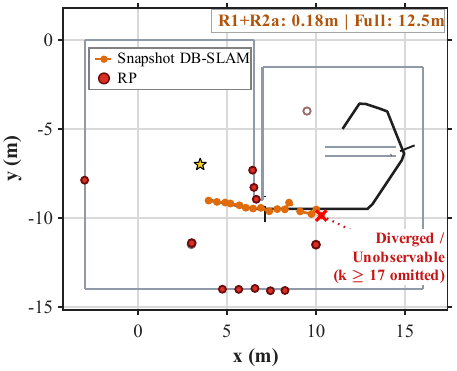}
\end{minipage}\hfill%
\begin{minipage}[t]{0.320\textwidth}
\noindent\hspace*{0.136\linewidth}\makebox[0.844\linewidth][c]{\footnotesize\bfseries (b) 1-b Batch SLAM}\\[2pt]
\noindent\includegraphics[width=\textwidth]{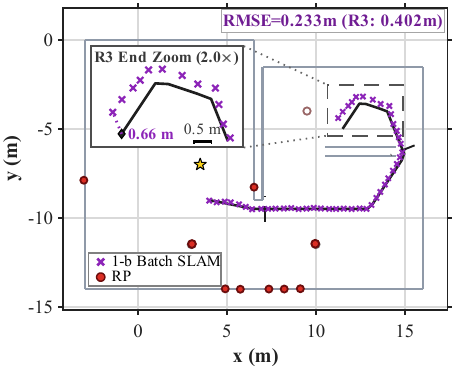}
\end{minipage}\hfill%
\begin{minipage}[t]{0.338\textwidth}
\noindent\hspace*{0.132\linewidth}\makebox[0.846\linewidth][c]{\footnotesize\bfseries (c) 2-b BP-SLAM}\\[2pt]
\noindent\includegraphics[width=\textwidth]{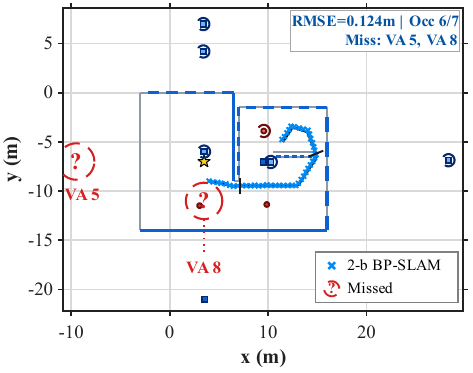}
\end{minipage}
\caption{Representative localization and mapping results of the three baselines in the indoor RT scenario under regime G1: (a)~Snapshot DB-SLAM (local view over $k = 1\text{--}16$ prior to rank deficiency); (b)~1-b Batch SLAM; and (c)~2-b BP-SLAM. Shared map and ground-truth symbols follow Fig.~\ref{fig:rt_proposed}.}
\label{fig:rt_baselines}
\end{figure*}

\subsection{Evaluation Setup, Baselines, and Metrics} \label{sec:exp_setup}

We first consider an indoor bistatic RT scenario containing a static BS at $\boldsymbol{p}_{\mathrm{BS}} = [3.5, -7.0]^{\mathsf{T}}\,\mathrm{m}$ ($\alpha_{\mathrm{BS}} = 0\,\mathrm{rad}$) and 12 identifiable landmarks: nine planar walls (VAs~1--9) and three cylindrical point scatterers (RPs~1--3), as shown in Fig.~\ref{fig:rt_proposed}. Five landmarks are directly observable via 1-bounce paths, whereas seven (VAs~1--3, 7--9 and RP~3) have no 1-bounce visibility along the trajectory and are observable only through 2-bounce reflections. The RT simulator operates at $60\,\mathrm{GHz}$ with transmit power $20\,\mathrm{dBm}$, unit-gain antennas ($0\,\mathrm{dBi}$), and a maximum reflection order of $3$, retaining multipath components above a minimum received-power threshold of $-120\,\mathrm{dBm}$ and within a $99.5\%$ cumulative multipath energy threshold. As the mobile RX moves over $K = 45$ steps around an interior partition, visibility transitions across three stages: \emph{Stage~R1} ($k = 1\text{--}8$, LoS available), \emph{Stage~R2} ($k = 9\text{--}31$, obstructed LoS with 1-bounce and higher-order paths), and \emph{Stage~R3} ($k = 32\text{--}45$, deep NLoS where LoS and 1-bounce paths are blocked and only 2-bounce and higher-order paths remain). Over the 45 scans, the RT simulator produces 228 paths up to second order (8 LoS, 88 1-bounce, and 132 2-bounce) and 22 third-order (3-bounce) reflections; independent Poisson clutter is added with a mean rate of $0.8$ measurements per scan.

Four noise regimes are evaluated over 50 Monte Carlo (MC) runs each: \emph{G1} ($\boldsymbol{R}, \boldsymbol{Q}$), \emph{G2} ($4\boldsymbol{R}, \boldsymbol{Q}$), \emph{G3} ($\boldsymbol{R}, 4\boldsymbol{Q}$), and \emph{G4} ($4\boldsymbol{R}, 4\boldsymbol{Q}$), with nominal $\boldsymbol{R} = \operatorname{diag}(0.05^2\,\mathrm{m}^2, (0.8^\circ)^2, (0.8^\circ)^2)$, $\boldsymbol{Q} = \operatorname{diag}(0.05^2\,\mathrm{m}^2, 0.05^2\,\mathrm{m}^2, (0.9^\circ)^2)$, and scan-$1$ prior $\boldsymbol{P}_{1,\mathrm{prior}} = \operatorname{diag}(0.025^2\,\mathrm{m}^2, 0.025^2\,\mathrm{m}^2, (0.5^\circ)^2)$. In-FoV ($48/68\,\mathrm{m}$ for 1-/2-bounce, valid half-plane) detection probabilities are $(p_{\mathrm{D},0}, p_{\mathrm{D},1}, p_{\mathrm{D},2}) = (0.62, 0.42, 0.04)$ ($10^{-3}$ outside); over the $2912\,\mathrm{m}^2$ region, $(\lambda_{\mathrm{VA}}, \lambda_{\mathrm{RP}}) = (6, 3)/2912\,\mathrm{m}^{-2}$ and $c(\boldsymbol{z}) = 1.4/(55\cdot 4\pi^2)$. We set $N_{\mathrm{burn}} = \Gamma = 10$, $T_{\mathrm{sweep}} = 1$, $\vartheta_{\mathrm{prune}} = 0.1$, $\vartheta_{\mathrm{det}} = 0.5$, and $\gamma_{\mathrm{merge}}$ as squared Mahalanobis distance $< 9.0$ (or $0.5\,\mathrm{m}$).

We compare the proposed method against three baselines: (i)~\emph{Snapshot DB-SLAM}~\cite{Zhang2026TWC}, which estimates single-scan RX pose and interaction points from up to 2-bounce paths without motion priors; (ii)~\emph{1-b Batch SLAM}~\cite{Ge2025TSP}, which performs batch MCMC DA sampling and GraphSLAM under a single-bounce RP model; and (iii)~\emph{2-b BP-SLAM}, a sequential particle-based belief propagation (BP) filter with $10{,}000$ particles built on the double-bounce BP framework of~\cite{Leitinger2023TSP} and extended with multi-type (VA and RP) modeling and 2-bounce occluded-landmark birth following~\cite{Li2024ICC}.

Localization accuracy is evaluated via RX position and heading RMSEs, and mapping accuracy via the generalized optimal subpattern assignment (GOSPA) distance~\cite{Rahmathullah2017GOSPA} ($c = 2\,\mathrm{m}$, $p = 2$, $\alpha = 2$). We also report each method's \emph{associated measurement count} (and ratio), measuring how many measurements are associated with LoS or landmark hypotheses rather than rejected as clutter: batch methods weight each measurement by its post-burn-in assignment frequency to LoS or confirmed landmarks ($\hat{r}_m \ge \vartheta_{\mathrm{det}}$), 2-b BP-SLAM by its online marginal association probability, and Snapshot DB-SLAM by LoS/1-bounce/2-bounce inlier classification. Associated measurements are stratified into five physical RT path categories. Hence, counts in unmodeled categories (e.g., 3-bounce/clutter, or 2-bounce returns under 1-b Batch SLAM) indicate false absorption (mis-association to modeled landmarks) rather than true physical modeling. Finally, inverting the parametric joint-SLAM Fisher information matrix (FIM) around the fixed RT ground-truth trajectory and map under clutter-free known data association with zero landmark priors~\cite{Kay1993,Ge2025TSP,Deutschmann2025Asilomar} yields the \emph{1-b Batch}, \emph{2-b Filtering} (causal prefix), and \emph{2-b Batch CRLBs} (Appendix~D).

\subsection{Ray-Tracing Results: Localization, Mapping, and Bounds} \label{sec:exp_rt_nominal}

Figs.~\ref{fig:rt_proposed} and~\ref{fig:rt_baselines} show a representative MC run under regime~G1. As shown in Fig.~\ref{fig:rt_proposed}, the proposed method reduces the trajectory position RMSE from the odometry prior of $0.835\,\mathrm{m}$ (final drift $1.34\,\mathrm{m}$) to $0.047\,\mathrm{m}$, reconstructing all 12 landmarks (including $7/7$ occluded landmarks) with zero false alarms. Compared with Snapshot DB-SLAM in Fig.~\ref{fig:rt_baselines}(a), which achieves a position RMSE of $0.18\,\mathrm{m}$ during R1 and early R2 ($k \le 16$) where sufficient LoS and 1-bounce rays are present but encounters single-scan observability rank deficiency in deep occlusion ($k \ge 17$) due to fewer 1-bounce paths, more higher-order reflections, and the absence of motion priors, the proposed method maintains continuous localization across all stages. Compared with 1-b Batch SLAM in Fig.~\ref{fig:rt_baselines}(b), which considers only 1-bounce paths and treats planar walls as point scatterers (RPs), thereby increasing RP false alarms and drifting in R3 once 1-bounce paths disappear (with a terminal position error of $0.66\,\mathrm{m}$), the proposed method jointly models VAs and RPs and constrains R3 via 2-bounce observations. Compared with 2-b BP-SLAM in Fig.~\ref{fig:rt_baselines}(c), which maintains R3 tracking via occluded-landmark birth (RMSE $0.124\,\mathrm{m}$) but misses VA~5 (whose existence weight decays under R3 occlusion) and VA~8 (observed only at $k = 9, 10$ without accumulating sufficient existence probability), the proposed batch formulation jointly exploits observations across all scans to avoid both missed detections, thereby improving localization and mapping accuracy.

\begin{figure}[!t]
\centering
\includegraphics[width=\columnwidth]{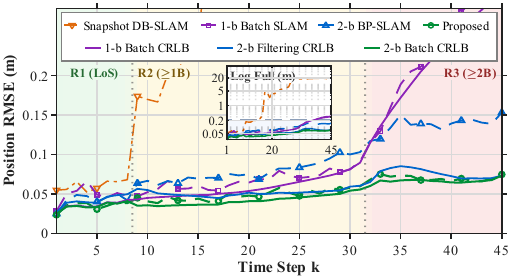}\\[-1.5pt]
\sublbl{0.543}{(a)}\\[2pt]
\begin{minipage}[t]{0.495\columnwidth}
\centering
\includegraphics[width=\textwidth]{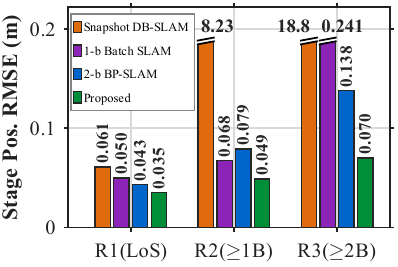}\\[-1.5pt]
\sublbl{0.574}{(b)}
\end{minipage}\hfill%
\begin{minipage}[t]{0.495\columnwidth}
\centering
\includegraphics[width=\textwidth]{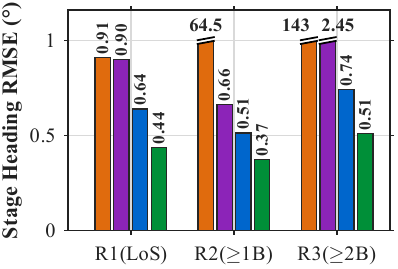}\\[-1.5pt]
\sublbl{0.574}{(c)}
\end{minipage}\\[2pt]
\includegraphics[width=\columnwidth]{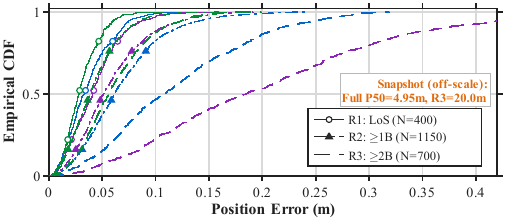}\\[-1.5pt]
\sublbl{0.539}{(d)}
\caption{Localization statistics under regime G1: (a)~per-scan position RMSE vs.\ CRLBs (inset: full log scale); (b)~stage-wise position RMSE; (c)~stage-wise heading RMSE; and (d)~stage-wise position error CDFs.}
\label{fig:rt_g1_stats}
\end{figure}

Fig.~\ref{fig:rt_g1_stats} reports the 50-MC statistical localization results under regime~G1. For RX localization in Figs.~\ref{fig:rt_g1_stats}(a)--(c), the proposed method stays close to the 2-b Batch CRLB across all stages, with stage-wise position RMSEs of $0.035$, $0.049$, and $0.070\,\mathrm{m}$ in R1, R2, and R3 (overall full-trajectory position RMSE $0.055\,\mathrm{m}$ vs.\ the bound of $0.050\,\mathrm{m}$, and heading RMSE $0.43^\circ$ vs.\ $0.40^\circ$). Comparing the error curves with the theoretical bounds yields two observations. Compared with 1-b Batch SLAM, when 1-bounce paths vanish in R3, both 1-b Batch SLAM and the 1-b Batch CRLB increase above $0.22\,\mathrm{m}$, whereas exploiting 2-bounce constraints lowers the R3 position RMSE by $71\%$ to $0.070\,\mathrm{m}$. Compared with 2-b BP-SLAM, because batch inference conditions on the full measurement sequence $\mathcal{Z}_{1:K}$ across all $K$ scans rather than only causal measurements $\mathcal{Z}_{1:k}$ up to scan $k$, the 2-b Batch CRLB lies below the 2-b Filtering CRLB ($0.050\,\mathrm{m}$ vs.\ $0.059\,\mathrm{m}$ overall); moreover, due to particle degeneracy and accumulated filtering errors, 2-b BP-SLAM exhibits an additional gap relative to the filtering bound in R3 ($0.138\,\mathrm{m}$ vs.\ $0.076\,\mathrm{m}$). The stage-wise empirical CDFs in Fig.~\ref{fig:rt_g1_stats}(d) further show that the proposed method achieves lower position errors than the baselines across all quantiles. For mapping, because Snapshot DB-SLAM and 1-b Batch SLAM model landmarks as single-type RPs, we compare the 50-MC heterogeneous (VA and RP) map GOSPA decomposition (localization/missed/false) of 2-b BP-SLAM and the proposed method. Both attain low RP GOSPA: $0.25\,\mathrm{m}$ ($0.19/0.00/0.06\,\mathrm{m}$) for 2-b BP-SLAM versus $0.13\,\mathrm{m}$ ($0.11/0.00/0.03\,\mathrm{m}$) for the proposed method. For VAs, 2-b BP-SLAM incurs a GOSPA of $2.29\,\mathrm{m}$ ($0.08/1.96/0.25\,\mathrm{m}$, $5.74/7$ occluded landmarks detected), mainly from missed detections, whereas the proposed method reduces VA GOSPA to $0.27\,\mathrm{m}$ ($0.24/0.03/0.00\,\mathrm{m}$, $6.98/7$ detected), cutting total map GOSPA by $84\%$ ($2.54$ to $0.40\,\mathrm{m}$).

\subsection{Robustness Across Noise Regimes and Path Association} \label{sec:exp_rt_robust}

\begin{figure}[!t]
\centering
\includegraphics[width=\columnwidth]{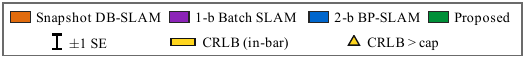}\\[2.5pt]
\begin{minipage}[t]{0.495\columnwidth}
\centering
\includegraphics[width=\textwidth]{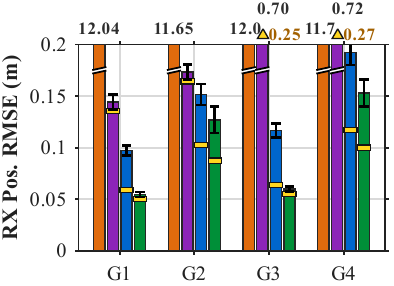}\\[-1pt]
\sublbl{0.591}{(a)}
\end{minipage}\hfill%
\begin{minipage}[t]{0.495\columnwidth}
\centering
\includegraphics[width=\textwidth]{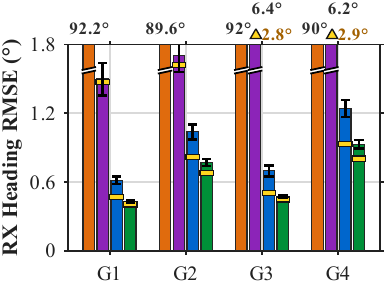}\\[-1pt]
\sublbl{0.577}{(b)}
\end{minipage}
\caption{Localization performance across noise regimes G1--G4: (a)~full-trajectory position RMSE and (b)~full-trajectory heading RMSE. Bar heights denote overall RMSEs (capped flush at the upper axis limit when overflowing), black vertical bars indicate $\pm 1\;\mathrm{SE}$, horizontal gold ticks denote in-bar theoretical CRLBs, and gold triangles (with numeric bound values) indicate CRLBs exceeding the axis ceiling.}
\label{fig:rt_4regimes}
\end{figure}

\begin{table}[!b]
\centering
\caption{Heterogeneous Map GOSPA (Mean $\pm$ 1 SE, m) Across G1--G4}
\label{tab:rt_4reg_gospa}
\scriptsize
\setlength{\tabcolsep}{3.8pt}
\begin{tabular}{@{}lcccc@{}}
\toprule
Method & G1 (m) & G2 (m) & G3 (m) & G4 (m) \\
\midrule
2-b BP-SLAM & $2.54 \pm 0.06$ & $3.32 \pm 0.10$ & $2.76 \pm 0.11$ & $3.52 \pm 0.10$ \\
\textbf{Proposed} & $\mathbf{0.40 \pm 0.05}$ & $\mathbf{0.97 \pm 0.09}$ & $\mathbf{0.43 \pm 0.06}$ & $\mathbf{1.13 \pm 0.11}$ \\
\bottomrule
\end{tabular}
\end{table}

\begin{figure}[!b]
\centering
\includegraphics[width=\columnwidth]{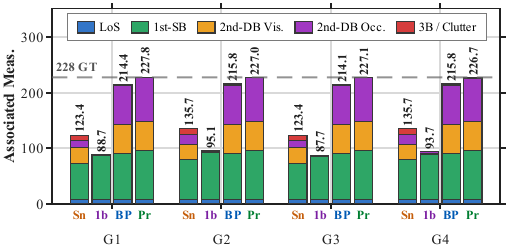}
\caption{Mean associated measurements per batch across G1--G4 ($\mathrm{Sn}, \mathrm{1b}, \mathrm{BP}, \mathrm{Pr}$: Snapshot, 1-b Batch, 2-b BP, Proposed) by RT path category (228 $\le 2$-bounce; 286 expected total with 22 3-bounce and 36 clutter).}
\label{fig:rt_4reg_assoc}
\end{figure}

Figs.~\ref{fig:rt_4regimes}(a)--(b) and Table~\ref{tab:rt_4reg_gospa} evaluate robustness across noise regimes G1--G4. Across all four regimes, the proposed method achieves the lowest localization and mapping errors and closely follows the 2-b Batch CRLB. Under the highest-noise regime G4 ($4\boldsymbol{R}, 4\boldsymbol{Q}$), the proposed method attains an overall position RMSE of $0.153\,\mathrm{m}$ (2-b Batch CRLB $0.100\,\mathrm{m}$), compared with $0.193\,\mathrm{m}$ for 2-b BP-SLAM, $0.721\,\mathrm{m}$ for 1-b Batch SLAM, and $11.65\,\mathrm{m}$ for Snapshot DB-SLAM; its overall heading RMSE is $0.93^\circ$ (2-b Batch CRLB $0.80^\circ$) versus $1.24^\circ$ for 2-b BP-SLAM and $6.23^\circ$ for 1-b Batch SLAM, and its mean total map GOSPA is $1.13\,\mathrm{m}$ versus $3.52\,\mathrm{m}$ for 2-b BP-SLAM.

Fig.~\ref{fig:rt_4reg_assoc} shows the mean associated measurement counts per batch across G1--G4. Out of the 228 valid ($\le 2$-bounce) RT paths, 1-b Batch SLAM associates only $88\text{--}95$ measurements (rejecting most 2-bounce returns as clutter while mis-associating a few as spurious 1-bounce RPs under higher noise), Snapshot DB-SLAM associates $123\text{--}136$ measurements while falsely absorbing $9\text{--}10$ 3-bounce/clutter returns, 2-b BP-SLAM associates $214\text{--}216$ paths, and the proposed method associates $227\text{--}228$ valid paths while rejecting almost all 3-bounce reflections and clutter. In MATLAB, the average runtimes per 45-scan batch over G1--G4 are $0.31\,\mathrm{s}$ for Snapshot DB-SLAM, $14.9\,\mathrm{s}$ for 1-b Batch SLAM, $41.4\,\mathrm{s}$ for 2-b BP-SLAM, and $116.4\,\mathrm{s}$ for the proposed method; higher runtime than snapshot and sequential filtering methods is an inherent trade-off of joint full-batch double-bounce inference.

\subsection{Real-World 60 GHz Experimental Validation} \label{sec:exp_real}

\begin{figure}[!t]
\centering
\includegraphics[width=\columnwidth]{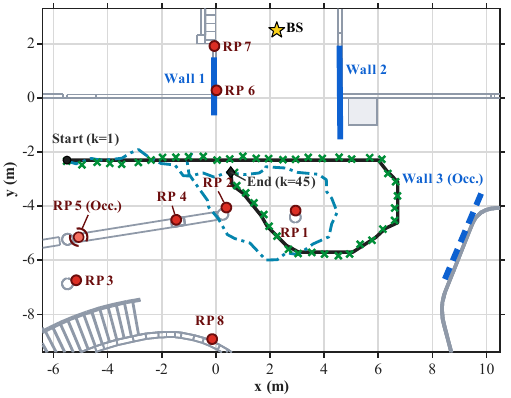}\\[2.5pt]
\includegraphics[width=\columnwidth]{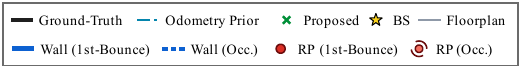}
\caption{Localization and mapping results of the proposed method on the Tampere 60~GHz dataset, showing the ground-truth trajectory, odometry prior, estimated trajectory, visible features (Walls~1--2 from estimated VAs, and RPs~1--4, 6--8), and occluded features recovered via 2-bounce paths (Wall~3 and RP~5).}
\label{fig:60ghz_traj_map}
\end{figure}

We further validate the proposed method on real 60~GHz indoor mmWave MIMO measurements from the Tampere University Kampusareena hallway dataset~\cite{RastorguevaFoi2024JSAC,Zhang2026TWC}. Selecting the $99.99\%$ multipath energy subset yields 362 clock-synchronized measurements $\boldsymbol{z} = [d, \theta, \phi]^{\mathsf{T}}$ over $K = 45$ scans along a closed-loop trajectory (Fig.~\ref{fig:60ghz_traj_map}). Over $50$ MC runs, odometry inputs are synthesized with per-step covariance $\boldsymbol{Q} = \operatorname{diag}(0.15^2\,\mathrm{m}^2, 0.15^2\,\mathrm{m}^2, (2.8^\circ)^2)$ (overall dead-reckoning position RMSE $1.41\,\mathrm{m}$), and measurement covariances are $\operatorname{diag}(0.16^2\,\mathrm{m}^2, (2.2^\circ)^2, (2.2^\circ)^2)$ for LoS, $\operatorname{diag}(0.18^2\,\mathrm{m}^2, (2.5^\circ)^2, (2.5^\circ)^2)$ for 1-bounce, and $\operatorname{diag}(0.28^2\,\mathrm{m}^2, (3.2^\circ)^2, (3.2^\circ)^2)$ for 2-bounce paths. In-FoV ($32/45\,\mathrm{m}$) parameters are $(p_{\mathrm{D},0}, p_{\mathrm{D},1}, p_{\mathrm{D},2}) = (0.84, 0.42, 0.02)$, $\lambda_{\mathrm{VA}} = \lambda_{\mathrm{RP}} = 8/2100\,\mathrm{m}^{-2}$, and $c(\boldsymbol{z}) = 3.0/(45\cdot 2\pi^2)$ (other settings as in Section~\ref{sec:exp_setup}).

\begin{figure}[!t]
\centering
\includegraphics[width=\columnwidth]{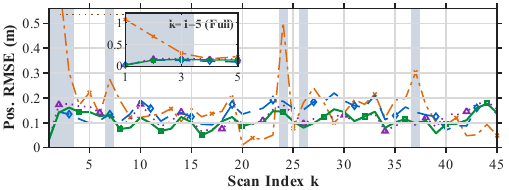}\\[-1.5pt]
\sublbl{0.536}{(a)}\\[2pt]
\includegraphics[width=\columnwidth]{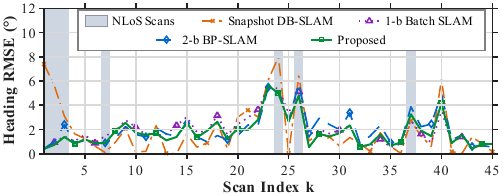}\\[-1.5pt]
\sublbl{0.523}{(b)}
\caption{Per-scan localization RMSEs on the 60~GHz dataset (shaded vertical bands indicate NLoS scans): (a)~2D position RMSE (inset: full-scale scans $k = 1\text{--}5$) and (b)~heading RMSE.}
\label{fig:60ghz_perscan}
\end{figure}

\begin{figure}[!b]
\centering
\includegraphics[width=\columnwidth]{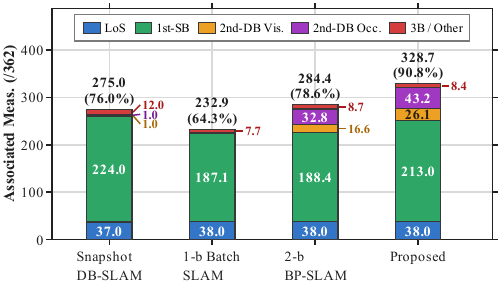}
\caption{Mean associated 60~GHz measurements (out of 362) by propagation category (``3B / Other'' comprises 3-bounce returns, clutter, and unmodeled minor scatterers absent from the floorplan).}
\label{fig:60ghz_assoc}
\end{figure}

\begin{table}[!b]
\centering
\caption{60~GHz Indoor Dataset Performance (Mean $\pm$ SD) and Runtime}
\label{tab:60ghz_benchmark}
\scriptsize
\setlength{\tabcolsep}{1.8pt}
\begin{tabular}{@{}lcccc@{}}
\toprule
Method & Pos.\ RMSE (m) & Head.\ ($^\circ$) & Assoc.\ (/362) & Time (s) \\
\midrule
Snapshot DB-SLAM & $0.2555 \pm 0.0000$ & $2.82 \pm 0.00$ & 275.0 (76.0\%) & 0.69 \\
1-b Batch SLAM & $0.1255 \pm 0.0055$ & $2.32 \pm 0.12$ & 232.9 (64.3\%) & 17.2 \\
2-b BP-SLAM & $0.1435 \pm 0.0180$ & $2.36 \pm 0.28$ & 284.4 (78.6\%) & 132.7 \\
\textbf{Proposed} & $\mathbf{0.1142 \pm 0.0042}$ & $\mathbf{2.15 \pm 0.06}$ & \textbf{328.7 (90.8\%)} & 359.0 \\
\bottomrule
\end{tabular}
\end{table}

Fig.~\ref{fig:60ghz_traj_map} shows the estimated trajectory and environment map obtained by the proposed method on the 60~GHz dataset. Along the closed-loop trajectory, the proposed method corrects the dead-reckoning drift and estimates both the directly visible structures (Walls~1--2 and RPs~1--4, 6--8) and the two occluded features observed through 2-bounce paths, namely the right-hand inclined wall (Wall~3) and the left-side pillar (RP~5), in agreement with the hallway floorplan geometry.

Figs.~\ref{fig:60ghz_perscan},~\ref{fig:60ghz_assoc} and Table~\ref{tab:60ghz_benchmark} summarize the 50-MC localization, association, and runtime results. Owing to the absence of ray-interaction and landmark ground truth, the LoS, 1-bounce, and 2-bounce categories in Fig.~\ref{fig:60ghz_assoc} are inferred offline from the floorplan, reference trajectory, and major structural reflectors (Walls~1--3 and RPs~1--8), whereas ``3B / Other'' comprises 3-bounce returns, clutter, and unmodeled minor scatterers absent from the floorplan. Snapshot DB-SLAM (deterministic without odometry) incurs larger NLoS errors ($0.2555\,\mathrm{m}$), outputs unassociated single-scan points, and fails to detect the occluded Wall~3 and RP~5. 1-b Batch SLAM achieves $0.1255 \pm 0.0055\,\mathrm{m}$ position RMSE by associating $64.3\%$ ($232.9/362$) of measurements via 1-bounce smoothing, yet cannot detect Wall~3 or RP~5. 2-b BP-SLAM attains $0.1435 \pm 0.0180\,\mathrm{m}$ position RMSE, associates $78.6\%$ ($284.4/362$) of measurements, and detects Wall~3 in $12/50$ ($24\%$) and RP~5 in $36/50$ ($72\%$) of runs. The proposed method achieves the lowest mean position RMSE ($0.1142 \pm 0.0042\,\mathrm{m}$, reducing error by $55.3\%$, $20.4\%$, and $9.0\%$ over Snapshot DB-SLAM, 2-b BP-SLAM, and 1-b Batch SLAM, respectively) and heading RMSE ($2.15 \pm 0.06^\circ$), associates $90.8\%$ of all measurements ($328.7/362$), and detects both occluded Wall~3 and RP~5 in $50/50$ ($100\%$) of runs.

\section{Conclusion} \label{sec:conclusion}

This paper proposed a double-bounce batch radio SLAM algorithm combining double-bounce MCMC data association sampling with conditional GraphSLAM for environments with multi-type reflectors and severe occlusion. Formulated within the RFS framework across LoS, 1-bounce, and 2-bounce paths, the algorithm alternates warm-start pre-clustering, two-stage data association and landmark type sampling, conditional graph-based optimization, and cross-sample multi-type PMB map fusion to jointly estimate the sensor trajectory and heterogeneous map over the full batch. Ray-tracing simulations and real indoor measurements demonstrate that the proposed method outperforms snapshot double-bounce, single-bounce batch, and sequential 2-bounce filtering baselines in localization and mapping accuracy, reliably recovers occluded landmarks, and closely approaches the theoretical performance bound even under deep occlusion. Future work includes computational complexity reduction, extended-target modeling, multimodal sensor fusion, and dynamic environments with moving targets.


\begin{thebibliography}{10}
\providecommand{\url}[1]{#1}
\csname url@samestyle\endcsname
\providecommand{\newblock}{\relax}
\providecommand{\bibinfo}[2]{#2}
\providecommand{\BIBentrySTDinterwordspacing}{\spaceskip=0pt\relax}
\providecommand{\BIBentryALTinterwordstretchfactor}{4}
\providecommand{\BIBentryALTinterwordspacing}{\spaceskip=\fontdimen2\font plus
\BIBentryALTinterwordstretchfactor\fontdimen3\font minus
  \fontdimen4\font\relax}
\providecommand{\BIBforeignlanguage}[2]{{%
\expandafter\ifx\csname l@#1\endcsname\relax
\typeout{** WARNING: IEEEtran.bst: No hyphenation pattern has been}%
\typeout{** loaded for the language `#1'. Using the pattern for}%
\typeout{** the default language instead.}%
\else
\language=\csname l@#1\endcsname
\fi
#2}}
\providecommand{\BIBdecl}{\relax}
\BIBdecl
\renewcommand{\BIBentryALTinterwordstretchfactor}{4}

\bibitem{DiTaranto2014SPM}
R.~Di~Taranto \emph{et~al.}, ``Location-aware communications for {5G} networks:
  {How} location information can improve scalability, latency, and robustness
  of {5G},'' \emph{IEEE Signal Process. Mag.}, vol.~31, no.~6, pp. 102--112,
  Nov. 2014.

\bibitem{Witrisal2016SPM}
K.~Witrisal \emph{et~al.}, ``High-accuracy localization for assisted living:
  {5G} systems will turn multipath channels from foe to friend,'' \emph{IEEE
  Signal Process. Mag.}, vol.~33, no.~2, pp. 59--70, Mar. 2016.

\bibitem{Ge2023ISAC}
Y.~Ge \emph{et~al.}, ``{MmWave} mapping and {SLAM} for {5G} and beyond,'' in
  \emph{Integrated Sensing and Communications}, F.~Liu, C.~Masouros, and Y.~C.
  Eldar, Eds.\hskip 1em plus 0.5em minus 0.4em\relax Singapore: Springer, 2023,
  pp. 445--475.

\bibitem{Ge2025ComMag}
Y.~Ge \emph{et~al.}, ``Sensing with mobile devices through radio {SLAM}:
  {Models}, methods, opportunities, and challenges,'' \emph{IEEE Commun. Mag.},
  vol.~63, no.~12, pp. 80--87, Dec. 2025.

\bibitem{Mendrzik2019JSTSP}
R.~Mendrzik \emph{et~al.}, ``Enabling situational awareness in millimeter wave
  massive {MIMO} systems,'' \emph{IEEE J. Sel. Topics Signal Process.},
  vol.~13, no.~5, pp. 1196--1211, Sep. 2019.

\bibitem{Li2019TWC}
X.~Li \emph{et~al.}, ``Massive {MIMO}-based localization and mapping exploiting
  phase information of multipath components,'' \emph{IEEE Trans. Wireless
  Commun.}, vol.~18, no.~9, pp. 4254--4267, Sep. 2019.

\bibitem{Ge2024TVT}
Y.~Ge \emph{et~al.}, ``Experimental validation of single {BS} {5G} {mmWave}
  positioning and mapping for intelligent transport,'' \emph{IEEE Trans. Veh.
  Technol.}, vol.~73, no.~11, pp. 16\,744--16\,757, Nov. 2024.

\bibitem{RastorguevaFoi2024JSAC}
E.~Rastorgueva-Foi \emph{et~al.}, ``Millimeter-wave radio {SLAM}: {End}-to-end
  processing methods and experimental validation,'' \emph{IEEE J. Sel. Areas
  Commun.}, vol.~42, no.~9, pp. 2550--2567, Sep. 2024.

\bibitem{Liang2025TSP}
M.~Liang, E.~Leitinger, and F.~Meyer, ``Direct multipath-based {SLAM},''
  \emph{IEEE Trans. Signal Process.}, vol.~73, pp. 2336--2352, Mar. 2025.

\bibitem{GarciaFernandez2018TAES}
{\'A}.~F. Garc{\'\i}a-Fern{\'a}ndez \emph{et~al.}, ``Poisson multi-{Bernoulli}
  mixture filter: {Direct} derivation and implementation,'' \emph{IEEE Trans.
  Aerosp. Electron. Syst.}, vol.~54, no.~4, pp. 1883--1901, Aug. 2018.

\bibitem{Kim2020TWC}
H.~Kim \emph{et~al.}, ``{5G} {mmWave} cooperative positioning and mapping using
  multi-model {PHD} filter and map fusion,'' \emph{IEEE Trans. Wireless
  Commun.}, vol.~19, no.~6, pp. 3782--3795, Jun. 2020.

\bibitem{Ge2020Sensors}
Y.~Ge \emph{et~al.}, ``{5G} {SLAM} using the clustering and assignment approach
  with diffuse multipath,'' \emph{Sensors}, vol.~20, no.~16, p. 4656, Aug.
  2020.

\bibitem{Ge2022JSAC}
Y.~Ge \emph{et~al.}, ``A computationally efficient {EK-PMBM} filter for
  bistatic {mmWave} radio {SLAM},'' \emph{IEEE J. Sel. Areas Commun.}, vol.~40,
  no.~7, pp. 2179--2192, Jul. 2022.

\bibitem{Kaltiokallio2024TRO}
O.~Kaltiokallio \emph{et~al.}, ``A multi-hypotheses importance density for
  {SLAM} in cluttered scenarios,'' \emph{IEEE Trans. Robot.}, vol.~40, pp.
  1019--1035, 2024.

\bibitem{Kim2024TSP}
H.~Kim \emph{et~al.}, ``Set-type belief propagation with applications to
  {Poisson} multi-{Bernoulli} {SLAM},'' \emph{IEEE Trans. Signal Process.},
  vol.~72, pp. 1989--2005, 2024.

\bibitem{Leitinger2019TWC}
E.~Leitinger \emph{et~al.}, ``A belief propagation algorithm for
  multipath-based {SLAM},'' \emph{IEEE Trans. Wireless Commun.}, vol.~18,
  no.~12, pp. 5613--5629, Dec. 2019.

\bibitem{Thrun2006IJRR}
S.~Thrun and M.~Montemerlo, ``The {GraphSLAM} algorithm with applications to
  large-scale mapping of urban structures,'' \emph{Int. J. Robot. Res.},
  vol.~25, no. 5--6, pp. 403--429, May 2006.

\bibitem{Ge2025TSP}
Y.~Ge \emph{et~al.}, ``Batch {SLAM} with {PMBM} data association sampling and
  graph-based optimization,'' \emph{IEEE Trans. Signal Process.}, vol.~73, pp.
  2139--2153, May 2025.

\bibitem{Xia2024TAES}
Y.~Xia, {\'A}.~F. Garc{\'\i}a-Fern{\'a}ndez, and L.~Svensson, ``Markov chain
  {Monte Carlo} multiscan data association for sets of trajectories,''
  \emph{IEEE Trans. Aerosp. Electron. Syst.}, vol.~60, no.~6, pp. 7804--7819,
  Dec. 2024.

\bibitem{Steinboeck2016TAP}
G.~Steinb{\"o}ck \emph{et~al.}, ``Hybrid model for reverberant indoor radio
  channels using rays and graphs,'' \emph{IEEE Trans. Antennas Propag.},
  vol.~64, no.~9, pp. 4036--4048, Sep. 2016.

\bibitem{Xu2025AWPL}
Y.~Xu \emph{et~al.}, ``Experimental analysis of multipath characteristics in
  indoor distributed massive {MIMO} channels,'' \emph{IEEE Antennas Wireless
  Propag. Lett.}, vol.~24, no.~11, pp. 3866--3870, Nov. 2025.

\bibitem{Nazari2023TVT}
M.~A. Nazari \emph{et~al.}, ``{MmWave} {6D} radio localization with a snapshot
  observation from a single {BS},'' \emph{IEEE Trans. Veh. Technol.}, vol.~72,
  no.~7, pp. 8914--8928, Jul. 2023.

\bibitem{Li2023WCL}
Z.~Li \emph{et~al.}, ``An iterative {5G} positioning and synchronization
  algorithm in {NLOS} environments with multi-bounce paths,'' \emph{IEEE
  Wireless Commun. Lett.}, vol.~12, no.~5, pp. 804--808, May 2023.

\bibitem{Zhang2026TWC}
X.~Zhang \emph{et~al.}, ``Exploiting double-bounce paths in snapshot radio
  {SLAM}: {Bounds}, algorithms, and experiments,'' \emph{IEEE Trans. Wireless
  Commun.}, vol.~25, pp. 20\,836--20\,851, 2026.

\bibitem{Leitinger2023TSP}
E.~Leitinger \emph{et~al.}, ``Data fusion for multipath-based {SLAM}:
  {Combining} information from multiple propagation paths,'' \emph{IEEE Trans.
  Signal Process.}, vol.~71, pp. 4011--4028, Sep. 2023.

\bibitem{Li2022TWC}
X.~Li \emph{et~al.}, ``Sequential detection and estimation of multipath channel
  parameters using belief propagation,'' \emph{IEEE Trans. Wireless Commun.},
  vol.~21, no.~10, pp. 8385--8402, Oct. 2022.

\bibitem{Li2026TWC}
X.~Li \emph{et~al.}, ``Adaptive multipath-based {SLAM} for distributed {MIMO}
  systems,'' \emph{IEEE Trans. Wireless Commun.}, vol.~25, pp.
  16\,931--16\,946, 2026.

\bibitem{Kim2022TVT}
H.~Kim \emph{et~al.}, ``{PMBM}-based {SLAM} filters in {5G} {mmWave} vehicular
  networks,'' \emph{IEEE Trans. Veh. Technol.}, vol.~71, no.~8, pp. 8646--8661,
  Aug. 2022.

\bibitem{Li2024ICC}
X.~Li \emph{et~al.}, ``A belief propagation algorithm for multipath-based
  {SLAM} with multiple map features: {A} {mmWave} {MIMO} application,'' in
  \emph{Proc. IEEE Int. Conf. Commun. Workshops (ICC Workshops)}, Aug. 2024,
  pp. 269--275.

\bibitem{Zhai2025TWC}
S.~Zhai \emph{et~al.}, ``Multipath-based {SLAM} exploiting extended object
  estimation and classification,'' \emph{IEEE Trans. Wireless Commun.},
  vol.~24, no.~8, pp. 7029--7045, Aug. 2025.

\bibitem{Venus2024TWC}
A.~Venus \emph{et~al.}, ``Graph-based simultaneous localization and bias
  tracking,'' \emph{IEEE Trans. Wireless Commun.}, vol.~23, no.~10, pp.
  13\,141--13\,158, Oct. 2024.

\bibitem{Fascista2025JSTSP}
A.~Fascista \emph{et~al.}, ``Joint localization, synchronization and mapping
  via phase-coherent distributed arrays,'' \emph{IEEE J. Sel. Topics Signal
  Process.}, vol.~19, no.~2, pp. 412--429, Mar. 2025.

\bibitem{Wen2021TWC}
F.~Wen \emph{et~al.}, ``{5G} positioning and mapping with diffuse multipath,''
  \emph{IEEE Trans. Wireless Commun.}, vol.~20, no.~2, pp. 1164--1174, Feb.
  2021.

\bibitem{Fatemi2017TSP}
M.~Fatemi \emph{et~al.}, ``Poisson multi-{Bernoulli} mapping using {Gibbs}
  sampling,'' \emph{IEEE Trans. Signal Process.}, vol.~65, no.~11, pp.
  2814--2827, Jun. 2017.

\bibitem{Rahmathullah2017GOSPA}
A.~S. Rahmathullah, {\'A}.~F. Garc{\'\i}a-Fern{\'a}ndez, and L.~Svensson,
  ``Generalized optimal sub-pattern assignment metric,'' in \emph{Proc. 20th
  Int. Conf. Inf. Fusion (FUSION)}, Jul. 2017, pp. 1--8.

\bibitem{Kay1993}
S.~M. Kay, \emph{Fundamentals of Statistical Signal Processing: Estimation
  Theory}.\hskip 1em plus 0.5em minus 0.4em\relax Englewood Cliffs, NJ, USA:
  Prentice Hall, 1993.

\bibitem{Deutschmann2025Asilomar}
B.~J.~B. Deutschmann \emph{et~al.}, ``Posterior {Cram{\'e}r-Rao} bounds on
  localization and mapping errors in distributed {MIMO} {SLAM},'' in
  \emph{Proc. Asilomar Conf. Signals, Syst., Comput.}, Pacific Grove, CA, USA,
  Oct. 2025, pp. 1--5.

\end{thebibliography}
\end{document}


\title{Double-Bounce Multi-Type Landmarks Batch Radio SLAM via MCMC and Graph-Based Optimization:\\Supplementary Material}

\author{Zhibo~Duan, Jiancun~Fan, Yulin~Fu, Shiyu~Zhai, and~Jie~Luo}

\maketitle
\thispagestyle{plain}
\pagestyle{plain}

\begin{quote}
\noindent\textit{This supporting document provides derivations, proofs, closed-form expressions, and algorithm pseudocode omitted from ``Double-Bounce Multi-Type Landmarks Batch Radio SLAM via MCMC and Graph-Based Optimization''~\cite{MainPaper}: motion and six-path observation models with analytical Jacobians (Appendix~\ref{app:A}), derivation of Theorem~1 (Appendix~\ref{app:B}), marginal likelihood approximations, MCMC weight ratios, and warm-start 2-bounce pre-clustering (Appendix~\ref{app:C}), geometric identifiability analysis, sampler convergence justification, fast coupled inference, and Cram\'er-Rao lower bounds (Appendix~\ref{app:D}), and algorithm pseudocode summaries (Appendix~\ref{app:E}, Algorithms~\ref{alg:S_mcmc}--\ref{alg:S_overall}).}
\end{quote}

\appendices

\section{Motion and Measurement Models and Analytical Jacobians} \label{app:A}

This appendix corresponds to Sections~II and~V-A of~\cite{MainPaper}, providing the explicit odometry motion transition function and its Jacobian, the Householder mirror-reflection operator, derivations and a summary table for the six multipath observation functions, and analytical Jacobians for nonlinear least-squares (LS) optimization and Fisher information matrix (FIM) evaluation.

\subsection{Sensor Motion Transition Function and Jacobian}

At discrete time $k \in \{1, \dots, K\}$, let $\boldsymbol{s}_{k-1} = [\boldsymbol{p}_{k-1}^{\mathsf{T}}, \alpha_{k-1}]^{\mathsf{T}} = [x_{k-1}, y_{k-1}, \alpha_{k-1}]^{\mathsf{T}} \in \mathbb{R}^2 \times (-\pi, \pi]$ be the receiver (RX) state and $\boldsymbol{u}_{k-1} = [\Delta x_{k-1}, \Delta y_{k-1}, \Delta\alpha_{k-1}]^{\mathsf{T}}$ the odometry displacement expressed in the body frame at scan $k-1$. Rotating $\boldsymbol{u}_{k-1}$ into the global coordinate frame gives the explicit transition function $\boldsymbol{v}(\boldsymbol{s}_{k-1}, \boldsymbol{u}_{k-1})$ in~\cite[Eq.~(1)]{MainPaper}:
\begin{align}
\boldsymbol{v}(\boldsymbol{s}_{k-1}, \boldsymbol{u}_{k-1}) =
\begin{bmatrix}
x_{k-1} + \cos\alpha_{k-1}\,\Delta x_{k-1} - \sin\alpha_{k-1}\,\Delta y_{k-1} \\
y_{k-1} + \sin\alpha_{k-1}\,\Delta x_{k-1} + \cos\alpha_{k-1}\,\Delta y_{k-1} \\
\mathrm{wrap}(\alpha_{k-1} + \Delta\alpha_{k-1})
\end{bmatrix}. \label{eq:S_motion_v}
\end{align}
Differentiating~\eqref{eq:S_motion_v} with respect to $\boldsymbol{s}_{k-1}$ yields the motion Jacobian
\begin{align}
\boldsymbol{F}_{k-1} \triangleq \frac{\partial \boldsymbol{v}(\boldsymbol{s}_{k-1}, \boldsymbol{u}_{k-1})}{\partial \boldsymbol{s}_{k-1}} =
\begin{bmatrix}
1 & 0 & -\sin\alpha_{k-1}\,\Delta x_{k-1} - \cos\alpha_{k-1}\,\Delta y_{k-1} \\
0 & 1 & \cos\alpha_{k-1}\,\Delta x_{k-1} - \sin\alpha_{k-1}\,\Delta y_{k-1} \\
0 & 0 & 1
\end{bmatrix} \in \mathbb{R}^{3\times 3}. \label{eq:S_motion_F}
\end{align}

\subsection{Mirror-Reflection Operator and Six-Path Geometric Observation Models}

Let the fixed base station (BS) be at $\boldsymbol{p}_{\mathrm{BS}} \in \mathbb{R}^2$ with array orientation $\alpha_{\mathrm{BS}} \in (-\pi, \pi]$. A planar wall $i$ ($\psi_i = \mathrm{VA}$) is parameterized by the first-order virtual anchor $\boldsymbol{x}_i \in \mathbb{R}^2 \setminus \{\boldsymbol{p}_{\mathrm{BS}}\}$ of $\boldsymbol{p}_{\mathrm{BS}}$ across the wall line, with displacement $\boldsymbol{d}_i \triangleq \boldsymbol{x}_i - \boldsymbol{p}_{\mathrm{BS}}$, unit wall normal $\boldsymbol{n}_i \triangleq \boldsymbol{d}_i / \|\boldsymbol{d}_i\|$, wall foot midpoint $\boldsymbol{m}_i \triangleq \frac{1}{2}(\boldsymbol{p}_{\mathrm{BS}} + \boldsymbol{x}_i)$, and Householder matrix $\boldsymbol{H}(\boldsymbol{d}_i) \triangleq \boldsymbol{I}_2 - 2\boldsymbol{n}_i\boldsymbol{n}_i^{\mathsf{T}}$.

For a 2D point $\boldsymbol{a} \in \mathbb{R}^2$, reflecting across the wall flips its normal offset relative to $\boldsymbol{m}_i$ while preserving its tangential offset. Substituting $\boldsymbol{H}(\boldsymbol{d}_i)\boldsymbol{d}_i = -\boldsymbol{d}_i$ and $\boldsymbol{m}_i - \boldsymbol{p}_{\mathrm{BS}} = \frac{1}{2}\boldsymbol{d}_i$ yields the mirror-reflection operator in~\cite[Eq.~(2)]{MainPaper}:
\begin{align}
\mathcal{M}(\boldsymbol{a}; \boldsymbol{x}_i)
= \boldsymbol{m}_i + \big(\boldsymbol{I}_2 - 2\boldsymbol{n}_i\boldsymbol{n}_i^{\mathsf{T}}\big)(\boldsymbol{a} - \boldsymbol{m}_i)
= \boldsymbol{x}_i + \boldsymbol{H}(\boldsymbol{d}_i)(\boldsymbol{a} - \boldsymbol{p}_{\mathrm{BS}}). \label{eq:S_mirror}
\end{align}

In~\cite[Eqs.~(4)--(6)]{MainPaper}, for each propagation path type $\eta \in \mathcal{H} = \{\mathrm{LoS}\} \cup \{\mathrm{VA}, \mathrm{RP}\} \cup \{(\mathrm{VA}, \mathrm{VA}), (\mathrm{VA}, \mathrm{RP}), (\mathrm{RP}, \mathrm{VA})\}$, the noiseless 3D measurement vector $[d, \theta, \phi]^{\mathsf{T}}$ is assembled from propagation distance $h_{d,\eta}$, RX-side incoming vector $\boldsymbol{d}_{\mathrm{RX},\eta}$ (pointing from RX position $\boldsymbol{p}_k$ toward the last interaction point or unfolded virtual source), and BS-side departure vector $\boldsymbol{d}_{\mathrm{TX},\eta}$ (pointing from $\boldsymbol{p}_{\mathrm{BS}}$ toward the first interaction point or unfolded virtual RX) as
\begin{align}
\boldsymbol{h}_\eta(\cdot) \triangleq
\begin{bmatrix}
h_{d,\eta} \\
\mathrm{wrap}\!\big(\angle\boldsymbol{d}_{\mathrm{RX},\eta} - \alpha_k\big) \\
\mathrm{wrap}\!\big(\angle\boldsymbol{d}_{\mathrm{TX},\eta} - \alpha_{\mathrm{BS}}\big)
\end{bmatrix} \in \mathbb{R}^3, \label{eq:S_h_eta}
\end{align}
where $\angle\boldsymbol{v} \triangleq \operatorname{atan2}([\boldsymbol{v}]_2, [\boldsymbol{v}]_1)$. Using image-source unfolding~\cite{Leitinger2023TSP,Li2026TWC} (dropping subscript $\eta$ below and summarizing all six path types in Table~\ref{tab:S_paths}):
\begin{enumerate}
    \item \emph{0-bounce LoS path ($\mathrm{LoS}$):} Direct path $\mathrm{BS} \to \mathrm{RX}$ gives $h_d = \|\boldsymbol{p}_{\mathrm{BS}} - \boldsymbol{p}_k\|$, $\boldsymbol{d}_{\mathrm{RX}} = \boldsymbol{p}_{\mathrm{BS}} - \boldsymbol{p}_k$, and $\boldsymbol{d}_{\mathrm{TX}} = \boldsymbol{p}_k - \boldsymbol{p}_{\mathrm{BS}}$.
    \item \emph{1-bounce wall path ($\mathrm{VA}$):} For $\mathrm{BS} \to \text{wall } i \to \mathrm{RX}$, unfolding the BS leg across wall $i$ yields $h_d = \|\boldsymbol{x}_i - \boldsymbol{p}_k\|$ and $\boldsymbol{d}_{\mathrm{RX}} = \boldsymbol{x}_i - \boldsymbol{p}_k$, while unfolding the RX leg places the virtual RX at $\mathcal{M}(\boldsymbol{p}_k; \boldsymbol{x}_i)$; since $\boldsymbol{H}(\boldsymbol{d}_i)\boldsymbol{d}_i = -\boldsymbol{d}_i$,
    \begin{align}
    \boldsymbol{d}_{\mathrm{TX}} = \mathcal{M}(\boldsymbol{p}_k; \boldsymbol{x}_i) - \boldsymbol{p}_{\mathrm{BS}}
    = \boldsymbol{d}_i + \boldsymbol{H}(\boldsymbol{d}_i)(\boldsymbol{p}_k - \boldsymbol{p}_{\mathrm{BS}})
    = \boldsymbol{H}(\boldsymbol{d}_i)(\boldsymbol{p}_k - \boldsymbol{x}_i). \label{eq:S_dtx_va}
    \end{align}
    \item \emph{1-bounce point scatterer ($\mathrm{RP}$):} For $\mathrm{BS} \to \text{scatterer } i \to \mathrm{RX}$ at fixed coordinate $\boldsymbol{x}_i$, $h_d = \|\boldsymbol{x}_i - \boldsymbol{p}_{\mathrm{BS}}\| + \|\boldsymbol{x}_i - \boldsymbol{p}_k\|$, $\boldsymbol{d}_{\mathrm{RX}} = \boldsymbol{x}_i - \boldsymbol{p}_k$, and $\boldsymbol{d}_{\mathrm{TX}} = \boldsymbol{x}_i - \boldsymbol{p}_{\mathrm{BS}}$.
    \item \emph{2-bounce wall--wall path ($(\mathrm{VA}, \mathrm{VA})$):} For $\mathrm{BS} \to \text{wall } u \to \text{wall } v \to \mathrm{RX}$, defining the second-order virtual anchor $\boldsymbol{x}_{uv}^{(2)} \triangleq \mathcal{M}(\boldsymbol{x}_u; \boldsymbol{x}_v)$, unfolding across walls $u$ and $v$ gives $h_d = \|\boldsymbol{x}_{uv}^{(2)} - \boldsymbol{p}_k\|$ and $\boldsymbol{d}_{\mathrm{RX}} = \boldsymbol{x}_{uv}^{(2)} - \boldsymbol{p}_k$; reflecting the final unit arrival direction back across wall $v$ then wall $u$ gives
    \begin{align}
    \boldsymbol{d}_{\mathrm{TX}} = \boldsymbol{H}(\boldsymbol{d}_u)\boldsymbol{H}(\boldsymbol{d}_v)(-\boldsymbol{d}_{\mathrm{RX}}) = -\boldsymbol{H}(\boldsymbol{d}_u)\boldsymbol{H}(\boldsymbol{d}_v)\boldsymbol{d}_{\mathrm{RX}}. \label{eq:S_dtx_vava}
    \end{align}
    \item \emph{2-bounce wall--point path ($(\mathrm{VA}, \mathrm{RP})$):} For $\mathrm{BS} \to \text{wall } u \to \text{scatterer } v \to \mathrm{RX}$, unfolding the first bounce across wall $u$ yields leg length $\|\boldsymbol{x}_v - \boldsymbol{x}_u\|$ and $\boldsymbol{d}_{\mathrm{TX}} = \mathcal{M}(\boldsymbol{x}_v; \boldsymbol{x}_u) - \boldsymbol{p}_{\mathrm{BS}} = \boldsymbol{H}(\boldsymbol{d}_u)(\boldsymbol{x}_v - \boldsymbol{x}_u)$, while the second leg is direct from $\boldsymbol{x}_v$ to $\boldsymbol{p}_k$, giving $h_d = \|\boldsymbol{x}_v - \boldsymbol{x}_u\| + \|\boldsymbol{x}_v - \boldsymbol{p}_k\|$ and $\boldsymbol{d}_{\mathrm{RX}} = \boldsymbol{x}_v - \boldsymbol{p}_k$.
    \item \emph{2-bounce point--wall path ($(\mathrm{RP}, \mathrm{VA})$):} For $\mathrm{BS} \to \text{scatterer } u \to \text{wall } v \to \mathrm{RX}$, the first leg gives $\boldsymbol{d}_{\mathrm{TX}} = \boldsymbol{x}_u - \boldsymbol{p}_{\mathrm{BS}}$; unfolding the second bounce across wall $v$ defines $\boldsymbol{x}_{uv}^{(2)} \triangleq \mathcal{M}(\boldsymbol{x}_u; \boldsymbol{x}_v)$, giving $\boldsymbol{d}_{\mathrm{RX}} = \boldsymbol{x}_{uv}^{(2)} - \boldsymbol{p}_k$ and $h_d = \|\boldsymbol{x}_u - \boldsymbol{p}_{\mathrm{BS}}\| + \|\boldsymbol{x}_{uv}^{(2)} - \boldsymbol{p}_k\|$.
\end{enumerate}
Under the infinite-line VA and point-scattering parameterization of~\cite[Sec.~II]{MainPaper} (which requires no finite wall-endpoint coordinates), a geometrically valid reflection path requires transceiver and scatterer endpoints to lie on the reflecting half-plane of each wall without penetration and each propagation leg length to be positive.

\begin{table}[!t]
\centering
\caption{Closed-Form Geometric Quantities Across the Six Propagation Path Types ($\boldsymbol{x}_{uv}^{(2)} \triangleq \mathcal{M}(\boldsymbol{x}_u; \boldsymbol{x}_v)$)}
\label{tab:S_paths}
\footnotesize
\setlength{\tabcolsep}{7pt}
\begin{tabular}{@{}cccccc@{}}
\toprule
Bounce & Path Type $\eta$ & Active States & Propagation Distance $h_{d,\eta}$ & RX Vector $\boldsymbol{d}_{\mathrm{RX},\eta}$ & BS Vector $\boldsymbol{d}_{\mathrm{TX},\eta}$ \\
\midrule
$0$ & $\mathrm{LoS}$ & $\boldsymbol{s}_k$ & $\|\boldsymbol{p}_{\mathrm{BS}} - \boldsymbol{p}_k\|$ & $\boldsymbol{p}_{\mathrm{BS}} - \boldsymbol{p}_k$ & $\boldsymbol{p}_k - \boldsymbol{p}_{\mathrm{BS}}$ \\[2pt]
$1$ & $\mathrm{VA}$ & $\boldsymbol{x}_i, \boldsymbol{s}_k$ & $\|\boldsymbol{x}_i - \boldsymbol{p}_k\|$ & $\boldsymbol{x}_i - \boldsymbol{p}_k$ & $\boldsymbol{H}(\boldsymbol{d}_i)(\boldsymbol{p}_k - \boldsymbol{x}_i)$ \\[2pt]
$1$ & $\mathrm{RP}$ & $\boldsymbol{x}_i, \boldsymbol{s}_k$ & $\|\boldsymbol{x}_i - \boldsymbol{p}_{\mathrm{BS}}\| + \|\boldsymbol{x}_i - \boldsymbol{p}_k\|$ & $\boldsymbol{x}_i - \boldsymbol{p}_k$ & $\boldsymbol{x}_i - \boldsymbol{p}_{\mathrm{BS}}$ \\[2pt]
$2$ & $(\mathrm{VA}, \mathrm{VA})$ & $\boldsymbol{x}_u, \boldsymbol{x}_v, \boldsymbol{s}_k$ & $\|\boldsymbol{x}_{uv}^{(2)} - \boldsymbol{p}_k\|$ & $\boldsymbol{x}_{uv}^{(2)} - \boldsymbol{p}_k$ & $-\boldsymbol{H}(\boldsymbol{d}_u)\boldsymbol{H}(\boldsymbol{d}_v)\boldsymbol{d}_{\mathrm{RX}}$ \\[2pt]
$2$ & $(\mathrm{VA}, \mathrm{RP})$ & $\boldsymbol{x}_u, \boldsymbol{x}_v, \boldsymbol{s}_k$ & $\|\boldsymbol{x}_v - \boldsymbol{x}_u\| + \|\boldsymbol{x}_v - \boldsymbol{p}_k\|$ & $\boldsymbol{x}_v - \boldsymbol{p}_k$ & $\boldsymbol{H}(\boldsymbol{d}_u)(\boldsymbol{x}_v - \boldsymbol{x}_u)$ \\[2pt]
$2$ & $(\mathrm{RP}, \mathrm{VA})$ & $\boldsymbol{x}_u, \boldsymbol{x}_v, \boldsymbol{s}_k$ & $\|\boldsymbol{x}_u - \boldsymbol{p}_{\mathrm{BS}}\| + \|\boldsymbol{x}_{uv}^{(2)} - \boldsymbol{p}_k\|$ & $\boldsymbol{x}_{uv}^{(2)} - \boldsymbol{p}_k$ & $\boldsymbol{x}_u - \boldsymbol{p}_{\mathrm{BS}}$ \\
\bottomrule
\end{tabular}
\end{table}

\subsection{Derivative Identities and Analytical Observation Jacobians}

Let $\boldsymbol{J}_\perp \triangleq \begin{bmatrix} 0 & -1 \\ 1 & 0 \end{bmatrix}$ denote $90^\circ$ counterclockwise rotation. For a nonzero 2D vector $\boldsymbol{d} \in \mathbb{R}^2 \setminus \{\boldsymbol{0}\}$,
\begin{align}
\boldsymbol{e}(\boldsymbol{d}) \triangleq \frac{\boldsymbol{d}}{\|\boldsymbol{d}\|}, \qquad
\boldsymbol{g}(\boldsymbol{d}) \triangleq \frac{\boldsymbol{J}_\perp\boldsymbol{d}}{\|\boldsymbol{d}\|^2}, \qquad
\frac{\partial \|\boldsymbol{d}\|}{\partial \boldsymbol{d}} = \boldsymbol{e}(\boldsymbol{d})^{\mathsf{T}}, \qquad
\frac{\partial (\angle\boldsymbol{d})}{\partial \boldsymbol{d}} = \boldsymbol{g}(\boldsymbol{d})^{\mathsf{T}}. \label{eq:S_vec_deriv}
\end{align}
For a wall virtual anchor $\boldsymbol{x}_u \in \mathbb{R}^2 \setminus \{\boldsymbol{p}_{\mathrm{BS}}\}$ ($\boldsymbol{d}_u = \boldsymbol{x}_u - \boldsymbol{p}_{\mathrm{BS}}$), fixed vector $\boldsymbol{b} \in \mathbb{R}^2$, and 2D point $\boldsymbol{a} \in \mathbb{R}^2$, differentiating $\boldsymbol{H}(\boldsymbol{d}_u)\boldsymbol{b}$ and $\mathcal{M}(\boldsymbol{a}; \boldsymbol{x}_u)$ gives
\begin{gather}
\left.\frac{\partial\big(\boldsymbol{H}(\boldsymbol{d}_u)\boldsymbol{b}\big)}{\partial\boldsymbol{x}_u}\right|_{\boldsymbol{b}\text{ fixed}} = 2\big(\boldsymbol{J}_\perp\boldsymbol{H}(\boldsymbol{d}_u)\boldsymbol{b}\big)\boldsymbol{g}(\boldsymbol{d}_u)^{\mathsf{T}}, \label{eq:S_Hb_deriv} \\
\frac{\partial \mathcal{M}(\boldsymbol{a}; \boldsymbol{x}_u)}{\partial \boldsymbol{a}} = \boldsymbol{H}(\boldsymbol{d}_u), \qquad
\boldsymbol{M}_x(\boldsymbol{a}; \boldsymbol{x}_u) \triangleq \frac{\partial \mathcal{M}(\boldsymbol{a}; \boldsymbol{x}_u)}{\partial \boldsymbol{x}_u} = \boldsymbol{I}_2 + 2\big(\boldsymbol{J}_\perp\boldsymbol{H}(\boldsymbol{d}_u)(\boldsymbol{a} - \boldsymbol{p}_{\mathrm{BS}})\big)\boldsymbol{g}(\boldsymbol{d}_u)^{\mathsf{T}}. \label{eq:S_M_deriv}
\end{gather}
Moreover, for a nonzero vector $\boldsymbol{w} \in \mathbb{R}^2$, specular reflection shifts direction angles according to
\begin{align}
\angle\big(\boldsymbol{H}(\boldsymbol{d}_u)\boldsymbol{w}\big) \equiv \pi + 2\angle\boldsymbol{d}_u - \angle\boldsymbol{w} \pmod{2\pi}, \qquad
\angle\big(-\boldsymbol{H}(\boldsymbol{d}_u)\boldsymbol{H}(\boldsymbol{d}_v)\boldsymbol{w}\big) \equiv \pi + 2\angle\boldsymbol{d}_u - 2\angle\boldsymbol{d}_v + \angle\boldsymbol{w} \pmod{2\pi}. \label{eq:S_angle_cong}
\end{align}
Applying the chain rule with~\eqref{eq:S_vec_deriv}--\eqref{eq:S_angle_cong} yields the observation Jacobians for the six path types:
\begin{enumerate}
    \item \emph{0-bounce LoS ($\boldsymbol{d}_{\mathrm{RX}} = \boldsymbol{p}_{\mathrm{BS}} - \boldsymbol{p}_k$):}
    \begin{align}
    \boldsymbol{H}_{0,k} \triangleq \frac{\partial \boldsymbol{h}_{\mathrm{LoS}}(\boldsymbol{s}_k)}{\partial \boldsymbol{s}_k} =
    \begin{bmatrix}
    -\boldsymbol{e}(\boldsymbol{d}_{\mathrm{RX}})^{\mathsf{T}} & 0 \\[2pt]
    -\boldsymbol{g}(\boldsymbol{d}_{\mathrm{RX}})^{\mathsf{T}} & -1 \\[2pt]
    -\boldsymbol{g}(\boldsymbol{d}_{\mathrm{RX}})^{\mathsf{T}} & 0
    \end{bmatrix} \in \mathbb{R}^{3\times 3}. \label{eq:S_jac_los}
    \end{align}
    \item \emph{1-bounce VA ($\boldsymbol{d}_i = \boldsymbol{x}_i - \boldsymbol{p}_{\mathrm{BS}}$, $\boldsymbol{d}_{\mathrm{RX}} = \boldsymbol{x}_i - \boldsymbol{p}_k$):}
    \begin{align}
    \boldsymbol{H}_{1,i,k,\mathrm{S}}^{(\mathrm{VA})} \triangleq \frac{\partial \boldsymbol{h}_{\mathrm{VA}}}{\partial \boldsymbol{s}_k} =
    \begin{bmatrix}
    -\boldsymbol{e}(\boldsymbol{d}_{\mathrm{RX}})^{\mathsf{T}} & 0 \\[2pt]
    -\boldsymbol{g}(\boldsymbol{d}_{\mathrm{RX}})^{\mathsf{T}} & -1 \\[2pt]
    \boldsymbol{g}(\boldsymbol{d}_{\mathrm{RX}})^{\mathsf{T}} & 0
    \end{bmatrix}, \qquad
    \boldsymbol{H}_{1,i,k,\mathrm{L}}^{(\mathrm{VA})} \triangleq \frac{\partial \boldsymbol{h}_{\mathrm{VA}}}{\partial \boldsymbol{x}_i} =
    \begin{bmatrix}
    \boldsymbol{e}(\boldsymbol{d}_{\mathrm{RX}})^{\mathsf{T}} \\[2pt]
    \boldsymbol{g}(\boldsymbol{d}_{\mathrm{RX}})^{\mathsf{T}} \\[2pt]
    2\boldsymbol{g}(\boldsymbol{d}_i)^{\mathsf{T}} - \boldsymbol{g}(\boldsymbol{d}_{\mathrm{RX}})^{\mathsf{T}}
    \end{bmatrix}. \label{eq:S_jac_va}
    \end{align}
    \item \emph{1-bounce RP ($\boldsymbol{d}_{\mathrm{RX}} = \boldsymbol{x}_i - \boldsymbol{p}_k$, $\boldsymbol{d}_{\mathrm{TX}} = \boldsymbol{x}_i - \boldsymbol{p}_{\mathrm{BS}}$):}
    \begin{align}
    \boldsymbol{H}_{1,i,k,\mathrm{S}}^{(\mathrm{RP})} \triangleq \frac{\partial \boldsymbol{h}_{\mathrm{RP}}}{\partial \boldsymbol{s}_k} =
    \begin{bmatrix}
    -\boldsymbol{e}(\boldsymbol{d}_{\mathrm{RX}})^{\mathsf{T}} & 0 \\[2pt]
    -\boldsymbol{g}(\boldsymbol{d}_{\mathrm{RX}})^{\mathsf{T}} & -1 \\[2pt]
    \boldsymbol{0}_{1\times 2} & 0
    \end{bmatrix}, \qquad
    \boldsymbol{H}_{1,i,k,\mathrm{L}}^{(\mathrm{RP})} \triangleq \frac{\partial \boldsymbol{h}_{\mathrm{RP}}}{\partial \boldsymbol{x}_i} =
    \begin{bmatrix}
    \boldsymbol{e}(\boldsymbol{d}_{\mathrm{TX}})^{\mathsf{T}} + \boldsymbol{e}(\boldsymbol{d}_{\mathrm{RX}})^{\mathsf{T}} \\[2pt]
    \boldsymbol{g}(\boldsymbol{d}_{\mathrm{RX}})^{\mathsf{T}} \\[2pt]
    \boldsymbol{g}(\boldsymbol{d}_{\mathrm{TX}})^{\mathsf{T}}
    \end{bmatrix}. \label{eq:S_jac_rp}
    \end{align}
    \item \emph{2-bounce $(\mathrm{VA}, \mathrm{VA})$ ($\boldsymbol{d}_u = \boldsymbol{x}_u - \boldsymbol{p}_{\mathrm{BS}}$, $\boldsymbol{d}_v = \boldsymbol{x}_v - \boldsymbol{p}_{\mathrm{BS}}$, $\boldsymbol{x}_{uv}^{(2)} = \mathcal{M}(\boldsymbol{x}_u; \boldsymbol{x}_v)$, $\boldsymbol{d}_{\mathrm{RX}} = \boldsymbol{x}_{uv}^{(2)} - \boldsymbol{p}_k$):}
    \begin{gather}
    \boldsymbol{H}_{2,uv,k,\mathrm{S}}^{(\mathrm{VA},\mathrm{VA})} =
    \begin{bmatrix}
    -\boldsymbol{e}(\boldsymbol{d}_{\mathrm{RX}})^{\mathsf{T}} & 0 \\[2pt]
    -\boldsymbol{g}(\boldsymbol{d}_{\mathrm{RX}})^{\mathsf{T}} & -1 \\[2pt]
    -\boldsymbol{g}(\boldsymbol{d}_{\mathrm{RX}})^{\mathsf{T}} & 0
    \end{bmatrix}, \qquad
    \boldsymbol{H}_{2,uv,k,u}^{(\mathrm{VA},\mathrm{VA})} =
    \begin{bmatrix}
    \boldsymbol{e}(\boldsymbol{d}_{\mathrm{RX}})^{\mathsf{T}}\boldsymbol{H}(\boldsymbol{d}_v) \\[2pt]
    \boldsymbol{g}(\boldsymbol{d}_{\mathrm{RX}})^{\mathsf{T}}\boldsymbol{H}(\boldsymbol{d}_v) \\[2pt]
    2\boldsymbol{g}(\boldsymbol{d}_u)^{\mathsf{T}} + \boldsymbol{g}(\boldsymbol{d}_{\mathrm{RX}})^{\mathsf{T}}\boldsymbol{H}(\boldsymbol{d}_v)
    \end{bmatrix}, \notag\\
    \boldsymbol{H}_{2,uv,k,v}^{(\mathrm{VA},\mathrm{VA})} =
    \begin{bmatrix}
    \boldsymbol{e}(\boldsymbol{d}_{\mathrm{RX}})^{\mathsf{T}}\boldsymbol{M}_x(\boldsymbol{x}_u; \boldsymbol{x}_v) \\[2pt]
    \boldsymbol{g}(\boldsymbol{d}_{\mathrm{RX}})^{\mathsf{T}}\boldsymbol{M}_x(\boldsymbol{x}_u; \boldsymbol{x}_v) \\[2pt]
    -2\boldsymbol{g}(\boldsymbol{d}_v)^{\mathsf{T}} + \boldsymbol{g}(\boldsymbol{d}_{\mathrm{RX}})^{\mathsf{T}}\boldsymbol{M}_x(\boldsymbol{x}_u; \boldsymbol{x}_v)
    \end{bmatrix}. \label{eq:S_jac_vava}
    \end{gather}
    \item \emph{2-bounce $(\mathrm{VA}, \mathrm{RP})$ ($\boldsymbol{d}_u = \boldsymbol{x}_u - \boldsymbol{p}_{\mathrm{BS}}$, $\boldsymbol{d}_{uv} \triangleq \boldsymbol{x}_v - \boldsymbol{x}_u$, $\boldsymbol{d}_{\mathrm{RX}} = \boldsymbol{x}_v - \boldsymbol{p}_k$):}
    \begin{align}
    \boldsymbol{H}_{2,uv,k,\mathrm{S}}^{(\mathrm{VA},\mathrm{RP})} =
    \begin{bmatrix}
    -\boldsymbol{e}(\boldsymbol{d}_{\mathrm{RX}})^{\mathsf{T}} & 0 \\[2pt]
    -\boldsymbol{g}(\boldsymbol{d}_{\mathrm{RX}})^{\mathsf{T}} & -1 \\[2pt]
    \boldsymbol{0}_{1\times 2} & 0
    \end{bmatrix}, \quad
    \boldsymbol{H}_{2,uv,k,u}^{(\mathrm{VA},\mathrm{RP})} =
    \begin{bmatrix}
    -\boldsymbol{e}(\boldsymbol{d}_{uv})^{\mathsf{T}} \\[2pt]
    \boldsymbol{0}_{1\times 2} \\[2pt]
    2\boldsymbol{g}(\boldsymbol{d}_u)^{\mathsf{T}} + \boldsymbol{g}(\boldsymbol{d}_{uv})^{\mathsf{T}}
    \end{bmatrix}, \quad
    \boldsymbol{H}_{2,uv,k,v}^{(\mathrm{VA},\mathrm{RP})} =
    \begin{bmatrix}
    \boldsymbol{e}(\boldsymbol{d}_{uv})^{\mathsf{T}} + \boldsymbol{e}(\boldsymbol{d}_{\mathrm{RX}})^{\mathsf{T}} \\[2pt]
    \boldsymbol{g}(\boldsymbol{d}_{\mathrm{RX}})^{\mathsf{T}} \\[2pt]
    -\boldsymbol{g}(\boldsymbol{d}_{uv})^{\mathsf{T}}
    \end{bmatrix}. \label{eq:S_jac_varp}
    \end{align}
    \item \emph{2-bounce $(\mathrm{RP}, \mathrm{VA})$ ($\boldsymbol{d}_v = \boldsymbol{x}_v - \boldsymbol{p}_{\mathrm{BS}}$, $\boldsymbol{x}_{uv}^{(2)} = \mathcal{M}(\boldsymbol{x}_u; \boldsymbol{x}_v)$, $\boldsymbol{d}_{\mathrm{RX}} = \boldsymbol{x}_{uv}^{(2)} - \boldsymbol{p}_k$, $\boldsymbol{d}_{\mathrm{TX}} = \boldsymbol{x}_u - \boldsymbol{p}_{\mathrm{BS}}$):}
    \begin{gather}
    \boldsymbol{H}_{2,uv,k,\mathrm{S}}^{(\mathrm{RP},\mathrm{VA})} =
    \begin{bmatrix}
    -\boldsymbol{e}(\boldsymbol{d}_{\mathrm{RX}})^{\mathsf{T}} & 0 \\[2pt]
    -\boldsymbol{g}(\boldsymbol{d}_{\mathrm{RX}})^{\mathsf{T}} & -1 \\[2pt]
    \boldsymbol{0}_{1\times 2} & 0
    \end{bmatrix}, \qquad
    \boldsymbol{H}_{2,uv,k,u}^{(\mathrm{RP},\mathrm{VA})} =
    \begin{bmatrix}
    \boldsymbol{e}(\boldsymbol{d}_{\mathrm{TX}})^{\mathsf{T}} + \boldsymbol{e}(\boldsymbol{d}_{\mathrm{RX}})^{\mathsf{T}}\boldsymbol{H}(\boldsymbol{d}_v) \\[2pt]
    \boldsymbol{g}(\boldsymbol{d}_{\mathrm{RX}})^{\mathsf{T}}\boldsymbol{H}(\boldsymbol{d}_v) \\[2pt]
    \boldsymbol{g}(\boldsymbol{d}_{\mathrm{TX}})^{\mathsf{T}}
    \end{bmatrix}, \notag\\
    \boldsymbol{H}_{2,uv,k,v}^{(\mathrm{RP},\mathrm{VA})} =
    \begin{bmatrix}
    \boldsymbol{e}(\boldsymbol{d}_{\mathrm{RX}})^{\mathsf{T}}\boldsymbol{M}_x(\boldsymbol{x}_u; \boldsymbol{x}_v) \\[2pt]
    \boldsymbol{g}(\boldsymbol{d}_{\mathrm{RX}})^{\mathsf{T}}\boldsymbol{M}_x(\boldsymbol{x}_u; \boldsymbol{x}_v) \\[2pt]
    \boldsymbol{0}_{1\times 2}
    \end{bmatrix}. \label{eq:S_jac_rpva}
    \end{gather}
\end{enumerate}
For brevity in later sections, given $\psi_i$ and $\eta_{uv} = (\psi_u, \psi_v) \in \mathcal{T}_2$, we denote $\boldsymbol{H}_{1,i,k,\mathrm{L}} \triangleq \boldsymbol{H}_{1,i,k,\mathrm{L}}^{(\psi_i)} \in \mathbb{R}^{3\times 2}$ and define the stacked 2-bounce landmark Jacobian
\begin{align}
\boldsymbol{H}_{2,uv,k,\mathrm{L}} \triangleq \begin{bmatrix} \boldsymbol{H}_{2,uv,k,u}^{(\psi_u,\psi_v)} & \boldsymbol{H}_{2,uv,k,v}^{(\psi_u,\psi_v)} \end{bmatrix} \in \mathbb{R}^{3\times 4}. \label{eq:S_jac_2L}
\end{align}

\section{Derivation of the Batch Joint Posterior Distribution (Proof of Theorem~1)} \label{app:B}

This appendix corresponds to Section~III-A of~\cite{MainPaper} and derives the batch double-bounce multi-type joint posterior representation in~\cite[Theorem~1, Eq.~(11)]{MainPaper} under the data association constraints and decoupling approximation stated therein.

\subsection{Single-Scan Multipath RFS Likelihood and Batch Expansion}

Given RX pose $\boldsymbol{s}_k$ and landmark map $\mathcal{X} = \{\tilde{\boldsymbol{x}}_1, \dots, \tilde{\boldsymbol{x}}_n\}$ with $\tilde{\boldsymbol{x}}_i = (\boldsymbol{x}_i, \psi_i)$, under the point-target and Poisson point process (PPP) clutter assumptions in~\cite[Sec.~II-D]{MainPaper}, each scan $\mathcal{Z}_k$ is partitioned into disjoint subsets assigned to clutter $\mathcal{Z}_k^{\mathrm{c}}$, 0-bounce LoS $\mathcal{Z}_{0,k}$, 1-bounce paths $\mathcal{Z}_{i,k}^{(1)}$, and ordered 2-bounce paths $\mathcal{Z}_{uv,k}^{(2)}$:
\begin{align}
\mathcal{Z}_k = \mathcal{Z}_k^{\mathrm{c}} \biguplus \mathcal{Z}_{0,k} \biguplus \left(\biguplus_{i=1}^n \mathcal{Z}_{i,k}^{(1)}\right) \biguplus \left(\biguplus_{\substack{u,v=1\\u \neq v}}^n \mathcal{Z}_{uv,k}^{(2)}\right). \label{eq:S_scan_part}
\end{align}
Let $\mathcal{P}(\mathcal{Z}_k)$ denote the set of valid single-scan assignments satisfying~\eqref{eq:S_scan_part} with at most one measurement per path. Using the single-scan likelihood factors $\ell_b(\mathcal{Z}|\cdot)$ in~\cite[Eq.~(8)]{MainPaper} (setting $\ell_2(\emptyset|\cdot) = 1$ and $\ell_2(\{\boldsymbol{z}\}|\cdot) = 0$ whenever $(\psi_u, \psi_v) \notin \mathcal{T}_2$), the single-scan RFS likelihood~\cite{Ge2025TSP} reads
\begin{align}
g(\mathcal{Z}_k|\boldsymbol{s}_k, \mathcal{X}) = e^{-\int c(\boldsymbol{z})\,\mathrm{d}\boldsymbol{z}} \sum_{\mathcal{P}(\mathcal{Z}_k)}
\left[\prod_{\boldsymbol{z} \in \mathcal{Z}_k^{\mathrm{c}}} c(\boldsymbol{z})\right]
\ell_0(\mathcal{Z}_{0,k}|\boldsymbol{s}_k)
\left[\prod_{i=1}^n \ell_1(\mathcal{Z}_{i,k}^{(1)}|\tilde{\boldsymbol{x}}_i, \boldsymbol{s}_k)\right]
\left[\prod_{\substack{u,v=1\\u \neq v}}^n \ell_2(\mathcal{Z}_{uv,k}^{(2)}|\tilde{\boldsymbol{x}}_u, \tilde{\boldsymbol{x}}_v, \boldsymbol{s}_k)\right]. \label{eq:S_scan_rfs}
\end{align}
Multiplying~\eqref{eq:S_scan_rfs} over $k = 1, \dots, K$, letting $\mathcal{P}(\mathcal{Z}_{1:K}) \triangleq \mathcal{P}(\mathcal{Z}_1) \times \dots \times \mathcal{P}(\mathcal{Z}_K)$ denote the joint set of batch assignments, and grouping time products per physical path yields
\begin{align}
\prod_{k=1}^K g(\mathcal{Z}_k|\boldsymbol{s}_k, \mathcal{X})
&= e^{-K\int c(\boldsymbol{z})\,\mathrm{d}\boldsymbol{z}} \sum_{\mathcal{P}(\mathcal{Z}_{1:K})}
\left[\prod_{\boldsymbol{z} \in \mathcal{Z}_{1:K}^{\mathrm{c}}} c(\boldsymbol{z})\right]
\left[\prod_{k=1}^K \ell_0(\mathcal{Z}_{0,k}|\boldsymbol{s}_k)\right] \notag\\
&\quad \times \left[\prod_{i=1}^n \prod_{k=1}^K \ell_1(\mathcal{Z}_{i,k}^{(1)}|\tilde{\boldsymbol{x}}_i, \boldsymbol{s}_k)\right]
\left[\prod_{\substack{u,v=1\\u \neq v}}^n \prod_{k=1}^K \ell_2(\mathcal{Z}_{uv,k}^{(2)}|\tilde{\boldsymbol{x}}_u, \tilde{\boldsymbol{x}}_v, \boldsymbol{s}_k)\right]. \label{eq:S_batch_rfs}
\end{align}

\subsection{Detected/Undetected Partition and Decoupling Approximation}

For a given batch assignment in~\eqref{eq:S_batch_rfs}, partition $\mathcal{X} = \mathcal{X}_{\mathrm{D}} \biguplus \mathcal{X}_{\mathrm{U}}$, where $\mathcal{X}_{\mathrm{D}}$ comprises landmarks generating at least one measurement (1-bounce or 2-bounce) across $1:K$ and $\mathcal{X}_{\mathrm{U}}$ comprises landmarks generating no measurements over $1:K$. With $p_{\mathrm{U}}(\tilde{\boldsymbol{x}}, \boldsymbol{s}_{1:K}) \triangleq \prod_{k=1}^K (1 - p_{\mathrm{D},1}(\tilde{\boldsymbol{x}}, \boldsymbol{s}_k))$, the 1-bounce product splits as
\begin{align}
\prod_{i=1}^n \prod_{k=1}^K \ell_1(\mathcal{Z}_{i,k}^{(1)}|\tilde{\boldsymbol{x}}_i, \boldsymbol{s}_k)
= \underbrace{\left[\prod_{\tilde{\boldsymbol{x}}_i \in \mathcal{X}_{\mathrm{D}}} \prod_{k=1}^K \ell_1(\mathcal{Z}_{i,k}^{(1)}|\tilde{\boldsymbol{x}}_i, \boldsymbol{s}_k)\right]}_{\text{detected landmarks (incl.\ }\mathcal{I}_{\mathrm{occ}}\text{ 1-B misses)}}
\underbrace{\left[\prod_{\tilde{\boldsymbol{x}} \in \mathcal{X}_{\mathrm{U}}} p_{\mathrm{U}}(\tilde{\boldsymbol{x}}, \boldsymbol{s}_{1:K})\right]}_{\text{undetected landmarks}}, \label{eq:S_split_1b}
\end{align}
whereas the 2-bounce product decomposes according to endpoint membership into four factors:
\begin{align}
\prod_{\substack{u,v=1\\u \neq v}}^n \prod_{k=1}^K \ell_2(\mathcal{Z}_{uv,k}^{(2)}|\tilde{\boldsymbol{x}}_u, \tilde{\boldsymbol{x}}_v, \boldsymbol{s}_k)
= &\underbrace{\left[\prod_{\substack{\tilde{\boldsymbol{x}}_u,\tilde{\boldsymbol{x}}_v \in \mathcal{X}_{\mathrm{D}}\\u \neq v}} \prod_{k=1}^K \ell_2(\mathcal{Z}_{uv,k}^{(2)}|\tilde{\boldsymbol{x}}_u, \tilde{\boldsymbol{x}}_v, \boldsymbol{s}_k)\right]}_{\text{retained (2-B factors on }\mathcal{X}_{\mathrm{D}}\text{)}}
\underbrace{\left[\prod_{\tilde{\boldsymbol{x}}_u \in \mathcal{X}_{\mathrm{D}},\,\tilde{\boldsymbol{x}}_v \in \mathcal{X}_{\mathrm{U}}} \prod_{k=1}^K \big(1 - p_{\mathrm{D},2}(\tilde{\boldsymbol{x}}_u, \tilde{\boldsymbol{x}}_v, \boldsymbol{s}_k)\big)\right]}_{\approx 1\text{ (neglected)}} \notag\\
&\times \underbrace{\left[\prod_{\tilde{\boldsymbol{x}}_u \in \mathcal{X}_{\mathrm{U}},\,\tilde{\boldsymbol{x}}_v \in \mathcal{X}_{\mathrm{D}}} \prod_{k=1}^K \big(1 - p_{\mathrm{D},2}(\tilde{\boldsymbol{x}}_u, \tilde{\boldsymbol{x}}_v, \boldsymbol{s}_k)\big)\right]}_{\approx 1\text{ (neglected)}}
\underbrace{\left[\prod_{\substack{\tilde{\boldsymbol{x}}_u,\tilde{\boldsymbol{x}}_v \in \mathcal{X}_{\mathrm{U}}\\u \neq v}} \prod_{k=1}^K \big(1 - p_{\mathrm{D},2}(\tilde{\boldsymbol{x}}_u, \tilde{\boldsymbol{x}}_v, \boldsymbol{s}_k)\big)\right]}_{\approx 1\text{ (neglected)}}. \label{eq:S_split_2b}
\end{align}
Following the modeling assumption in~\cite[Sec.~III-A]{MainPaper}, when landmarks in $\mathcal{X}_{\mathrm{U}}$ (which generate no measurements over $1{:}K$) have small effective 2-bounce detection probability ($1 - p_{\mathrm{D},2} \approx 1$), approximating the trailing three products in~\eqref{eq:S_split_2b} by $1$ decouples $\mathcal{X}_{\mathrm{U}}$ from $\mathcal{X}_{\mathrm{D}}$ and retains a tractable marked PPP form for $\mathcal{X}_{\mathrm{U}}$. In the $\mathcal{X}_{\mathrm{D}}$ factor,~\cite[Constraints~(ii)--(iii)]{MainPaper} restrict the 2-bounce product to admissible identifiable pairs.

\subsection{Association Correspondence and Derivation of Theorem~1}

Substituting~\eqref{eq:S_split_1b} and the decoupled approximation of~\eqref{eq:S_split_2b} into~\eqref{eq:S_batch_rfs}, and following the subset-reindexing step of~\cite[Theorem~1]{Ge2025TSP}, the approximated batch RFS likelihood can be rewritten as
\begin{align}
\prod_{k=1}^K g(\mathcal{Z}_k|\boldsymbol{s}_k, \mathcal{X})
&= e^{-K\int c(\boldsymbol{z})\,\mathrm{d}\boldsymbol{z}} \sum_{j \in \mathbb{J}} L_0(\mathcal{Z}_0^j|\boldsymbol{s}_{1:K}) \notag\\
&\quad \times \sum_{\mathcal{X}_{\mathrm{U}} \biguplus \mathcal{Y}_1^j \biguplus \dots \biguplus \mathcal{Y}_{N^j}^j = \mathcal{X}}
\left[\prod_{\tilde{\boldsymbol{x}} \in \mathcal{X}_{\mathrm{U}}} p_{\mathrm{U}}(\tilde{\boldsymbol{x}}, \boldsymbol{s}_{1:K})\right]
\prod_{i=1}^{N^j} L_1(\mathcal{Z}_i^{j,(1)}|\boldsymbol{s}_{1:K}, \mathcal{Y}_i^j)
\prod_{u \neq v}^{N^j} L_2(\mathcal{Z}_{uv}^{j,(2)}|\boldsymbol{s}_{1:K}, \mathcal{Y}_u^j, \mathcal{Y}_v^j). \label{eq:S_batch_reindex}
\end{align}
For a given static map $\mathcal{X} = \{\tilde{\boldsymbol{x}}_1, \dots, \tilde{\boldsymbol{x}}_n\}$, the nonzero batch assignments in~\eqref{eq:S_batch_rfs} (after Appendix~\ref{app:B}-B) and the nonzero summands in~\eqref{eq:S_batch_reindex} stand in one-to-one correspondence with identical factor values:
\begin{itemize}
    \item \emph{Forward map:} Each nonzero batch assignment in~\eqref{eq:S_batch_rfs} specifies the LoS sequence $\mathcal{Z}_0$, $n_{\mathrm{c}}$ clutter singletons, $m_{\mathrm{vis}}$ detected landmarks with non-empty 1-bounce sequences, $m_{\mathrm{occ}}$ detected occluded landmarks observed solely via a single 2-bounce cluster, and $\mathcal{X}_{\mathrm{U}} = \mathcal{X} \setminus \mathcal{X}_{\mathrm{D}}$. Sorting the $N_{\mathrm{vis}}^j = m_{\mathrm{vis}} + n_{\mathrm{c}}$ non-empty 1-bounce clusters and $N_{\mathrm{occ}}^j = m_{\mathrm{occ}}$ non-empty occluded 2-bounce clusters by the global index of their earliest measurement (ordered by scan $k$ and then within-scan index $m$) defines $\mathcal{I}^j = \mathcal{I}_{\mathrm{vis}}^j \biguplus \mathcal{I}_{\mathrm{occ}}^j$, a corresponding global DA $j \in \mathbb{J}$, and---by assigning $\mathcal{Y}_i^j = \{\tilde{\boldsymbol{x}}\}$ to detected landmarks and $\mathcal{Y}_i^j = \emptyset$ to clutter---a corresponding partition $\mathcal{X}_{\mathrm{U}} \biguplus \mathcal{Y}_1^j \biguplus \dots \biguplus \mathcal{Y}_{N^j}^j = \mathcal{X}$.
    \item \emph{Inverse map:} In the double sum of~\eqref{eq:S_batch_reindex}, by~\cite[Eqs.~(9)--(10)]{MainPaper}, a subset partition with $|\mathcal{Y}_i^j| \ge 2$, with $\mathcal{Y}_i^j = \emptyset$ on a cluster of cardinality $|\mathcal{Z}_i^{j,(1)}| \neq 1$, or violating constraints~(i)--(iii) evaluates to zero, whereas each remaining nonzero summand maps back to its corresponding valid batch assignment in~\eqref{eq:S_batch_rfs} with matching factor values.
\end{itemize}
Factorizing the marked PPP prior~\cite[Eq.~(3)]{MainPaper} over the same disjoint partition (with $\lambda(\emptyset) = 1$, $\lambda(\{\tilde{\boldsymbol{x}}_i\}) = \lambda_{\psi_i}(\boldsymbol{x}_i)$, and $0$ otherwise) gives
\begin{align}
f_{\mathrm{P}}(\mathcal{X}) = e^{-\sum_{\psi \in \mathcal{T}_1} \int \lambda_\psi(\boldsymbol{x})\,\mathrm{d}\boldsymbol{x}}
\left[\prod_{\tilde{\boldsymbol{x}} \in \mathcal{X}_{\mathrm{U}}} \lambda_\psi(\boldsymbol{x})\right]
\prod_{i=1}^{N^j} \lambda(\mathcal{Y}_i^j). \label{eq:S_ppp_split}
\end{align}
Substituting~\eqref{eq:S_batch_reindex} and~\eqref{eq:S_ppp_split} into Bayes' rule~\cite[Eq.~(7)]{MainPaper} yields~\cite[Theorem~1, Eq.~(11)]{MainPaper}. \hfill$\blacksquare$

\section{Marginal Likelihood Approximations, MCMC Weight Ratios, and Warm-Start Pre-Clustering} \label{app:C}

This appendix corresponds to Section~IV of~\cite{MainPaper}, detailing the pairwise expectation factorization, Laplace marginal likelihood approximations, local MCMC weight ratios, and the warm-start 2-bounce matching and geometric inversion procedure (Algorithm~\ref{alg:S_warmstart}, Steps~4--5).

\subsection{Pairwise Expectation Factorization of the DA Weight}

When integrating the $\mathcal{A}$-term of~\cite[Eq.~(11)]{MainPaper} over $\mathcal{X} = \mathcal{X}_{\mathrm{U}} \biguplus \mathcal{Y}_1 \biguplus \dots \biguplus \mathcal{Y}_N$, $\mathcal{X}_{\mathrm{U}}$ is decoupled from $\mathcal{Y}_{1:N}$, so its set integral depends only on $\boldsymbol{s}_{1:K}$ and enters the proportionality constant of~\cite[Eq.~(16)]{MainPaper}. Substituting $L_1(\mathcal{Z}_i^{(1)}|\boldsymbol{s}_{1:K}, \mathcal{Y}_i)\lambda(\mathcal{Y}_i) = \mathcal{L}_1(\mathcal{Z}_i^{(1)})\,p(\mathcal{Y}_i|\mathcal{Z}_i^{(1)}, \boldsymbol{s}_{1:K})$ from~\cite[Eq.~(17)]{MainPaper} into~\cite[Eq.~(16)]{MainPaper} and defining $q(\mathcal{Y}_{1:N}) \triangleq \prod_{i=1}^N p(\mathcal{Y}_i|\mathcal{Z}_i^{(1)}, \boldsymbol{s}_{1:K})$ yields
\begin{align}
w(\mathcal{A}) = \mathcal{L}_0(\mathcal{Z}_0) \left[\prod_{i=1}^N \mathcal{L}_1(\mathcal{Z}_i^{(1)})\right] \mathbb{E}_q\!\left[\prod_{\substack{u,v=1\\u \neq v}}^N L_2(\mathcal{Z}_{uv}^{(2)}|\boldsymbol{s}_{1:K}, \mathcal{Y}_u, \mathcal{Y}_v)\right]. \label{eq:S_weight_exp}
\end{align}

In~\eqref{eq:S_weight_exp}, by~\cite[Constraint~(ii)]{MainPaper}, each $v \in \mathcal{I}_{\mathrm{occ}}$ enters only a single non-empty occluded 2-bounce cluster while all other $L_2$ terms involving $v$ equal $1$, so integration over $\mathcal{Y}_v$ acts on a single factor. When a visible candidate $u \in \mathcal{I}_{\mathrm{vis}}$ receives informative 3D range--angle measurements, $p(\mathcal{Y}_u|\mathcal{Z}_u^{(1)}, \boldsymbol{s}_{1:K})$ is typically concentrated around $\hat{\boldsymbol{x}}_{u,\psi_u}$, so coupling across distinct 2-bounce factors sharing $u$ remains modest. For computational tractability, approximating the expectation of the product by the product of pairwise expectations,
\begin{align}
\mathbb{E}_q\!\left[\prod_{u \neq v} L_2(\mathcal{Z}_{uv}^{(2)}|\boldsymbol{s}_{1:K}, \mathcal{Y}_u, \mathcal{Y}_v)\right]
\approx \prod_{u \neq v} \mathbb{E}_{p_u p_v}\!\left[L_2(\mathcal{Z}_{uv}^{(2)}|\boldsymbol{s}_{1:K}, \mathcal{Y}_u, \mathcal{Y}_v)\right]
= \prod_{u \neq v} \mathcal{L}_2(\mathcal{Z}_{uv}^{(2)}|\mathcal{Z}_u^{(1)}, \mathcal{Z}_v^{(1)}), \label{eq:S_exp_fact}
\end{align}
gives~\cite[Eqs.~(18)--(19)]{MainPaper}.

\subsection{Laplace Marginal Likelihood Approximations and Regularization}

Let $\mathcal{K}_{\mathrm{det},i}$ ($M_i = |\mathcal{K}_{\mathrm{det},i}| \ge 1$) and $\mathcal{K}_{\mathrm{det},uv}$ ($M_{uv} = |\mathcal{K}_{\mathrm{det},uv}| \ge 1$) denote the active detection scan sets of a 1-bounce cluster and a 2-bounce cluster, respectively.
\begin{enumerate}
    \item \emph{1-bounce marginal likelihood $\mathcal{L}_1(\mathcal{Z}_i^{(1)}; \psi)$ (\cite[Eqs.~(20)--(21)]{MainPaper}):} For $\psi \in \mathcal{T}_1$, define the weighted LS objective
    \begin{align}
    S_{1,\psi}(\boldsymbol{x}_i) \triangleq \sum_{k \in \mathcal{K}_{\mathrm{det},i}} \big(\boldsymbol{z}_k - \boldsymbol{h}_\psi(\boldsymbol{x}_i, \boldsymbol{s}_k)\big)^{\mathsf{T}}\boldsymbol{R}_\psi^{-1}\big(\boldsymbol{z}_k - \boldsymbol{h}_\psi(\boldsymbol{x}_i, \boldsymbol{s}_k)\big). \label{eq:S_cost_1b}
    \end{align}
    Minimizing~\eqref{eq:S_cost_1b} via damped Gauss-Newton yields $\hat{\boldsymbol{x}}_{i,\psi}$ and the local landmark information matrix $\boldsymbol{\Omega}_{i,\psi} \in \mathbb{R}^{2\times 2}$ given by $\boldsymbol{\Omega}_{i,\psi} = \sum_{k \in \mathcal{K}_{\mathrm{det},i}} \boldsymbol{H}_{1,i,k,\mathrm{L}}^{\mathsf{T}}\boldsymbol{R}_\psi^{-1}\boldsymbol{H}_{1,i,k,\mathrm{L}}$ ($\hat{\boldsymbol{\Sigma}}_{i,\psi} = \boldsymbol{\Omega}_{i,\psi}^{-1}$). Approximating the integral around $\hat{\boldsymbol{x}}_{i,\psi}$ yields
    \begin{align}
    \mathcal{L}_1(\mathcal{Z}_i^{(1)}; \psi) \approx &\,\lambda_\psi(\hat{\boldsymbol{x}}_{i,\psi}) \left[\prod_{k \notin \mathcal{K}_{\mathrm{det},i}} \big(1 - p_{\mathrm{D},1}(\hat{\boldsymbol{x}}_{i,\psi}, \psi, \boldsymbol{s}_k)\big)\right] \left[\prod_{k \in \mathcal{K}_{\mathrm{det},i}} p_{\mathrm{D},1}(\hat{\boldsymbol{x}}_{i,\psi}, \psi, \boldsymbol{s}_k)\right] \notag\\
    &\times (2\pi)^{\frac{2 - 3M_i}{2}} |\boldsymbol{R}_\psi|^{-\frac{M_i}{2}} |\boldsymbol{\Omega}_{i,\psi}|^{-\frac{1}{2}} \exp\!\left(-\frac{1}{2} S_{1,\psi}(\hat{\boldsymbol{x}}_{i,\psi})\right). \label{eq:S_L1_lap}
    \end{align}
    \item \emph{Non-empty visible 2-bounce factor $\mathcal{L}_2(\mathcal{Z}_{uv}^{(2)}|\mathcal{Z}_u^{(1)}, \mathcal{Z}_v^{(1)}; \eta_{uv})$ (\cite[Eq.~(23)]{MainPaper}):} For $u, v \in \mathcal{I}_{\mathrm{vis}}$ and $\eta_{uv} = (\psi_u, \psi_v) \in \mathcal{T}_2$, stack $\boldsymbol{x}_{uv} \triangleq [\boldsymbol{x}_u^{\mathsf{T}}, \boldsymbol{x}_v^{\mathsf{T}}]^{\mathsf{T}} \in \mathbb{R}^4$ and form the 4D objective using the 1-bounce Gaussian posteriors ($\hat{\boldsymbol{x}}_{uv,0} = [\hat{\boldsymbol{x}}_{u,\psi_u}^{\mathsf{T}}, \hat{\boldsymbol{x}}_{v,\psi_v}^{\mathsf{T}}]^{\mathsf{T}}$, $\boldsymbol{\Omega}_{uv,0} = \operatorname{blkdiag}(\boldsymbol{\Omega}_{u,\psi_u}, \boldsymbol{\Omega}_{v,\psi_v})$):
    \begin{align}
    S_2(\boldsymbol{x}_{uv}) \triangleq (\boldsymbol{x}_{uv} - \hat{\boldsymbol{x}}_{uv,0})^{\mathsf{T}}\boldsymbol{\Omega}_{uv,0}(\boldsymbol{x}_{uv} - \hat{\boldsymbol{x}}_{uv,0}) + \sum_{k \in \mathcal{K}_{\mathrm{det},uv}} \big(\boldsymbol{z}_k - \boldsymbol{h}_{\eta_{uv}}(\boldsymbol{x}_u, \boldsymbol{x}_v, \boldsymbol{s}_k)\big)^{\mathsf{T}}\boldsymbol{R}_{\eta_{uv}}^{-1}\big(\boldsymbol{z}_k - \boldsymbol{h}_{\eta_{uv}}(\boldsymbol{x}_u, \boldsymbol{x}_v, \boldsymbol{s}_k)\big). \label{eq:S_cost_2v}
    \end{align}
    Minimizing~\eqref{eq:S_cost_2v} from $\hat{\boldsymbol{x}}_{uv,0}$ gives $\hat{\boldsymbol{x}}_{uv} = [\hat{\boldsymbol{x}}_u^{\mathsf{T}}, \hat{\boldsymbol{x}}_v^{\mathsf{T}}]^{\mathsf{T}}$ and $\boldsymbol{\Omega}_{uv} = \boldsymbol{\Omega}_{uv,0} + \sum_{k \in \mathcal{K}_{\mathrm{det},uv}} \boldsymbol{H}_{2,uv,k,\mathrm{L}}^{\mathsf{T}}\boldsymbol{R}_{\eta_{uv}}^{-1}\boldsymbol{H}_{2,uv,k,\mathrm{L}}$ using~\eqref{eq:S_jac_2L}; integrating around $\hat{\boldsymbol{x}}_{uv}$ gives
    \begin{align}
    \mathcal{L}_2(\mathcal{Z}_{uv}^{(2)}|\mathcal{Z}_u^{(1)}, \mathcal{Z}_v^{(1)}; \eta_{uv}) \approx &\left[\prod_{k \notin \mathcal{K}_{\mathrm{det},uv}} \big(1 - p_{\mathrm{D},2}(\hat{\boldsymbol{x}}_u, \psi_u, \hat{\boldsymbol{x}}_v, \psi_v, \boldsymbol{s}_k)\big)\right] \left[\prod_{k \in \mathcal{K}_{\mathrm{det},uv}} p_{\mathrm{D},2}(\hat{\boldsymbol{x}}_u, \psi_u, \hat{\boldsymbol{x}}_v, \psi_v, \boldsymbol{s}_k)\right] \notag\\
    &\times (2\pi)^{-\frac{3M_{uv}}{2}} |\boldsymbol{R}_{\eta_{uv}}|^{-\frac{M_{uv}}{2}} \frac{|\boldsymbol{\Omega}_{uv,0}|^{1/2}}{|\boldsymbol{\Omega}_{uv}|^{1/2}} \exp\!\left(-\frac{1}{2} S_2(\hat{\boldsymbol{x}}_{uv})\right). \label{eq:S_L2v_lap}
    \end{align}
    \item \emph{Non-empty occluded 2-bounce factor $\mathcal{L}_{2,\mathrm{occ}}(\mathcal{Z}_{uv}^{(2)}|\mathcal{Z}_u^{(1)}; \eta_{uv})$ (\cite[Eq.~(24)]{MainPaper}):} For $u \in \mathcal{I}_{\mathrm{vis}}, v \in \mathcal{I}_{\mathrm{occ}}$ (when $u$ is the 2nd bounce in cluster $(v, u)$, the factor is denoted $\mathcal{L}_{2,\mathrm{occ}}(\mathcal{Z}_{vu}^{(2)}|\mathcal{Z}_u^{(1)}; \eta_{vu})$ with $(\boldsymbol{x}_v, \psi_v)$ and $(\boldsymbol{x}_u, \psi_u)$ swapped), define
    \begin{align}
    S_{2,\mathrm{occ}}(\boldsymbol{x}_{uv}) \triangleq (\boldsymbol{x}_u - \hat{\boldsymbol{x}}_{u,\psi_u})^{\mathsf{T}}\boldsymbol{\Omega}_{u,\psi_u}(\boldsymbol{x}_u - \hat{\boldsymbol{x}}_{u,\psi_u}) + \sum_{k \in \mathcal{K}_{\mathrm{det},uv}} \big(\boldsymbol{z}_k - \boldsymbol{h}_{\eta_{uv}}(\boldsymbol{x}_u, \boldsymbol{x}_v, \boldsymbol{s}_k)\big)^{\mathsf{T}}\boldsymbol{R}_{\eta_{uv}}^{-1}\big(\boldsymbol{z}_k - \boldsymbol{h}_{\eta_{uv}}(\boldsymbol{x}_u, \boldsymbol{x}_v, \boldsymbol{s}_k)\big). \label{eq:S_cost_2o}
    \end{align}
    Initializing $\boldsymbol{x}_v$ from the closed-form inversion in Appendix~\ref{app:C}-D and minimizing~\eqref{eq:S_cost_2o} yields $\hat{\boldsymbol{x}}_{uv}$ and $\boldsymbol{\Omega}_{uv} = \operatorname{blkdiag}(\boldsymbol{\Omega}_{u,\psi_u}, \boldsymbol{0}_{2\times 2}) +\allowbreak \sum_{k \in \mathcal{K}_{\mathrm{det},uv}}\allowbreak \boldsymbol{H}_{2,uv,k,\mathrm{L}}^{\mathsf{T}}\allowbreak \boldsymbol{R}_{\eta_{uv}}^{-1}\allowbreak \boldsymbol{H}_{2,uv,k,\mathrm{L}}$, approximating the integral as
    \begin{align}
    \mathcal{L}_{2,\mathrm{occ}}(\mathcal{Z}_{uv}^{(2)}|\mathcal{Z}_u^{(1)}; \eta_{uv}) \approx &\,\lambda_{\psi_v}(\hat{\boldsymbol{x}}_v)\,p_{\mathrm{U}}(\hat{\boldsymbol{x}}_v, \psi_v, \boldsymbol{s}_{1:K}) \left[\prod_{k \notin \mathcal{K}_{\mathrm{det},uv}} \big(1 - p_{\mathrm{D},2}(\hat{\boldsymbol{x}}_u, \psi_u, \hat{\boldsymbol{x}}_v, \psi_v, \boldsymbol{s}_k)\big)\right] \notag\\
    &\times \left[\prod_{k \in \mathcal{K}_{\mathrm{det},uv}} p_{\mathrm{D},2}(\hat{\boldsymbol{x}}_u, \psi_u, \hat{\boldsymbol{x}}_v, \psi_v, \boldsymbol{s}_k)\right] \notag\\
    &\times (2\pi)^{\frac{2 - 3M_{uv}}{2}} |\boldsymbol{R}_{\eta_{uv}}|^{-\frac{M_{uv}}{2}} \frac{|\boldsymbol{\Omega}_{u,\psi_u}|^{1/2}}{|\boldsymbol{\Omega}_{uv}|^{1/2}} \exp\!\left(-\frac{1}{2} S_{2,\mathrm{occ}}(\hat{\boldsymbol{x}}_{uv})\right). \label{eq:S_L2o_lap}
    \end{align}
\end{enumerate}

When substituting $\mathcal{L}_2(\mathcal{Z}_{uv}^{(2)}|\mathcal{Z}_u^{(1)}, \emptyset) = \mathcal{L}_{2,\mathrm{occ}}(\mathcal{Z}_{uv}^{(2)}|\mathcal{Z}_u^{(1)})/\mathcal{L}_1(\emptyset)$ (\cite[Eqs.~(21), (24)]{MainPaper}) into~\cite[Eq.~(18)]{MainPaper} to obtain~\cite[Eq.~(25)]{MainPaper}, the denominator $\mathcal{L}_1(\emptyset)$ cancels against the empty-cluster factor $\mathcal{L}_1(\emptyset)$ in~\cite[Eq.~(20)]{MainPaper}. Two numerical regularizations are used in evaluation: under unknown blockage boundaries, $p_{\mathrm{U}}(\hat{\boldsymbol{x}}_v, \psi_v, \boldsymbol{s}_{1:K}) = \max\big(p_{\mathrm{U},\min}, \prod_{k=1}^K (1 - p_{\mathrm{D},1})\big)$ alleviates excessive missed-detection penalties on occluded candidates over long trajectories; for ill-conditioned landmark blocks with few scans or degenerate geometry, a weak prior $\boldsymbol{\Omega}_{\mathrm{env}} = \frac{2\pi}{A_{\mathrm{env}}}\boldsymbol{I}_2$ over surveillance area $A_{\mathrm{env}}$ is added to regularize local information matrices against near-singularity.

\subsection{Detailed Balance of Step~1 Sub-Moves and Local MCMC Weight Ratios}

\subsubsection{Why Step~1 Is Split Into Two Conditional Sub-Steps}
Consider two same-scan measurements $\boldsymbol{m} \in C_\beta$ and $\boldsymbol{m}' \in C_\gamma$. Combining single-point moves (fixing $\boldsymbol{m}' \in C_\gamma$) and the swap $\mathcal{A}' = \mathcal{A}_{\beta\leftrightarrow\gamma}(\boldsymbol{m})$ inside a single normalizing denominator $S(\mathcal{A})$ causes the forward and reverse move branches to condition on different placements ($\boldsymbol{m}' \in C_\gamma$ vs.\ $\boldsymbol{m}' \in C_\beta$), yielding $S(\mathcal{A}) \neq S(\mathcal{A}')$ and breaking detailed balance. Splitting Step~1 into sequential conditional Gibbs moves (\texttt{move} in~\cite[Eq.~(27)]{MainPaper} and \texttt{swap} in~\cite[Eq.~(28)]{MainPaper}) maintains shared forward and reverse candidate sets and normalizing constants within each sub-step.

\subsubsection{Local Weight Ratio Formulas (\cite[Eqs.~(25)--(29)]{MainPaper})}
Using the invariant-cancellation grouping $\Phi_u(C_u)$ in~\cite[Eq.~(26)]{MainPaper}:
\begin{itemize}
    \item \emph{Step~1 (\texttt{move}, \cite[Eq.~(27)]{MainPaper}):} Let $\mathcal{A}_{-\boldsymbol{m}}$ be the partition obtained after removing $\boldsymbol{m}$ (with visible 1-bounce index set $\mathcal{I}_{\mathrm{vis}}(\mathcal{A}_{-\boldsymbol{m}})$), and let $\Delta_+(\gamma; \boldsymbol{m}, \mathcal{A}_{-\boldsymbol{m}})$ denote the multiplicative weight increment when placing $\boldsymbol{m}$ into target $\gamma$ (with $C_\gamma' = C_\gamma \cup \{\boldsymbol{m}\}$). Then $\rho_{\beta\to\gamma}^{(\iota)}(\boldsymbol{m}) = \Delta_+(\gamma; \boldsymbol{m}, \mathcal{A}_{-\boldsymbol{m}}) / \Delta_+(\beta; \boldsymbol{m}, \mathcal{A}_{-\boldsymbol{m}})$, with
    \begin{align}
    \Delta_+(\gamma; \boldsymbol{m}, \mathcal{A}_{-\boldsymbol{m}}) =
    \begin{cases}
    \dfrac{\mathcal{L}_0(C_0')}{\mathcal{L}_0(C_0)}, & \tcond{into LoS cluster $0$,} \\[9pt]
    \dfrac{\Phi_u(C_u')}{\Phi_u(C_u)}, & \tcond{into existing visible 1-B cluster $u \in \mathcal{I}_{\mathrm{vis}}$,} \\[9pt]
    \dfrac{\mathcal{L}_2(C_{uv}'|C_u, C_v)}{\mathcal{L}_2(C_{uv}|C_u, C_v)}, & \tcond{into visible 2-B cluster $(u,v)$ ($u \neq v \in \mathcal{I}_{\mathrm{vis}}$),} \\[9pt]
    \dfrac{\mathcal{L}_{2,\mathrm{occ}}(C_{uv}'|C_u)}{\mathcal{L}_{2,\mathrm{occ}}(C_{uv}|C_u)} \text{ or } \dfrac{\mathcal{L}_{2,\mathrm{occ}}(C_{vu}'|C_u)}{\mathcal{L}_{2,\mathrm{occ}}(C_{vu}|C_u)}, & \tcond{into existing occluded 2-B cluster $(u,v)$ or $(v,u)$,} \\[10pt]
    \mathcal{L}_1(\{\boldsymbol{m}\}) \!\displaystyle\prod_{v \in \mathcal{I}_{\mathrm{vis}}(\mathcal{A}_{-\boldsymbol{m}})}\! \mathcal{L}_2(\emptyset|\{\boldsymbol{m}\}, C_v)\mathcal{L}_2(\emptyset|C_v, \{\boldsymbol{m}\}), & \tcond{new 1-B singleton $u_{\mathrm{new}}$,} \\[10pt]
    \mathcal{L}_{2,\mathrm{occ}}(\{\boldsymbol{m}\}|C_u), & \tcond{new occluded 2-B singleton $(u, v_{\mathrm{new}})$ or $(v_{\mathrm{new}}, u)$,}
    \end{cases} \label{eq:S_delta_move}
    \end{align}
    where the last branch evaluates the occluded 2-bounce factor matching the bounce order of $(u, v_{\mathrm{new}})$ or $(v_{\mathrm{new}}, u)$.
    \item \emph{Step~1 (\texttt{swap}, \cite[Eq.~(28)]{MainPaper}):} Denote non-visible-1-bounce cluster likelihoods uniformly by $g_\gamma(C_\gamma)$ ($\mathcal{L}_0$ for LoS, $\mathcal{L}_2$ for visible 2-bounce, $\mathcal{L}_{2,\mathrm{occ}}$ for occluded 2-bounce). Swapping $\boldsymbol{m} \in C_\beta$ with $\boldsymbol{m}' \in C_\gamma$ yields $C_\beta' = (C_\beta \setminus \{\boldsymbol{m}\}) \cup \{\boldsymbol{m}'\}$ and $C_\gamma' = (C_\gamma \setminus \{\boldsymbol{m}'\}) \cup \{\boldsymbol{m}\}$; whenever $\Phi_u(C_u')$ appears, it is evaluated holding all other clusters at their pre-swap states, giving five cases:
    \begin{align}
    \rho_{\beta\leftrightarrow\gamma}(\boldsymbol{m}) =
    \begin{cases}
    \dfrac{g_\beta(C_\beta')}{g_\beta(C_\beta)}\cdot\dfrac{g_\gamma(C_\gamma')}{g_\gamma(C_\gamma)}, & \tcond{neither cluster is 1-B visible,} \\[10pt]
    \dfrac{\Phi_u(C_u')}{\Phi_u(C_u)}\cdot\dfrac{g_\gamma(C_\gamma')}{g_\gamma(C_\gamma)}, & \tcond{$u \in \mathcal{I}_{\mathrm{vis}}$ and non-adjacent cluster $\gamma$,} \\[10pt]
    \dfrac{\Phi_u(C_u')}{\Phi_u(C_u)}\cdot\dfrac{\mathcal{L}_2(C_{uv}'|C_u', C_v)}{\mathcal{L}_2(C_{uv}|C_u', C_v)}, & \tcond{$u \in \mathcal{I}_{\mathrm{vis}}$ and adjacent visible 2-B $(u,v)$ (sym.\ $(v,u)$),} \\[10pt]
    \dfrac{\Phi_u(C_u')}{\Phi_u(C_u)}\cdot\dfrac{\mathcal{L}_{2,\mathrm{occ}}(C_{uv}'|C_u')}{\mathcal{L}_{2,\mathrm{occ}}(C_{uv}|C_u')}, & \tcond{$u \in \mathcal{I}_{\mathrm{vis}}$ and adjacent occluded 2-B $(u,v)$ (sym.\ $(v,u)$),} \\[10pt]
    \dfrac{\Phi_{u_1}(C_{u_1}')}{\Phi_{u_1}(C_{u_1})}\cdot\dfrac{\Phi_{u_2}(C_{u_2}')}{\Phi_{u_2}(C_{u_2})}\cdot \zeta_{u_1 u_2}, & \tcond{two visible 1-B clusters $u_1 \neq u_2$,}
    \end{cases} \label{eq:S_rho_swap}
    \end{align}
    where the cross-pair correction $\zeta_{u_1 u_2}$ removes the double counting between $\Phi_{u_1}$ and $\Phi_{u_2}$:
    \begin{align}
    \zeta_{u_1 u_2} \triangleq
    \frac{\mathcal{L}_2(C_{u_1 u_2}|C_{u_1}', C_{u_2}')\,\mathcal{L}_2(C_{u_2 u_1}|C_{u_2}', C_{u_1}')\,\mathcal{L}_2(C_{u_1 u_2}|C_{u_1}, C_{u_2})\,\mathcal{L}_2(C_{u_2 u_1}|C_{u_2}, C_{u_1})}{\mathcal{L}_2(C_{u_1 u_2}|C_{u_1}', C_{u_2})\,\mathcal{L}_2(C_{u_2 u_1}|C_{u_2}, C_{u_1}')\,\mathcal{L}_2(C_{u_1 u_2}|C_{u_1}, C_{u_2}')\,\mathcal{L}_2(C_{u_2 u_1}|C_{u_2}', C_{u_1})}. \label{eq:S_zeta_swap}
    \end{align}
    \item \emph{Step~2 (\texttt{relabel}, \cite[Eq.~(29)]{MainPaper}):} Keeping measurements inside $C_{(l)}$ fixed, the relative weight $\tilde{w}(\gamma) = \hat{w}(\mathcal{A}_\gamma) / \hat{w}(\mathcal{A}_{-l})$ across $\gamma \in \mathcal{K}_l$ is
    \begin{align}
    \tilde{w}(\gamma) =
    \begin{cases}
    \dfrac{\mathcal{L}_0(C_{(l)})}{\mathcal{L}_0(\emptyset)}, & \tcond{$\gamma = \text{LoS cluster } 0$ ($\mathcal{Z}_0^{(-l)} = \emptyset$),} \\[9pt]
    \Phi_{u_{\mathrm{new}}}(C_{(l)}), & \tcond{$\gamma = \text{standalone 1-B cluster } u_{\mathrm{new}}$,} \\[9pt]
    \dfrac{\mathcal{L}_2(C_{(l)}|\mathcal{Z}_u^{(1)}, \mathcal{Z}_v^{(1)})}{\mathcal{L}_2(\emptyset|\mathcal{Z}_u^{(1)}, \mathcal{Z}_v^{(1)})}, & \tcond{$\gamma = \text{visible 2-B cluster } (u,v)$ ($u\neq v \in \mathcal{I}_{\mathrm{vis}}$, $\mathcal{Z}_{uv}^{(2)(-l)} = \emptyset$),} \\[9pt]
    \mathcal{L}_{2,\mathrm{occ}}(C_{(l)}|\mathcal{Z}_u^{(1)}), & \tcond{$\gamma = \text{occluded 2-B cluster } (u, v_{\mathrm{new}})$ or $(v_{\mathrm{new}}, u)$ ($u \in \mathcal{I}_{\mathrm{vis}}$).}
    \end{cases} \label{eq:S_w_relabel}
    \end{align}
    \item \emph{Step~3 (burn-in merge-and-split MH, $n \le N_{\mathrm{burn}}$):} Following~\cite{Ge2025TSP}, proposals are accepted with simplified burn-in probability $\min[1, \allowbreak \hat{w}(\mathcal{A}')/\allowbreak \hat{w}(\mathcal{A})]$ in~\cite[Sec.~IV-B]{MainPaper}:
    \begin{enumerate}
        \item \emph{General scan-disjoint merge/split:} Merging two scan-disjoint clusters $C_\beta, C_\gamma$ into $C_{\beta\cup\gamma} = C_\beta \cup C_\gamma$ yields
        \begin{align}
        \frac{\hat{w}(\mathcal{A}_{\mathrm{merge}})}{\hat{w}(\mathcal{A})} &= \frac{\Phi_{\beta\cup\gamma}(C_{\beta\cup\gamma})\,\mathcal{L}_2(\emptyset|C_\beta, C_\gamma)\,\mathcal{L}_2(\emptyset|C_\gamma, C_\beta)}{\Phi_\beta(C_\beta)\,\Phi_\gamma(C_\gamma)} \quad \text{(two 1-B clusters)}, \notag\\
        \frac{\hat{w}(\mathcal{A}_{\mathrm{merge}})}{\hat{w}(\mathcal{A})} &= \frac{g_{\beta\cup\gamma}(C_{\beta\cup\gamma})}{g_\beta(C_\beta)\,g_\gamma(C_\gamma)} \quad \text{(2-B/LoS clusters)}, \label{eq:S_merge_gen}
        \end{align}
        paired with a reverse \texttt{k-means++} bisection split operator on multi-scan clusters $|C| \ge 2$.
        \item \emph{Occluded 2-bounce pair merge/split:} Merging two scan-disjoint 1-bounce clusters $C_b, C_g$ without active children into $(u, v_{\mathrm{new}}) = C_b \cup C_g$ (or $(v_{\mathrm{new}}, u)$) anchored to $u \in \mathcal{I}_{\mathrm{vis}}$ yields
        \begin{align}
        \frac{\hat{w}(\mathcal{A}_{\mathrm{b\text{-}merge}})}{\hat{w}(\mathcal{A})} = \frac{\mathcal{L}_{2,\mathrm{occ}}(C_b \cup C_g|\mathcal{Z}_u^{(1)})\,\mathcal{L}_2(\emptyset|C_b, C_g)\,\mathcal{L}_2(\emptyset|C_g, C_b)}{\Phi_b(C_b)\,\Phi_g(C_g)}. \label{eq:S_merge_occ}
        \end{align}
    \end{enumerate}
\end{itemize}

\subsection{Warm-Start 2-Bounce Cluster Matching and Geometric Inversion}

In Algorithm~\ref{alg:S_warmstart}, Steps~1--3 use the odometry initial poses $\boldsymbol{s}_k^{(0)} = [(\boldsymbol{p}_k^{(0)})^{\mathsf{T}}, \alpha_k^{(0)}]^{\mathsf{T}}$ to extract the LoS cluster $\mathcal{Z}_0$ and visible 1-bounce clusters (with estimated landmark positions and most likely types $\{(\hat{\boldsymbol{x}}_u, \hat{\psi}_u)\}$), placing remaining measurements in the 2-bounce candidate pool $\mathcal{P}_{\mathrm{DB}}$. For a scan-$k$ measurement $\boldsymbol{z}_m = [d, \theta, \phi]^{\mathsf{T}} \in \mathcal{P}_{\mathrm{DB}}$, define unit arrival and departure vectors $\boldsymbol{e}_{\mathrm{RX}} \triangleq [\cos(\alpha_k^{(0)}+\theta), \sin(\alpha_k^{(0)}+\theta)]^{\mathsf{T}}$ and $\boldsymbol{e}_{\mathrm{TX}} \triangleq [\cos(\alpha_{\mathrm{BS}}+\phi), \sin(\alpha_{\mathrm{BS}}+\phi)]^{\mathsf{T}}$. Steps~4 and~5 extract visible and occluded 2-bounce clusters sequentially as follows:
\begin{enumerate}
    \item \emph{Step~4---Visible 2-bounce cluster matching (both endpoints known):} Iterate over ordered visible pairs satisfying $(\hat{\psi}_u, \hat{\psi}_v) \in \mathcal{T}_2$ ($u \neq v$). For each scan, measurements $\boldsymbol{z}_m \in \mathcal{P}_{\mathrm{DB}}$ satisfying valid reflection geometry are tested via the predicted Mahalanobis distance:
    \begin{align}
    d_{2,uv}^2(m) \triangleq \big(\boldsymbol{z}_m - \boldsymbol{h}_{\hat{\psi}_u,\hat{\psi}_v}(\hat{\boldsymbol{x}}_u, \hat{\boldsymbol{x}}_v, \boldsymbol{s}_k^{(0)})\big)^{\mathsf{T}}\boldsymbol{R}_{\hat{\psi}_u,\hat{\psi}_v}^{-1}\big(\boldsymbol{z}_m - \boldsymbol{h}_{\hat{\psi}_u,\hat{\psi}_v}(\hat{\boldsymbol{x}}_u, \hat{\boldsymbol{x}}_v, \boldsymbol{s}_k^{(0)})\big) < \gamma_{\mathrm{reproj}}, \label{eq:S_ws_vis}
    \end{align}
    retaining at most the minimum-distance match per scan; candidates spanning at least $2$ scans and passing the likelihood check in~\cite[Eq.~(23)]{MainPaper} are initialized as visible 2-bounce clusters $\mathcal{Z}_{uv}^{(2)}$ and removed from $\mathcal{P}_{\mathrm{DB}}$.

    \item \emph{Step~5---Occluded 2-bounce inversion and clustering (one endpoint known, one unknown):} For each visible landmark $(\hat{\boldsymbol{x}}_u, \hat{\psi}_u)$, iterate over compatible bounce orders ($u$ as 1st or 2nd bounce) and unknown endpoint types $\psi_v \in \mathcal{T}_1$ forming $\eta \in \mathcal{T}_2$, processing remaining measurements in $\mathcal{P}_{\mathrm{DB}}$ in three sub-steps:
    \begin{itemize}
        \item \emph{Step~5a (Single-measurement closed-form inversion):} Define two elementary geometric operators---(i) for symmetric points $\boldsymbol{a} \neq \boldsymbol{b}$, the BS virtual anchor $\mathcal{V}(\boldsymbol{a}, \boldsymbol{b})$ across the perpendicular bisector of segment $\boldsymbol{a}\boldsymbol{b}$ (with unit normal $\boldsymbol{n} = (\boldsymbol{b}-\boldsymbol{a})/\|\boldsymbol{b}-\boldsymbol{a}\|$); and (ii) for a path departing $\boldsymbol{p}_0$ along unit ray $\boldsymbol{e}$ and folding back to $\boldsymbol{q}$ with total length $d$, the scatterer intersection $\mathcal{R}(\boldsymbol{p}_0, \boldsymbol{e}, \boldsymbol{q}, d)$ (matching~\cite[Eq.~(33)]{MainPaper}, with $\boldsymbol{b} = \boldsymbol{p}_0 - \boldsymbol{q}$):
        \begin{align}
        \mathcal{V}(\boldsymbol{a}, \boldsymbol{b}) \triangleq \boldsymbol{p}_{\mathrm{BS}} + 2\Big(\big(\tfrac{\boldsymbol{a}+\boldsymbol{b}}{2} - \boldsymbol{p}_{\mathrm{BS}}\big)^{\mathsf{T}}\boldsymbol{n}\Big)\boldsymbol{n}, \qquad
        \mathcal{R}(\boldsymbol{p}_0, \boldsymbol{e}, \boldsymbol{q}, d) \triangleq \boldsymbol{p}_0 + \frac{d^2 - \|\boldsymbol{b}\|^2}{2(d + \boldsymbol{b}^{\mathsf{T}}\boldsymbol{e})}\,\boldsymbol{e}. \label{eq:S_ws_ops}
        \end{align}
        Under positive segment lengths and non-degenerate denominators, the six bounce-order and type configurations give candidate coordinates $\hat{\boldsymbol{x}}_v \in \mathbb{R}^2$ via~\eqref{eq:S_ws_ops} and $\mathcal{M}$, as listed in Table~\ref{tab:S_inv}.
        \item \emph{Step~5b (Reprojection gating):} Because inverting 2D $\hat{\boldsymbol{x}}_v$ from 3D $\boldsymbol{z}_m$ leaves $1$ redundant geometric degree of freedom, substitute $\hat{\boldsymbol{x}}_v$ back into the observation model to test the reprojection Mahalanobis distance (swapping $\hat{\boldsymbol{x}}_u, \hat{\boldsymbol{x}}_v$ when $v$ is the 1st bounce):
        \begin{align}
        d_{\mathrm{reproj}}^2(m) \triangleq \big(\boldsymbol{z}_m - \boldsymbol{h}_\eta(\hat{\boldsymbol{x}}_u, \hat{\boldsymbol{x}}_v, \boldsymbol{s}_k^{(0)})\big)^{\mathsf{T}}\boldsymbol{R}_\eta^{-1}\big(\boldsymbol{z}_m - \boldsymbol{h}_\eta(\hat{\boldsymbol{x}}_u, \hat{\boldsymbol{x}}_v, \boldsymbol{s}_k^{(0)})\big) < \gamma_{\mathrm{reproj}}, \label{eq:S_ws_reproj}
        \end{align}
        filtering out inconsistent candidates that violate~\eqref{eq:S_ws_reproj}.
        \item \emph{Step~5c (Spatial clustering and validation):} Group gated points $\hat{\boldsymbol{x}}_v$ within radius $r_{\mathrm{clust}}$ (at most one measurement per scan); clusters validated by $\mathcal{L}_{2,\mathrm{occ}}$ are initialized as occluded 2-bounce clusters and removed from $\mathcal{P}_{\mathrm{DB}}$.
    \end{itemize}
\end{enumerate}

\begin{table}[!t]
\centering
\caption{Closed-Form Inversion Formulas for Occluded Landmark $\hat{\boldsymbol{x}}_v$ Across the Six Configurations in Algorithm~\ref{alg:S_warmstart}, Step~5}
\label{tab:S_inv}
\footnotesize
\setlength{\tabcolsep}{7pt}
\begin{tabular}{@{}ccccc@{}}
\toprule
Path Type $\eta$ & Known Bounce $u$ & Unknown Bounce $v$ & Auxiliary Point & Closed-Form Estimate $\hat{\boldsymbol{x}}_v$ \\
\midrule
$(\mathrm{VA}, \mathrm{VA})$ & 1st ($\mathrm{VA}$) & 2nd ($\mathrm{VA}$) & 2nd-order VA $\hat{\boldsymbol{x}}^{(2)} = \boldsymbol{p}_k^{(0)} + d\boldsymbol{e}_{\mathrm{RX}}$ & $\mathcal{V}(\hat{\boldsymbol{x}}_u, \hat{\boldsymbol{x}}^{(2)})$ \\[2pt]
$(\mathrm{VA}, \mathrm{VA})$ & 2nd ($\mathrm{VA}$) & 1st ($\mathrm{VA}$) & 2nd-order VA $\hat{\boldsymbol{x}}^{(2)} = \boldsymbol{p}_k^{(0)} + d\boldsymbol{e}_{\mathrm{RX}}$ & $\mathcal{M}(\hat{\boldsymbol{x}}^{(2)}; \hat{\boldsymbol{x}}_u)$ \\[2pt]
$(\mathrm{VA}, \mathrm{RP})$ & 1st ($\mathrm{VA}$) & 2nd ($\mathrm{RP}$) & --- & $\mathcal{R}(\boldsymbol{p}_k^{(0)}, \boldsymbol{e}_{\mathrm{RX}}, \hat{\boldsymbol{x}}_u, d)$ \\[2pt]
$(\mathrm{VA}, \mathrm{RP})$ & 2nd ($\mathrm{RP}$) & 1st ($\mathrm{VA}$) & Unfolded point $\tilde{\boldsymbol{x}} = \boldsymbol{p}_{\mathrm{BS}} + (d - \|\hat{\boldsymbol{x}}_u - \boldsymbol{p}_k^{(0)}\|)\boldsymbol{e}_{\mathrm{TX}}$ & $\mathcal{V}(\hat{\boldsymbol{x}}_u, \tilde{\boldsymbol{x}})$ \\[2pt]
$(\mathrm{RP}, \mathrm{VA})$ & 1st ($\mathrm{RP}$) & 2nd ($\mathrm{VA}$) & Unfolded point $\tilde{\boldsymbol{x}} = \boldsymbol{p}_k^{(0)} + (d - \|\hat{\boldsymbol{x}}_u - \boldsymbol{p}_{\mathrm{BS}}\|)\boldsymbol{e}_{\mathrm{RX}}$ & $\mathcal{V}(\hat{\boldsymbol{x}}_u, \tilde{\boldsymbol{x}})$ \\[2pt]
$(\mathrm{RP}, \mathrm{VA})$ & 2nd ($\mathrm{VA}$) & 1st ($\mathrm{RP}$) & Mirrored RX $\boldsymbol{p}_k' = \mathcal{M}(\boldsymbol{p}_k^{(0)}; \hat{\boldsymbol{x}}_u)$ & $\mathcal{R}(\boldsymbol{p}_{\mathrm{BS}}, \boldsymbol{e}_{\mathrm{TX}}, \boldsymbol{p}_k', d)$ \\
\bottomrule
\end{tabular}
\end{table}

\section{Geometric Identifiability, Sampler Convergence, Fast Type/Information Evaluation, and Cram\'er-Rao Lower Bounds} \label{app:D}

This appendix corresponds to Sections~III-A, IV-B, IV-C, V-A, and~VI of~\cite{MainPaper}, presenting the noiseless geometric identifiability analysis, convergence justification of Theorem~2, occluded-landmark type elimination and block-diagonal information structure, and formulations of the 1-b Batch, 2-b Filtering, and 2-b Batch Cram\'er-Rao lower bounds (CRLBs).

\subsection{Geometric Identifiability of 2-Bounce Landmark Pairs (\cite[Sec.~III-A, Constraint~(iii)]{MainPaper})}

\subsubsection{Unidentifiability When Both Endpoints Lack 1-Bounce Observations}
\begin{itemize}
    \item \emph{Wall--wall paths $(\mathrm{VA}, \mathrm{VA})$:} Let $\beta_u = \angle\boldsymbol{d}_u$, $\beta_v = \angle\boldsymbol{d}_v$, and $\gamma_k(\boldsymbol{x}_{uv}^{(2)}) \triangleq \angle(\boldsymbol{x}_{uv}^{(2)} - \boldsymbol{p}_k)$. Combining Table~\ref{tab:S_paths} with~\eqref{eq:S_angle_cong} writes the scan-$k$ observation vector as
    \begin{align}
    \boldsymbol{h}_{\mathrm{VA},\mathrm{VA}}(\boldsymbol{x}_u, \boldsymbol{x}_v, \boldsymbol{s}_k) =
    \begin{bmatrix}
    \|\boldsymbol{x}_{uv}^{(2)} - \boldsymbol{p}_k\| \\[2pt]
    \mathrm{wrap}\!\big(\gamma_k(\boldsymbol{x}_{uv}^{(2)}) - \alpha_k\big) \\[2pt]
    \mathrm{wrap}\!\big(\vartheta_{uv} + \gamma_k(\boldsymbol{x}_{uv}^{(2)}) - \alpha_{\mathrm{BS}}\big)
    \end{bmatrix}, \qquad \vartheta_{uv} \triangleq \mathrm{wrap}\!\big(\pi + 2(\beta_u - \beta_v)\big). \label{eq:S_id_vava}
    \end{align}
    Equation~\eqref{eq:S_id_vava} shows that $(\boldsymbol{x}_u, \boldsymbol{x}_v) \in \mathbb{R}^4$ enters all scans $k$ solely through the scan-invariant 3D vector $\boldsymbol{\zeta}(\boldsymbol{x}_u, \boldsymbol{x}_v) \triangleq [(\boldsymbol{x}_{uv}^{(2)})^{\mathsf{T}}, \vartheta_{uv}]^{\mathsf{T}} \in \mathbb{R}^3$. Stacking $M_{\mathrm{obs}} \ge 1$ scans yields
    \begin{align}
    \frac{\partial \boldsymbol{h}_{(\mathrm{VA},\mathrm{VA}),1:M_{\mathrm{obs}}}}{\partial (\boldsymbol{x}_u, \boldsymbol{x}_v)}
    = \underbrace{\frac{\partial \boldsymbol{h}_{(\mathrm{VA},\mathrm{VA}),1:M_{\mathrm{obs}}}}{\partial \boldsymbol{\zeta}}}_{3M_{\mathrm{obs}}\times 3} \cdot \underbrace{\frac{\partial \boldsymbol{\zeta}(\boldsymbol{x}_u, \boldsymbol{x}_v)}{\partial (\boldsymbol{x}_u, \boldsymbol{x}_v)}}_{3\times 4}, \label{eq:S_id_rank}
    \end{align}
    whose rank is at most $3 < 4$ for $M_{\mathrm{obs}} \ge 1$, so $(\boldsymbol{x}_u, \boldsymbol{x}_v)$ cannot be recovered from $(\mathrm{VA}, \mathrm{VA})$ measurements alone.
    \item \emph{Hybrid paths $(\mathrm{VA}, \mathrm{RP})$ vs.\ $(\mathrm{RP}, \mathrm{VA})$:} Given a $(\mathrm{VA}, \mathrm{RP})$ pair $(\boldsymbol{x}_u, \boldsymbol{x}_v)$, set $\tilde{\boldsymbol{x}}_v \triangleq \mathcal{M}(\boldsymbol{x}_v; \boldsymbol{x}_u)$. By orthogonality and involution of $\mathcal{M}(\cdot; \boldsymbol{x}_u)$, $\|\tilde{\boldsymbol{x}}_v - \boldsymbol{p}_{\mathrm{BS}}\| = \|\boldsymbol{H}(\boldsymbol{d}_u)(\boldsymbol{x}_v - \boldsymbol{x}_u)\| = \|\boldsymbol{x}_v - \boldsymbol{x}_u\|$ and $\mathcal{M}(\tilde{\boldsymbol{x}}_v; \boldsymbol{x}_u) = \boldsymbol{x}_v$. Substituting $(\tilde{\boldsymbol{x}}_v, \boldsymbol{x}_u)$ into the $(\mathrm{RP}, \mathrm{VA})$ row of Table~\ref{tab:S_paths} produces identical $(h_d, \boldsymbol{d}_{\mathrm{RX}}, \boldsymbol{d}_{\mathrm{TX}})$ across RX poses $\boldsymbol{s}_k$, causing bounce-order and mirror ambiguity.
\end{itemize}

\subsubsection{Identifiability When At Least One Endpoint Has 1-Bounce Observations}
Under noiseless observations with $(\boldsymbol{x}_u, \psi_u)$ known from 1-bounce paths, the remaining unknowns are $\boldsymbol{x}_v \in \mathbb{R}^2$ and $\psi_v \in \mathcal{T}_1$. For a given $\psi_v$, Appendix~\ref{app:C}-D determines $\boldsymbol{x}_v$ under non-degenerate geometry.

When $\psi_v$ is unknown: if $\psi_u = \mathrm{RP}$, restricting 2-bounce types to $\mathcal{T}_2$ (which omits $(\mathrm{RP}, \mathrm{RP})$) gives $\psi_v = \mathrm{VA}$; if $\psi_u = \mathrm{VA}$, a planar wall shifts its reflection point across RX positions whereas a point scatterer remains fixed, allowing $(\boldsymbol{x}_v, \psi_v)$ to be distinguished from two or more non-collinear scans under generic non-degenerate geometry. When the bounce order $(u, v)$ versus $(v, u)$ is also unknown, swapping the first and second bounces alters the correspondence between BS departure and RX arrival angles and is uniquely resolved across two or more non-collinear scans except in the orthogonal-wall case $\beta_u - \beta_v \equiv \pm\frac{\pi}{2}\pmod\pi$ ($90^\circ$ corner). In that orthogonal case, commuting Householder reflections yield $\boldsymbol{x}_{uv}^{(2)} = \boldsymbol{x}_{vu}^{(2)} = \boldsymbol{x}_u + \boldsymbol{x}_v - \boldsymbol{p}_{\mathrm{BS}}$ and $\vartheta_{uv} = \vartheta_{vu} = 0$ in~\eqref{eq:S_id_vava}, so $\boldsymbol{h}_{\mathrm{VA},\mathrm{VA}}(\boldsymbol{x}_u, \boldsymbol{x}_v, \boldsymbol{s}_k) \equiv \boldsymbol{h}_{\mathrm{VA},\mathrm{VA}}(\boldsymbol{x}_v, \boldsymbol{x}_u, \boldsymbol{s}_k)$ across all RX poses and the bounce order itself is indistinguishable under infinite-line VAs; nevertheless, both $(\mathrm{VA}, \mathrm{VA})$ rows of Table~\ref{tab:S_inv} still recover the identical coordinate $\mathcal{V}(\boldsymbol{x}_u, \boldsymbol{x}_{uv}^{(2)}) = \mathcal{M}(\boldsymbol{x}_{uv}^{(2)}; \boldsymbol{x}_u) = \boldsymbol{x}_v$. Thus $(\boldsymbol{x}_v, \mathrm{VA})$ remains uniquely identifiable even for $90^\circ$ corners (samples assigned to $(u, v)$ or $(v, u)$ yield the same VA estimate and are merged in~\cite[Sec.~V-B]{MainPaper}). \hfill$\blacksquare$

\subsection{Justification of Theorem~2 (Asymptotic Convergence of the Inner DA Chain)}

For a fixed trajectory $\boldsymbol{s}_{0:K}$, finiteness of total measurements implies that $\mathbb{J}$ is finite; under $c(\boldsymbol{z}) > 0$, $\lambda_\psi(\boldsymbol{x}) > 0$, and $p_{\mathrm{D},b} \in (0, 1)$, the approximated weight satisfies $\hat{w}(\mathcal{A}) > 0$ for each $\mathcal{A} \in \mathbb{J}$~\cite[Theorem~2]{MainPaper}. Let $\pi(\mathcal{A}) \equiv \hat{p}(\mathcal{A}|\boldsymbol{s}_{0:K}, \mathcal{Z}_{1:K}) \triangleq \hat{w}(\mathcal{A}) / \sum_{\mathcal{A}'\in\mathbb{J}} \hat{w}(\mathcal{A}')$ denote the conditional target distribution and $\mathcal{K} = \mathcal{K}_{\mathrm{relabel}}\mathcal{K}_{\mathrm{swap}}\mathcal{K}_{\mathrm{move}}$ the post-burn-in transition kernel.
\begin{enumerate}
    \item \emph{Stationarity:} In \texttt{move} (\cite[Eq.~(27)]{MainPaper}), singleton anchors with non-empty 2-bounce children permit only operation~(i) (self-transition with probability $1$); otherwise removing $\boldsymbol{m}$ leaves the same remainder $\mathcal{A}_{-\boldsymbol{m}}$, so $\mathcal{M}_{\mathrm{move}}(\boldsymbol{m}; \mathcal{A}) = \mathcal{M}_{\mathrm{move}}(\boldsymbol{m}; \mathcal{A}')$ and $\pi(\mathcal{A})P_{\mathrm{move}}(\mathcal{A}\to\mathcal{A}') = \pi(\mathcal{A}')P_{\mathrm{move}}(\mathcal{A}'\to\mathcal{A})$. In \texttt{swap} (\cite[Eq.~(28)]{MainPaper}), the symmetric denominator $\hat{w}(\mathcal{A}) + \hat{w}(\mathcal{A}')$ satisfies detailed balance. In \texttt{relabel} (\cite[Eq.~(29)]{MainPaper}), fixing $\mathcal{A}_{-l}$ yields identical candidate label sets $\mathcal{K}_l$ and normalizing sums in both directions. Hence $\mathcal{K}$ admits $\pi$ as its stationary distribution.
    \item \emph{Irreducibility:} Let $\mathcal{A}_0 \in \mathbb{J}$ denote the singleton partition where each measurement forms an independent 1-bounce cluster. From a given $\mathcal{A} \in \mathbb{J}$, applying \texttt{move} empties non-empty 2-bounce clusters (releasing active 2-bounce anchor constraints) and splits remaining LoS and multi-scan 1-bounce clusters into singletons, reaching $\mathcal{A}_0$ in finitely many steps with positive probability. Reversing the sequence (forming $C_0'$, 1-bounce clusters, visible 2-bounce clusters, and occluded 2-bounce clusters) reaches a target $\mathcal{A}' \in \mathbb{J}$ with positive probability.
    \item \emph{Aperiodicity:} At each $\mathcal{A} \in \mathbb{J}$, each local Gibbs step retains the current state with positive probability, so $\mathcal{K}(\mathcal{A}\to\mathcal{A}) > 0$.
\end{enumerate}
By the finite-state Markov chain ergodic theorem~\cite{Xia2024TAES,Roberts1994SPA}, $\mathcal{K}^t(\mathcal{A}^{(0)}\to\cdot)$ converges asymptotically in total variation to $\pi(\cdot)$ as $t \to \infty$ from $\mathcal{A}^{(0)} \in \mathbb{J}$. Note that this property characterizes the inner DA Markov chain conditioned on $\boldsymbol{s}_{0:K}$ under the approximated weight $\pi(\mathcal{A})$; running a finite sweep count $T_{\mathrm{sweep}}$ in alternation with conditional MAP GraphSLAM constitutes a tractable approximate inference scheme. \hfill$\blacksquare$

\subsection{Occluded Type Elimination, Tree DP, and Block-Diagonal Landmark Information Matrix}

\subsubsection{Occluded Type Elimination and $O(n_m)$ Tree DP (\cite[Sec.~IV-C]{MainPaper})}
Within each coupled subset $\mathcal{G}_m$, the participating landmarks belong to non-empty 2-bounce clusters and satisfy $\psi \in \mathcal{T}_1$. Since each $v \in \mathcal{G}_m \cap \mathcal{I}_{\mathrm{occ}}$ connects to a single visible landmark $u \in \mathcal{G}_m^{\mathrm{vis}} \triangleq \mathcal{G}_m \cap \mathcal{I}_{\mathrm{vis}}$, collect the occluded set anchored at $u$ as $\mathcal{B}_u \subseteq \mathcal{I}_{\mathrm{occ}}$ and let $\mathcal{L}_{2,\mathrm{occ}}(v; \psi_u, \psi_v)$ denote~\eqref{eq:S_L2o_lap} for the active bounce order $(u, v)$ or $(v, u)$, where the first and second arguments always denote the anchor type $\psi_u$ and occluded type $\psi_v$, respectively (vanishing if $\psi_u = \psi_v = \mathrm{RP}$). Summing out $\mathcal{B}_u$ analytically in $O(N_{\mathrm{occ}})$ total time yields the collapsed vertex weights
\begin{align}
\bar{\mathcal{L}}_1(u; \psi_u)
&\triangleq \mathcal{L}_1(\mathcal{Z}_u^{(1)}; \psi_u) \prod_{v \in \mathcal{B}_u} \sum_{\psi_v \in \mathcal{T}_1} \mathcal{L}_{2,\mathrm{occ}}(v; \psi_u, \psi_v) \notag\\
&= \begin{cases}
\mathcal{L}_1(\mathcal{Z}_u^{(1)}; \mathrm{VA}) \displaystyle\prod_{v \in \mathcal{B}_u} \Big(\mathcal{L}_{2,\mathrm{occ}}(v; \mathrm{VA}, \mathrm{VA}) + \mathcal{L}_{2,\mathrm{occ}}(v; \mathrm{VA}, \mathrm{RP})\Big), & \psi_u = \mathrm{VA}, \\[10pt]
\mathcal{L}_1(\mathcal{Z}_u^{(1)}; \mathrm{RP}) \displaystyle\prod_{v \in \mathcal{B}_u} \mathcal{L}_{2,\mathrm{occ}}(v; \mathrm{RP}, \mathrm{VA}), & \psi_u = \mathrm{RP}.
\end{cases} \label{eq:S_occ_elim}
\end{align}

After sampling $\boldsymbol{\psi}_{\mathcal{G}_m^{\mathrm{vis}}}$ from the reduced $n_m = |\mathcal{G}_m^{\mathrm{vis}}|$-node visible core, each occluded type is drawn independently via $P(\psi_v|\psi_u, \mathcal{A}, \boldsymbol{s}_{0:K}, \mathcal{Z}_{1:K}) = \mathcal{L}_{2,\mathrm{occ}}(v; \psi_u, \psi_v) / \sum_{\psi_v'\in\mathcal{T}_1} \mathcal{L}_{2,\mathrm{occ}}(v; \psi_u, \psi_v')$ (with $\psi_v = \mathrm{VA}$ when $\psi_u = \mathrm{RP}$). When $\mathcal{G}_m^{\mathrm{vis}}$ forms an acyclic tree rooted at $r_{\mathrm{root}}$ with edge factors $\xi_{uc}(\psi_u, \psi_c)$ ($\xi_{uc}(\mathrm{RP}, \mathrm{RP}) = 0$), backward dynamic programming evaluates subtree partition sums from leaves to root in $O(n_m)$ time:
\begin{align}
Z_u(\psi_u) = \bar{\mathcal{L}}_1(u; \psi_u) \prod_{c \in \mathrm{children}(u)} \Big(\xi_{uc}(\psi_u, \mathrm{VA})\,Z_c(\mathrm{VA}) + \xi_{uc}(\psi_u, \mathrm{RP})\,Z_c(\mathrm{RP})\Big), \qquad \psi_u \in \mathcal{T}_1, \label{eq:S_tree_dp}
\end{align}
followed by top-down sampling of $\psi_{r_{\mathrm{root}}}$ via $Z_{r_{\mathrm{root}}}(\psi)$ and each $\psi_c$ via $\xi_{uc}(\psi_u, \psi_c)Z_c(\psi_c)$.

For non-tree subgraphs, unicyclic components reduce to two tree passes in $O(n_m)$ time by conditioning on one loop vertex, and cliques permit $n_m + 1$ valid assignments (at most one $\mathrm{RP}$). General cyclic partition-function summation is $\texttt{\#P}$-complete~\cite{Dyer2000RSA}; in the indoor scenarios evaluated in~\cite{MainPaper}, double-bounce path loss and geometric occlusion keep coupled visible components small ($n_m < 10$), allowing direct enumeration, whereas larger cyclic components in denser scenes can be approximated via local Gibbs sampling.

\subsubsection{Block-Diagonal Structure of $\boldsymbol{\Omega}_{\mathrm{xx}}$ (\cite[Sec.~V-A]{MainPaper})}
In the conditional GraphSLAM LS cost~\cite[Eq.~(36)]{MainPaper}, landmark priors and 1-bounce residuals depend only on a single landmark $\boldsymbol{x}_i$, contributing solely to the $2\times 2$ diagonal blocks of $\boldsymbol{\Omega}_{\mathrm{xx}}$. Only non-empty 2-bounce factors $(u, v) \in \mathcal{E}_2$ produce off-diagonal $2\times 2$ blocks $\sum_k \boldsymbol{H}_{2,uv,k,u}^{\mathsf{T}}\boldsymbol{R}_{\psi_u,\psi_v}^{-1}\boldsymbol{H}_{2,uv,k,v}$. Because distinct 2-bounce coupled subsets $\mathcal{G}_{m_1}^+ \neq \mathcal{G}_{m_2}^+$ share no pairs in $\mathcal{E}_2$, all off-diagonal blocks between distinct subsets vanish, rendering
\begin{align}
\boldsymbol{\Omega}_{\mathrm{xx}} = \operatorname{blkdiag}\!\left(\boldsymbol{\Omega}_{\mathrm{xx}}^{(1)},\, \dots,\, \boldsymbol{\Omega}_{\mathrm{xx}}^{(M_+)}\right), \label{eq:S_blkdiag}
\end{align}
when confirmed landmarks are ordered by $\mathcal{G}_1^+, \dots, \mathcal{G}_{M_+}^+$. Thus inverting $\boldsymbol{\Omega}_{\mathrm{xx}}$ in the Schur complement scales as $O\big(\sum_{m=1}^{M_+} |\mathcal{G}_m^+|^3\big)$ rather than $O(\kappa^3)$. \hfill$\blacksquare$

\subsection{Cram\'er-Rao Lower Bounds (1-b Batch, 2-b Filtering, and 2-b Batch)}

This subsection corresponds to Section~VI of~\cite{MainPaper} and derives the expressions and information-monotonicity properties of the \emph{2-b Batch CRLB}, \emph{1-b Batch CRLB}, and \emph{2-b Filtering CRLB}. In the Monte Carlo (MC) ray-tracing experiments of~\cite[Sec.~VI]{MainPaper}, the sensor trajectory $\boldsymbol{s}_{0:K}$ and landmarks $\boldsymbol{x}_{1:n_{\mathrm{obs}}}$ ($n_{\mathrm{obs}} = n_{\mathrm{vis}} + n_{\mathrm{occ}}$) are deterministic unknown parameters, whereas the initial-pose error, body-frame odometry increments, and multipath channel measurements are subject to independent zero-mean Gaussian noise.

Under clutter-free measurements with known data association and known landmark types~\cite{Kay1993,Ge2025TSP,Deutschmann2025Asilomar}, order the $n_{\mathrm{vis}}$ 1-bounce-visible landmarks $\boldsymbol{x}_{1:n_{\mathrm{vis}}}$ ahead of the $n_{\mathrm{occ}}$ 1-bounce-occluded landmarks $\boldsymbol{x}_{n_{\mathrm{vis}}+1:n_{\mathrm{obs}}}$. For a prefix horizon $1{:}t$ ($t \in \{1, \dots, K\}$), stack the joint parameter vector as
\begin{align}
\boldsymbol{q}_{1:t} \triangleq \big[\boldsymbol{s}_{0:t}^{\mathsf{T}},\, \boldsymbol{x}_1^{\mathsf{T}},\, \dots,\, \boldsymbol{x}_{n_{\mathrm{obs}}}^{\mathsf{T}}\big]^{\mathsf{T}} \in \mathbb{R}^{3(t+1) + 2n_{\mathrm{obs}}}, \label{eq:S_crlb_state}
\end{align}
recovering the full-batch parameter vector at $t = K$. When the initial pose prior is specified at the first measurement scan $k = 1$ (as in~\cite[Sec.~VI]{MainPaper}), the trajectory subvector starts at $\boldsymbol{s}_1$ with covariance $\boldsymbol{P}_{1,\mathrm{prior}}$ and transitions over $k = 2, \dots, t$, yielding the same block structure below.

\subsubsection{Explicit Fisher Information Matrix (FIM) Construction}
Over prefix $1{:}t$, the full double-bounce joint-SLAM FIM evaluated at the true parameter vector $\boldsymbol{q}_{1:t}$ is the sum of four symmetric positive-semidefinite (PSD) matrices:
\begin{align}
\boldsymbol{J}_{1:t} = \boldsymbol{J}_{\mathrm{odom},1:t} + \boldsymbol{J}_{\mathrm{LoS},1:t} + \boldsymbol{J}_{\mathrm{SB},1:t} + \boldsymbol{J}_{\mathrm{DB},1:t}. \label{eq:S_crlb_sum}
\end{align}
Let $\boldsymbol{S}_{0,k}$, $\boldsymbol{S}_{\mathrm{m},k}$, $\boldsymbol{S}_{1,i,k}$, and $\boldsymbol{S}_{2,uv,k}$ denote binary selection matrices extracting $\boldsymbol{s}_k$, $[\boldsymbol{s}_{k-1}^{\mathsf{T}}, \boldsymbol{s}_k^{\mathsf{T}}]^{\mathsf{T}}$, $[\boldsymbol{s}_k^{\mathsf{T}}, \boldsymbol{x}_i^{\mathsf{T}}]^{\mathsf{T}}$, and $[\boldsymbol{s}_k^{\mathsf{T}}, \boldsymbol{x}_u^{\mathsf{T}}, \boldsymbol{x}_v^{\mathsf{T}}]^{\mathsf{T}}$ from $\boldsymbol{q}_{1:t}$, respectively. Using the motion Jacobian $\boldsymbol{F}_{k-1}$ in~\eqref{eq:S_motion_F} and the observation Jacobians $\boldsymbol{H}_{0,k}$, $\boldsymbol{H}_{1,i,k} \triangleq [\boldsymbol{H}_{1,i,k,\mathrm{S}}, \boldsymbol{H}_{1,i,k,\mathrm{L}}] \in \mathbb{R}^{3\times 5}$, and $\boldsymbol{H}_{2,uv,k} \triangleq [\boldsymbol{H}_{2,uv,k,\mathrm{S}}, \boldsymbol{H}_{2,uv,k,\mathrm{L}}] \in \mathbb{R}^{3\times 7}$ from Appendix~\ref{app:A}, the four FIM terms are given by
\begin{align}
\boldsymbol{J}_{\mathrm{odom},1:t} &= \boldsymbol{S}_{0,0}^{\mathsf{T}}\boldsymbol{P}_{0,\mathrm{prior}}^{-1}\boldsymbol{S}_{0,0} + \sum_{k=1}^{t} \boldsymbol{S}_{\mathrm{m},k}^{\mathsf{T}}\,\boldsymbol{G}_{k-1}^{\mathsf{T}}\boldsymbol{Q}_{k-1}^{-1}\boldsymbol{G}_{k-1}\,\boldsymbol{S}_{\mathrm{m},k}, \qquad \boldsymbol{G}_{k-1} \triangleq \big[-\boldsymbol{F}_{k-1},\, \boldsymbol{I}_3\big], \label{eq:S_crlb_odom} \\
\boldsymbol{J}_{\mathrm{LoS},1:t} &= \sum_{k=1}^{t} \omega_{0,k}\,\boldsymbol{S}_{0,k}^{\mathsf{T}}\boldsymbol{H}_{0,k}^{\mathsf{T}}\boldsymbol{R}_{\mathrm{LoS}}^{-1}\boldsymbol{H}_{0,k}\boldsymbol{S}_{0,k}, \label{eq:S_crlb_los} \\
\boldsymbol{J}_{\mathrm{SB},1:t} &= \sum_{k=1}^{t}\sum_{i=1}^{n_{\mathrm{vis}}} \omega_{1,i,k}\,\boldsymbol{S}_{1,i,k}^{\mathsf{T}}\boldsymbol{H}_{1,i,k}^{\mathsf{T}}\boldsymbol{R}_{\psi_i}^{-1}\boldsymbol{H}_{1,i,k}\boldsymbol{S}_{1,i,k}, \label{eq:S_crlb_sb} \\
\boldsymbol{J}_{\mathrm{DB},1:t} &= \sum_{k=1}^{t}\sum_{\substack{u,v=1,\, u\neq v\\ (\psi_u,\psi_v)\in\mathcal{T}_2}}^{n_{\mathrm{obs}}} \omega_{2,uv,k}\,\boldsymbol{S}_{2,uv,k}^{\mathsf{T}}\boldsymbol{H}_{2,uv,k}^{\mathsf{T}}\boldsymbol{R}_{\psi_u,\psi_v}^{-1}\boldsymbol{H}_{2,uv,k}\boldsymbol{S}_{2,uv,k}. \label{eq:S_crlb_db}
\end{align}

In~\eqref{eq:S_crlb_odom}, $\boldsymbol{J}_{\mathrm{odom},1:t}$ captures the trajectory information from the initial prior and odometry increments, with zero landmark block $[\boldsymbol{J}_{\mathrm{odom},1:t}]_{\boldsymbol{x},\boldsymbol{x}} = \boldsymbol{0}$ since landmarks carry no prior distribution (unlike unconditional Bayesian bounds averaged over random state priors~\cite{Tichavsky1998TSP}). The weights $\omega_{0,k}, \omega_{1,i,k}, \omega_{2,uv,k} \in [0, 1]$ equal the detection probabilities $p_{\mathrm{D},0}(\boldsymbol{s}_k)$, $p_{\mathrm{D},1}(\tilde{\boldsymbol{x}}_i, \boldsymbol{s}_k)$, and $p_{\mathrm{D},2}(\tilde{\boldsymbol{x}}_u, \tilde{\boldsymbol{x}}_v, \boldsymbol{s}_k)$ under a parametric field-of-view model~\cite{Deutschmann2025Asilomar}, and take binary visibility values $\omega \in \{0, 1\}$ under the ray-tracing channel in~\cite[Sec.~VI]{MainPaper}. Because data association and landmark types are known without clutter, $\boldsymbol{J}_{1:t}$ upper-bounds the Fisher information under unknown data association ($\boldsymbol{J}_{\mathrm{known\text{-}DA}} \succeq \boldsymbol{J}_{\mathrm{unknown\text{-}DA}}$), so its inverse serves as a reference estimation-error lower bound for unbiased estimators under standard regularity conditions~\cite{Kay1993}.

\subsubsection{Identifiable Subspace Screening and the Three Error Bounds}
Because no prior distribution is imposed on landmarks, combining with the identifiability conditions of Appendix~\ref{app:D}-A defines the identifiable landmark subset over prefix $1{:}t$ as
\begin{align}
\mathcal{I}_{\mathrm{solv}}(\boldsymbol{J}_{1:t}) &\triangleq \mathcal{I}_{\mathrm{solv},1:t}^{(1)} \cup \Big\{v > n_{\mathrm{vis}} : \lambda_{\min}\!\big([\boldsymbol{J}_{1:t}]_{\boldsymbol{x}_v,\boldsymbol{x}_v}\big) > 0\text{ with an anchor } u \in \mathcal{I}_{\mathrm{solv},1:t}^{(1)}\Big\}, \label{eq:S_crlb_solv}
\end{align}
where $\mathcal{I}_{\mathrm{solv},1:t}^{(1)} \triangleq \big\{i \le n_{\mathrm{vis}} : \lambda_{\min}\!\big([\boldsymbol{J}_{\mathrm{SB},1:t}]_{\boldsymbol{x}_i,\boldsymbol{x}_i}\big) > 0\big\}$ denotes visible landmarks with full-rank 1-bounce information. Let $\bar{\boldsymbol{J}}_{1:t}$ denote the nonsingular FIM constructed from observations involving $\mathcal{I}_{\mathrm{solv}}(\boldsymbol{J}_{1:t})$ and restricted to $(\boldsymbol{s}_{0:t}, \mathcal{I}_{\mathrm{solv}}(\boldsymbol{J}_{1:t}))$. The three CRLBs in~\cite[Sec.~VI]{MainPaper} are defined as follows:
\begin{enumerate}
    \item \emph{2-b Batch CRLB:} Set $t = K$, form $\boldsymbol{J}_{1:K}^{(2\text{-}\mathrm{b})} \triangleq \boldsymbol{J}_{1:K}$, and invert $\boldsymbol{\Sigma}^{(2\text{-}\mathrm{b})} \triangleq (\bar{\boldsymbol{J}}_{1:K}^{(2\text{-}\mathrm{b})})^{-1}$ (which covers the $n_{\mathrm{obs}}$ observable landmarks over the full batch in~\cite[Sec.~VI]{MainPaper}). The per-scan position error bound (PEB), heading error bound (HEB), and landmark error bound (LEB) are
    \begin{align}
    \mathrm{PEB}_k^{(2\text{-}\mathrm{b})} = \sqrt{\operatorname{tr}\!\big([\boldsymbol{\Sigma}^{(2\text{-}\mathrm{b})}]_{\boldsymbol{p}_k,\boldsymbol{p}_k}\big)}, \qquad
    \mathrm{HEB}_k^{(2\text{-}\mathrm{b})} = \sqrt{[\boldsymbol{\Sigma}^{(2\text{-}\mathrm{b})}]_{\alpha_k,\alpha_k}}, \qquad
    \mathrm{LEB}_i^{(2\text{-}\mathrm{b})} = \sqrt{\operatorname{tr}\!\big([\boldsymbol{\Sigma}^{(2\text{-}\mathrm{b})}]_{\boldsymbol{x}_i,\boldsymbol{x}_i}\big)}. \label{eq:S_crlb_2b}
    \end{align}

    \item \emph{1-b Batch CRLB:} Omit the 2-bounce contributions ($\boldsymbol{J}_{\mathrm{DB},1:K} = \boldsymbol{0}$) and occluded landmarks $\boldsymbol{x}_2 \triangleq \boldsymbol{x}_{n_{\mathrm{vis}}+1:n_{\mathrm{obs}}}$, forming the 1-bounce batch FIM on $\boldsymbol{q}_1 \triangleq [\boldsymbol{s}_{0:K}^{\mathsf{T}}, \boldsymbol{x}_{1:n_{\mathrm{vis}}}^{\mathsf{T}}]^{\mathsf{T}}$:
    \begin{align}
    \boldsymbol{J}_{1:K}^{(1\text{-}\mathrm{b})} \triangleq \big[\boldsymbol{J}_{\mathrm{odom},1:K} + \boldsymbol{J}_{\mathrm{LoS},1:K} + \boldsymbol{J}_{\mathrm{SB},1:K}\big]_{\boldsymbol{q}_1,\boldsymbol{q}_1}, \qquad
    \boldsymbol{\Sigma}^{(1\text{-}\mathrm{b})} \triangleq \big(\bar{\boldsymbol{J}}_{1:K}^{(1\text{-}\mathrm{b})}\big)^{-1}. \label{eq:S_crlb_1b}
    \end{align}
    Extracting the diagonal blocks of $\boldsymbol{\Sigma}^{(1\text{-}\mathrm{b})}$ as in~\eqref{eq:S_crlb_2b} yields $\mathrm{PEB}_k^{(1\text{-}\mathrm{b})}$, $\mathrm{HEB}_k^{(1\text{-}\mathrm{b})}$, and $\mathrm{LEB}_i^{(1\text{-}\mathrm{b})}$ (in deep-NLoS scans where $\omega_{0,k} = \omega_{1,i,k} = 0$, $\boldsymbol{J}_{1:K}^{(1\text{-}\mathrm{b})}$ relies solely on odometry transitions).

    \item \emph{2-b Filtering CRLB:} For causal estimators conditioned on $\mathcal{Z}_{1:k}$, evaluate $\boldsymbol{J}_{1:k}^{(2\text{-}\mathrm{b})} \triangleq \boldsymbol{J}_{1:k}$ for $k = 1, \dots, K$, restrict to $\boldsymbol{s}_{0:k}$ and $\mathcal{I}_{\mathrm{solv}}(\boldsymbol{J}_{1:k}^{(2\text{-}\mathrm{b})})$, invert $\boldsymbol{\Sigma}_{1:k}^{\mathrm{filt}} \triangleq (\bar{\boldsymbol{J}}_{1:k}^{(2\text{-}\mathrm{b})})^{-1}$, and extract the scan-$k$ pose block:
    \begin{align}
    \mathrm{PEB}_k^{\mathrm{filt}} = \sqrt{\operatorname{tr}\!\big([\boldsymbol{\Sigma}_{1:k}^{\mathrm{filt}}]_{\boldsymbol{p}_k,\boldsymbol{p}_k}\big)}, \qquad
    \mathrm{HEB}_k^{\mathrm{filt}} = \sqrt{[\boldsymbol{\Sigma}_{1:k}^{\mathrm{filt}}]_{\alpha_k,\alpha_k}}. \label{eq:S_crlb_filt}
    \end{align}
    Eliminating past poses $\boldsymbol{s}_{0:k-1}$ and landmarks from $\bar{\boldsymbol{J}}_{1:k}^{(2\text{-}\mathrm{b})}$ via the Schur complement yields the recursive filtering information matrix evaluated along the deterministic trajectory~\cite{Tichavsky1998TSP}. Landmarks becoming observable over time are included progressively through $\mathcal{I}_{\mathrm{solv}}(\boldsymbol{J}_{1:k}^{(2\text{-}\mathrm{b})})$.
\end{enumerate}
For a scan subset $\mathcal{K} \subseteq \{1, \dots, K\}$ (the full trajectory or a visibility stage R1--R3), the root-mean-square (RMS) aggregated bounds in~\cite[Figs.~7 and~8]{MainPaper} are
\begin{align}
\mathrm{PEB}_{\mathcal{K}}^{(\cdot)} = \sqrt{\frac{1}{|\mathcal{K}|}\sum_{k\in\mathcal{K}} \big(\mathrm{PEB}_k^{(\cdot)}\big)^2}, \qquad
\mathrm{HEB}_{\mathcal{K}}^{(\cdot)} = \sqrt{\frac{1}{|\mathcal{K}|}\sum_{k\in\mathcal{K}} \big(\mathrm{HEB}_k^{(\cdot)}\big)^2}. \label{eq:S_crlb_rms}
\end{align}

\subsubsection{Information Monotonicity (2-Bounce Gain and Batch-vs.-Filtering Ordering)}
\paragraph{2-Bounce Information Gain With Unknown Occluded Landmarks}
Partition $\boldsymbol{q}_{1:K} = [\boldsymbol{q}_1^{\mathsf{T}}, \boldsymbol{x}_2^{\mathsf{T}}]^{\mathsf{T}}$ into $\boldsymbol{q}_1 = [\boldsymbol{s}_{0:K}^{\mathsf{T}}, \boldsymbol{x}_{1:n_{\mathrm{vis}}}^{\mathsf{T}}]^{\mathsf{T}}$ and occluded landmarks $\boldsymbol{x}_2 = \boldsymbol{x}_{n_{\mathrm{vis}}+1:n_{\mathrm{obs}}}$, and split $\boldsymbol{J}_{\mathrm{DB},1:K}$ into visible-only 2-bounce information $\operatorname{blkdiag}(\boldsymbol{D}_{\mathrm{vis}}, \boldsymbol{0})$ ($\boldsymbol{D}_{\mathrm{vis}} \succeq \boldsymbol{0}$) and occluded-involving information $\boldsymbol{J}_{\mathrm{DB}}^{\mathrm{occ}} = \begin{bmatrix} \boldsymbol{A}_{11} & \boldsymbol{A}_{12} \\ \boldsymbol{A}_{12}^{\mathsf{T}} & \boldsymbol{A}_{22} \end{bmatrix} \succeq \boldsymbol{0}$ (with $\boldsymbol{J}_{22} = \boldsymbol{A}_{22} \succ \boldsymbol{0}$ under Appendix~\ref{app:D}-A):
\begin{align}
\boldsymbol{J}_{1:K}^{(2\text{-}\mathrm{b})} = \begin{bmatrix} \boldsymbol{J}_{1:K}^{(1\text{-}\mathrm{b})} + \boldsymbol{D}_{\mathrm{vis}} + \boldsymbol{A}_{11} & \boldsymbol{A}_{12} \\ \boldsymbol{A}_{12}^{\mathsf{T}} & \boldsymbol{A}_{22} \end{bmatrix}. \label{eq:S_crlb_part}
\end{align}
Taking the Schur complement of $\boldsymbol{x}_2$ gives the marginal FIM on $\boldsymbol{q}_1$, $\tilde{\boldsymbol{J}}_{11}^{(2\text{-}\mathrm{b})} \triangleq \boldsymbol{J}_{11} - \boldsymbol{J}_{12}\boldsymbol{J}_{22}^{-1}\boldsymbol{J}_{12}^{\mathsf{T}}$, which satisfies $\tilde{\boldsymbol{J}}_{11}^{(2\text{-}\mathrm{b})} - \boldsymbol{J}_{1:K}^{(1\text{-}\mathrm{b})} = \boldsymbol{D}_{\mathrm{vis}} + (\boldsymbol{A}_{11} - \boldsymbol{A}_{12}\boldsymbol{A}_{22}^{-1}\boldsymbol{A}_{12}^{\mathsf{T}})$. For a vector $\boldsymbol{v}$ conformable with $\boldsymbol{q}_1$, positive semidefiniteness of $\boldsymbol{J}_{\mathrm{DB}}^{\mathrm{occ}}$ implies
\begin{align}
\boldsymbol{v}^{\mathsf{T}}\big(\boldsymbol{A}_{11} - \boldsymbol{A}_{12}\boldsymbol{A}_{22}^{-1}\boldsymbol{A}_{12}^{\mathsf{T}}\big)\boldsymbol{v}
= \begin{bmatrix} \boldsymbol{v}^{\mathsf{T}} & -\boldsymbol{v}^{\mathsf{T}}\boldsymbol{A}_{12}\boldsymbol{A}_{22}^{-1} \end{bmatrix}
\begin{bmatrix} \boldsymbol{A}_{11} & \boldsymbol{A}_{12} \\ \boldsymbol{A}_{12}^{\mathsf{T}} & \boldsymbol{A}_{22} \end{bmatrix}
\begin{bmatrix} \boldsymbol{v} \\ -\boldsymbol{A}_{22}^{-1}\boldsymbol{A}_{12}^{\mathsf{T}}\boldsymbol{v} \end{bmatrix} \ge 0. \label{eq:S_crlb_psd}
\end{align}
Hence $\tilde{\boldsymbol{J}}_{11}^{(2\text{-}\mathrm{b})} \succeq \boldsymbol{J}_{1:K}^{(1\text{-}\mathrm{b})}$ and $(\tilde{\boldsymbol{J}}_{11}^{(2\text{-}\mathrm{b})})^{-1} \preceq (\boldsymbol{J}_{1:K}^{(1\text{-}\mathrm{b})})^{-1}$ whenever nonsingular, indicating that, in terms of Fisher information, incorporating multi-scan 3D 2-bounce measurements does not increase the theoretical PEB, HEB, or visible-landmark LEBs even when unknown occluded landmarks $\boldsymbol{x}_2$ are jointly estimated.

\paragraph{Batch-vs.-Filtering Ordering and Terminal Equality}
For scan $k \in \{1, \dots, K\}$, partition the identifiable full-batch variables into the causal prefix subset $\boldsymbol{q}_{\le k}$ (comprising $\boldsymbol{s}_{0:k}$ and $\mathcal{I}_{\mathrm{solv}}(\boldsymbol{J}_{1:k}^{(2\text{-}\mathrm{b})})$) and the complementary future subset $\boldsymbol{q}_{>k}$ (comprising $\boldsymbol{s}_{k+1:K}$ and landmarks first becoming identifiable after scan $k$). Since motion factors and multipath observations over scans $k+1{:}K$, together with earlier prefix observations involving landmarks in $\boldsymbol{q}_{>k}$, contribute a PSD increment $\boldsymbol{\Delta}\boldsymbol{J}_{>k} = \begin{bmatrix} \boldsymbol{B}_{11} & \boldsymbol{B}_{12} \\ \boldsymbol{B}_{12}^{\mathsf{T}} & \boldsymbol{B}_{22} \end{bmatrix} \succeq \boldsymbol{0}$ (with $\boldsymbol{B}_{22} \succ \boldsymbol{0}$ because $\bar{\boldsymbol{J}}_{1:K}^{(2\text{-}\mathrm{b})} \succ \boldsymbol{0}$), eliminating $\boldsymbol{q}_{>k}$ via the Schur complement yields
\begin{align}
\tilde{\boldsymbol{J}}_{\le k}^{(2\text{-}\mathrm{b})} = \bar{\boldsymbol{J}}_{1:k}^{(2\text{-}\mathrm{b})} + \big(\boldsymbol{B}_{11} - \boldsymbol{B}_{12}\boldsymbol{B}_{22}^{-1}\boldsymbol{B}_{12}^{\mathsf{T}}\big) \succeq \bar{\boldsymbol{J}}_{1:k}^{(2\text{-}\mathrm{b})}, \label{eq:S_crlb_bf}
\end{align}
by the same quadratic-form argument as~\eqref{eq:S_crlb_psd}. Inverting both sides gives
\begin{align}
\mathrm{PEB}_k^{(2\text{-}\mathrm{b})} \le \mathrm{PEB}_k^{\mathrm{filt}}, \qquad \mathrm{HEB}_k^{(2\text{-}\mathrm{b})} \le \mathrm{HEB}_k^{\mathrm{filt}}, \qquad \forall k \in \{1, \dots, K\}, \label{eq:S_crlb_order}
\end{align}
with equality at the terminal scan $k = K$ ($\boldsymbol{q}_{>K} = \emptyset$, $\tilde{\boldsymbol{J}}_{\le K}^{(2\text{-}\mathrm{b})} = \bar{\boldsymbol{J}}_{1:K}^{(2\text{-}\mathrm{b})}$), matching the convergence of the 2-b Batch and 2-b Filtering CRLB curves at $k = K$ in~\cite[Fig.~7(a)]{MainPaper}. \hfill$\blacksquare$

\section{Algorithm Pseudocode} \label{app:E}

This appendix summarizes the procedural pseudocode for the three algorithms in~\cite[Secs.~IV and~V]{MainPaper}: double-bounce data association and landmark type sampling (Algorithm~\ref{alg:S_mcmc}), warm-start pre-clustering (Algorithm~\ref{alg:S_warmstart}), and the overall double-bounce multi-type batch radio SLAM procedure (Algorithm~\ref{alg:S_overall}).

\begin{ieeealg}{Double-Bounce DA and Landmark Type Sampling}{alg:S_mcmc}
\textbf{Input:} $\mathcal{Z}_{1:K}$, $\boldsymbol{s}_{0:K}$, initial $\mathcal{A}^{\mathrm{in}}$, sweep count $T_{\mathrm{sweep}}$, burn-in indicator $b_{\mathrm{burn}} \in \{0, 1\}$.\\
\textbf{Output:} Augmented DA sample $\tilde{\mathcal{A}} = (\mathcal{A}^\star, \boldsymbol{\psi}^\star)$.
\begin{algsteps}
    \item[1:] Set $\mathcal{A} \leftarrow \mathcal{A}^{\mathrm{in}}$;
    \item[2:] \textbf{for} $sw = 1 : T_{\mathrm{sweep}}$ \textbf{do}
    \item[3:] \quad Run single-measurement Gibbs sweep via~\cite[Eqs.~(27)--(28)]{MainPaper};
    \item[4:] \quad Run cluster-relabeling Gibbs sweep via~\cite[Eq.~(29)]{MainPaper};
    \item[5:] \quad \textbf{if} $b_{\mathrm{burn}} = 1$ \textbf{then} Run merge-and-split MH sweep (\cite[Sec.~IV-B, Step~3]{MainPaper}); \textbf{end if}
    \item[6:] \textbf{end for}
    \item[7:] Set $\mathcal{A}^\star \leftarrow \mathcal{A}$ and sample $\boldsymbol{\psi}^\star$ via~\cite[Eqs.~(31)--(32)]{MainPaper}.
\end{algsteps}
\end{ieeealg}

\begin{ieeealg}{Warm-Start Pre-Clustering}{alg:S_warmstart}
\textbf{Input:} $\mathcal{Z}_{1:K}$, $\boldsymbol{s}_{0:K}^{(0)}$, thresholds $(\tau_{\phi,\mathrm{SB}}, \gamma_{\mathrm{reproj}}, r_{\mathrm{clust}})$.\\
\textbf{Output:} Initial DA partition $\mathcal{A}^{(0)}$.
\begin{algsteps}
    \item[1:] Gate LoS measurements per scan into $\mathcal{Z}_0$;
    \item[2:] Split remaining measurements via~\cite[Eq.~(33)]{MainPaper} into 1-bounce candidates $\mathcal{P}_{\mathrm{SB}}$ ($e_{\phi,\mathrm{SB}} < \tau_{\phi,\mathrm{SB}}$) and 2-bounce candidates $\mathcal{P}_{\mathrm{DB}}$;
    \item[3:] Cluster $\mathcal{P}_{\mathrm{SB}}$ within radius $r_{\mathrm{clust}}$ and select visible 1-bounce clusters $\{\mathcal{Z}_u^{(1)}\}$ via $\mathcal{L}_1$; move leftovers to $\mathcal{P}_{\mathrm{DB}}$;
    \item[4:] Match $\mathcal{P}_{\mathrm{DB}}$ across visible cluster pairs via $\mathcal{L}_2$ to extract visible 2-bounce clusters $\{\mathcal{Z}_{uv}^{(2)}\}$ (Appendix~\ref{app:C}-D);
    \item[5:] Invert and cluster $\mathcal{P}_{\mathrm{DB}}$ per visible anchor $u$, selecting occluded 2-bounce clusters via $\mathcal{L}_{2,\mathrm{occ}}$ (Appendix~\ref{app:C}-D and Table~\ref{tab:S_inv});
    \item[6:] Set remaining points in $\mathcal{P}_{\mathrm{DB}}$ as 1-bounce singletons and index $\mathcal{A}^{(0)}$.
\end{algsteps}
\end{ieeealg}

\begin{ieeealg}{Double-Bounce Multi-Type Batch Radio SLAM}{alg:S_overall}
\textbf{Input:} $\mathcal{Z}_{1:K}$, $\boldsymbol{u}_{0:K-1}$, priors, $N_{\mathrm{burn}}$, $\Gamma$, $T_{\mathrm{sweep}}$.\\
\textbf{Output:} $(\boldsymbol{\epsilon}_{0:K}, \boldsymbol{P}_{0:K})$ and multi-type PMB map.
\begin{algsteps}
    \item[1:] Propagate $\boldsymbol{s}_{0:K}^{(0)}$ via~\cite[Eq.~(1)]{MainPaper} and construct $\mathcal{A}^{(0)}$ via Algorithm~\ref{alg:S_warmstart};
    \item[2:] \textbf{for} $n = 1 : N_{\mathrm{burn}} + \Gamma$ \textbf{do}
    \item[3:] \quad Set $b_{\mathrm{burn}} \leftarrow \mathbb{I}(n \le N_{\mathrm{burn}})$ and draw $\tilde{\mathcal{A}}^{(n)} = (\mathcal{A}^{(n)}, \boldsymbol{\psi}^{(n)})$ from $(\boldsymbol{s}_{0:K}^{(n-1)}, \mathcal{A}^{(n-1)}, T_{\mathrm{sweep}}, b_{\mathrm{burn}})$ via Algorithm~\ref{alg:S_mcmc};
    \item[4:] \quad Minimize~\cite[Eq.~(36)]{MainPaper} to obtain $(\boldsymbol{\epsilon}_{0:K}^{(n)}, \boldsymbol{P}_{0:K}^{(n)}, \{(\boldsymbol{\mu}_i^{(n)}, \boldsymbol{C}_i^{(n)})\})$ (\cite[Sec.~V-A]{MainPaper}), and set $\boldsymbol{s}_{0:K}^{(n)} \leftarrow \boldsymbol{\epsilon}_{0:K}^{(n)}$;
    \item[5:] \textbf{end for}
    \item[6:] Fuse the last $\Gamma$ samples via~\cite[Eqs.~(38)--(40)]{MainPaper} to output final estimates.
\end{algsteps}
\end{ieeealg}

\noindent\emph{Implementation Note:} In Algorithm~\ref{alg:S_warmstart}, the default warm-start thresholds are $\tau_{\phi,\mathrm{SB}} = 8^\circ$ (expanding smoothly up to $15^\circ$ with a $1.5\sigma_\phi$ noise margin along the trajectory to accommodate dead-reckoning drift), $\gamma_{\mathrm{reproj}} = 25$, and $r_{\mathrm{clust}} \in [1.2, 1.8]\,\mathrm{m}$ ($1.6\,\mathrm{m}$ for VAs, $1.2\,\mathrm{m}$ for RPs, and $1.5\,\mathrm{m}$ and $1.8\,\mathrm{m}$ for occluded 2-bounce inversions in the RT simulation and 60~GHz dataset, respectively). Following standard simulated-annealing warm-up practice~\cite{Neal2001StatComput}, during the first $3\text{--}4$ burn-in DA sampling steps of Algorithm~\ref{alg:S_overall}, the measurement covariance $\boldsymbol{R}_\eta$ is multiplied by fixed decreasing factors ($[4.5, 2.7, 1.8, 1.3]$ in RT regimes G1/G3, $[1.5, 1.3, 1.2, 1.1]$ in G2/G4, and $[3.0, 2.0, 1.4]$ in the 60~GHz dataset), with the identical warm-up schedule applied to the 1-b Batch SLAM baseline, before strictly restoring nominal covariances for the remaining burn-in iterations, all GraphSLAM updates, and all $\Gamma$ post-burn-in samples.